\pdfoutput=1
\documentclass[sigconf,nonacm]{acmart}

\usepackage{CJKutf8}
\usepackage{tikz}
\usetikzlibrary{arrows.meta,positioning,fit,calc,shapes.geometric}
\usepackage{subcaption}
\usepackage{booktabs}
\usepackage{multirow}
\usepackage{enumitem}
\usepackage{balance}
\pdfmapfile{=gbsnu.map}
\pdfmapfile{=gkaiu.map}
\newcommand{\zh}[1]{\begingroup\normalfont\begin{CJK*}{UTF8}{gbsn}#1\end{CJK*}\endgroup}
\newcommand{\kai}[1]{\begingroup\normalfont\begin{CJK*}{UTF8}{gkai}#1\end{CJK*}\endgroup}

\definecolor{navy}{HTML}{0B2036}
\definecolor{gold}{HTML}{D3A557}
\definecolor{cream}{HTML}{F6EFE2}
\definecolor{rose}{HTML}{C74B5F}
\definecolor{meadow}{HTML}{5F9E5A}
\definecolor{sky}{HTML}{3F7FB3}
\definecolor{merred}{HTML}{D7102F}
\definecolor{jpurple}{HTML}{7B4FD8}
\definecolor{ygreen}{HTML}{3E9A4E}
\definecolor{dink}{HTML}{2B2A27}
\definecolor{dpaper}{HTML}{F6F5F1}
\definecolor{dred}{HTML}{E34948}
\usetikzlibrary{shadows,backgrounds}
\definecolor{sInk}{HTML}{1F2A37}
\definecolor{sSoft}{HTML}{5B6472}
\definecolor{sLine}{HTML}{8A93A3}
\definecolor{sNavy}{HTML}{2F5D8A}
\definecolor{sGreen}{HTML}{1B8F6A}
\definecolor{sVio}{HTML}{7A5FC0}
\definecolor{sRed}{HTML}{C2554F}
\definecolor{sGold}{HTML}{A9781F}
\definecolor{sOrF}{HTML}{FDEBD3}\definecolor{sOrB}{HTML}{E59A45}
\definecolor{sYeF}{HTML}{FFF6CC}\definecolor{sYeB}{HTML}{D9B13B}
\definecolor{sTeF}{HTML}{DDF5F0}\definecolor{sTeB}{HTML}{2BB7A3}
\definecolor{sPuF}{HTML}{E9E2F8}\definecolor{sPuB}{HTML}{8E6CC9}
\definecolor{sBlF}{HTML}{EAF2FB}
\definecolor{sCrF}{HTML}{FBF7EE}\definecolor{sCrB}{HTML}{C9A35A}
\definecolor{sReF}{HTML}{FBE6E4}
\definecolor{sPnBlF}{HTML}{F4F8FD}\definecolor{sPnBlB}{HTML}{9DB7D5}
\definecolor{sPnGrF}{HTML}{F1F9F5}\definecolor{sPnGrB}{HTML}{8CC9B1}
\definecolor{sPnGoF}{HTML}{FBF8F1}\definecolor{sPnGoB}{HTML}{D2B77A}
\tikzset{
  sflat/.style={draw=sLine, line width=0.7pt, fill=white, rounded corners=3pt, align=center, text=sInk},
  scard/.style={sflat, drop shadow={opacity=0.2, shadow xshift=0.6pt, shadow yshift=-0.9pt}},
  sorange/.style={fill=sOrF, draw=sOrB}, syellow/.style={fill=sYeF, draw=sYeB},
  steal/.style={fill=sTeF, draw=sTeB}, spurple/.style={fill=sPuF, draw=sPuB},
  sblue/.style={fill=sBlF, draw=sNavy}, scream/.style={fill=sCrF, draw=sCrB},
  sredc/.style={fill=sReF, draw=sRed}, snavyc/.style={fill=white, draw=sNavy},
  spanel/.style={draw=sPnBlB, fill=sPnBlF, dashed, line width=0.7pt, rounded corners=4pt},
  spill/.style={fill=sNavy, text=white, font=\tiny\sffamily\bfseries, rounded corners=1.3mm, inner xsep=4pt, inner ysep=1.5pt, minimum height=2.8mm, text height=1.25ex, text depth=0.3ex},
  sarr/.style={-{Triangle[length=1.5mm, width=1.4mm]}, sNavy, line width=1pt},
}
\newcommand{\dsub}[1]{{\tiny\color{sSoft}#1}}

\newsavebox{\heartbox}
\sbox{\heartbox}{\tikz[baseline=-0.32em,x=1.5em,y=1.5em]{\fill[merred]
  (0,-0.42) .. controls (-0.58,0.02) and (-0.56,0.5) .. (-0.27,0.5)
  .. controls (-0.11,0.5) and (0,0.39) .. (0,0.27)
  .. controls (0,0.39) and (0.11,0.5) .. (0.27,0.5)
  .. controls (0.56,0.5) and (0.58,0.02) .. (0,-0.42);}}
\DeclareRobustCommand{\redheart}{\usebox{\heartbox}}
\DeclareRobustCommand{\ini}[1]{\textcolor{merred}{\textbf{#1}}}

\newcommand{\yt}{Yi Town}
\newcommand{\code}[1]{\textnormal{\texttt{\small #1}}}
\newcommand{\dc}[1]{\textsc{D#1}}

\renewcommand\footnotetextcopyrightpermission[1]{}
\setcopyright{none}

\begin{document}

\title[MERIDIAN]{\texorpdfstring{\textcolor{merred}{MERIDIAN}\hspace{0.18em}\redheart}{MERIDIAN}}
\subtitle{\ini{M}emory-\ini{E}ncoded \ini{R}ituals for \ini{I}ntimacy across \ini{D}istance
in an \ini{I}nteractive, \ini{A}synchronous \ini{N}eighbourhood}

\author{Jingyi Chen}
\authornote{Both authors approved this manuscript and the disclosure of the personal material it describes.}
\affiliation{\institution{Independent Researcher}\city{Shenzhen}\country{China}}
\email{jingyi@jingyifan.love}

\author{Yifan Yang}
\authornote{Corresponding author.}
\affiliation{\institution{Independent Researcher}\city{College Station, Texas}\country{USA}}
\email{yifan@jingyifan.love}

\renewcommand{\shortauthors}{J. Chen and Y. Yang}

\begin{abstract}
We report a research-through-design case of \yt{}, a bilingual browser game that one member of a
long-distance couple made for the other, and what its maker could verify about it without using
data from the recipient. Nine levels turn shared memories into tasks in a walkable pixel town;
within four weeks the site grew an infirmary for the days after a quarrel, a page for a booked
flight, and a request box that a pipeline of language models can answer. We reconstruct the
design decisions from the project history and test selected interaction rules against the
shipped software in twelve machine-run checks. After one online visit, 18 of 21 routes rendered
offline in desktop Chromium. The progress decoder refused all 140,840 single-character
substitutions and read 207 of 331,530 mutated codes as another state; 186 of those states
cannot arise in ordered play, and of 5,664 mutations of the 20 codes that play can produce it
accepted 24, all as such states. In a synthetic replay of the triage step, the model used in
deployment passed 2 of 80 requests labelled for a person and escalated 21 of 40 in-scope ones.
The case shows how such checks expose gaps between the rules a maker states and what a personal
artefact does. The recipient's experience remains an open question. The site is at
\url{https://jingyifan.love} and its source at \url{https://github.com/rayford295/yi-town}.
\end{abstract}



\begin{teaserfigure}
  \centering
  \begin{subfigure}[b]{0.25\textwidth}
    \includegraphics[width=\linewidth]{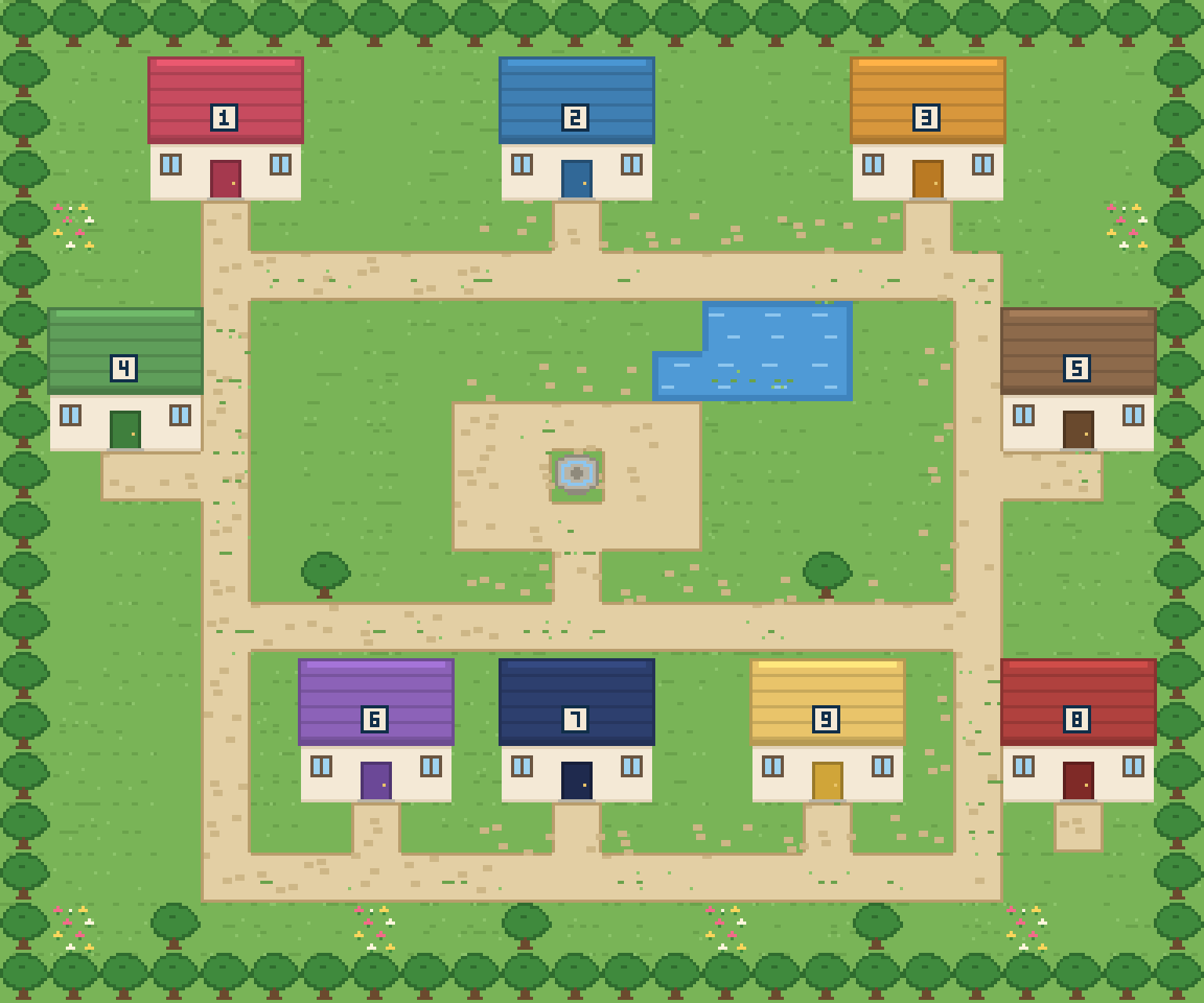}
    \caption{The whole town, drawn in code}\label{fig:teaser-map}
  \end{subfigure}\hfill
  \begin{subfigure}[b]{0.235\textwidth}
    \includegraphics[width=\linewidth]{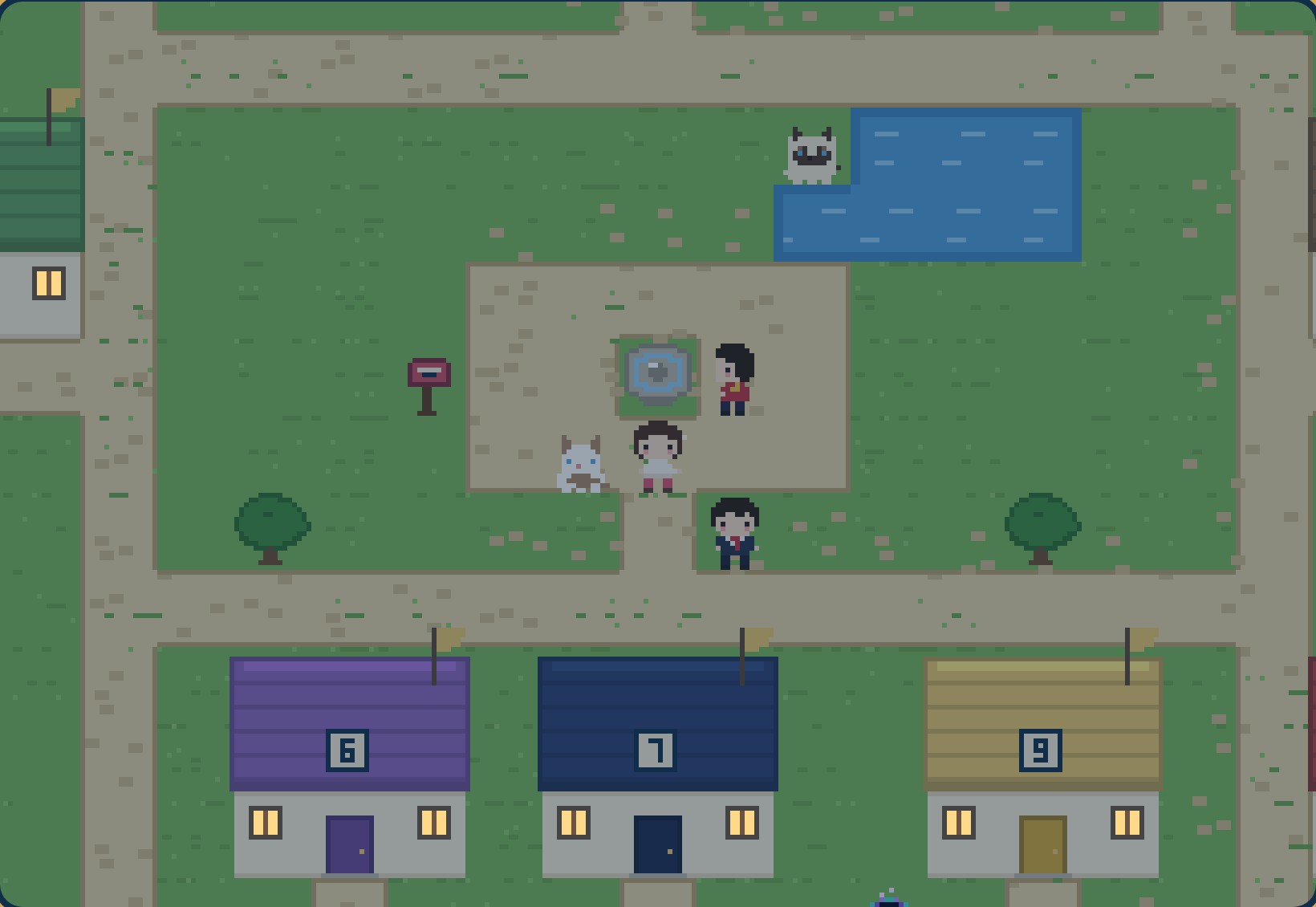}\\[2pt]
    \includegraphics[width=\linewidth]{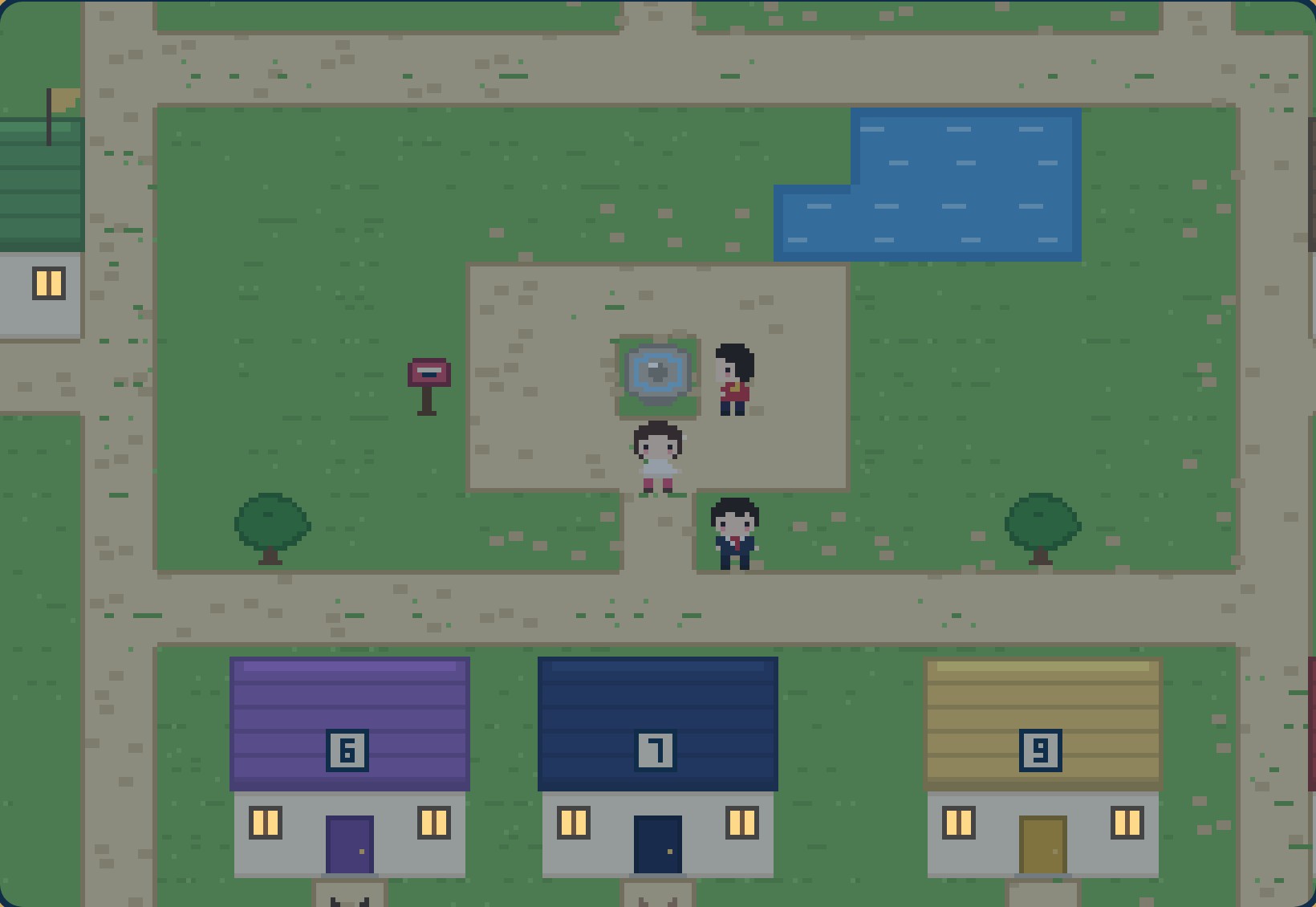}
    \caption{The square: guide, partner, cats}\label{fig:teaser-square}
  \end{subfigure}\hfill
  \begin{subfigure}[b]{0.245\textwidth}
    \includegraphics[width=\linewidth]{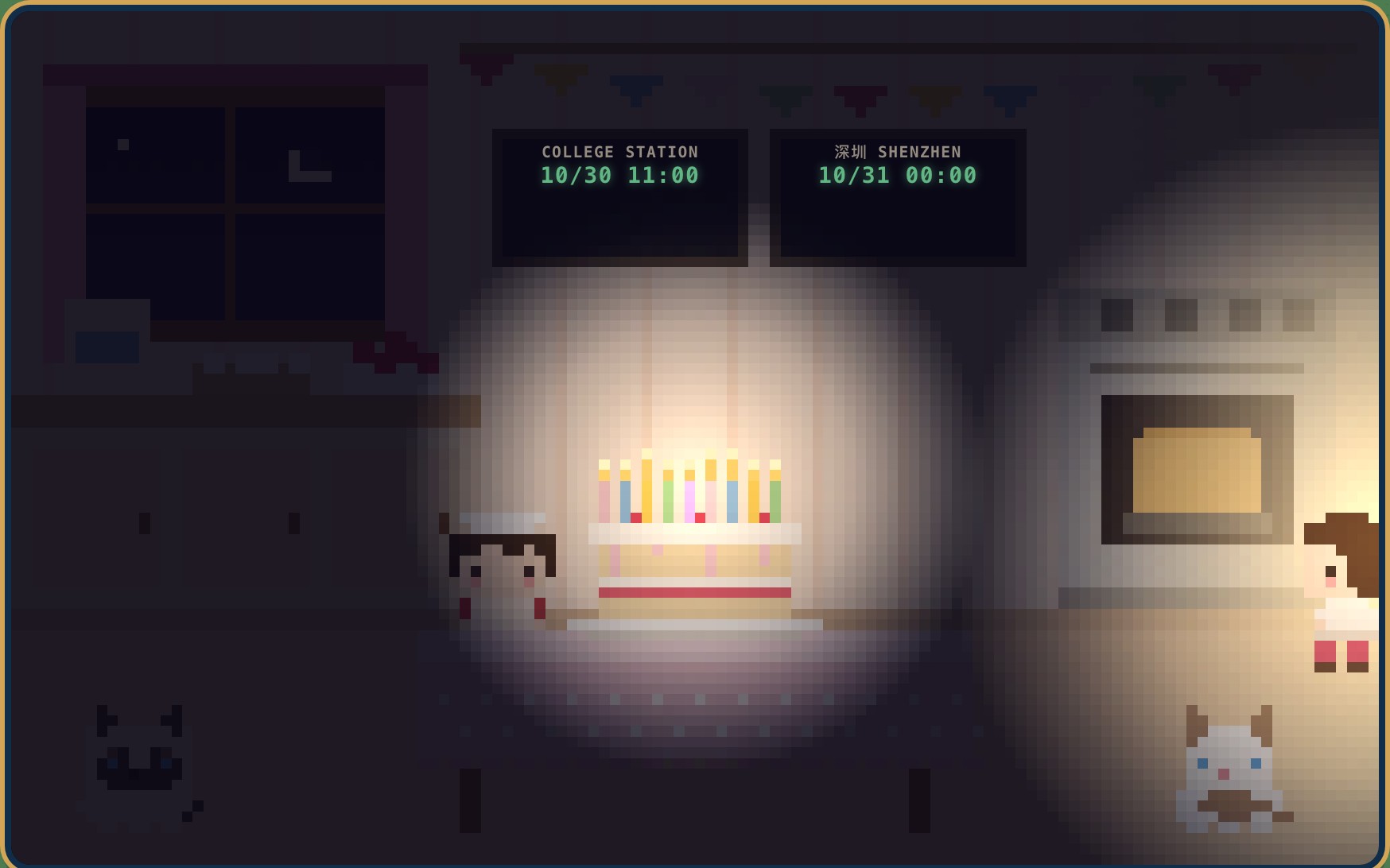}\\[2pt]
    \includegraphics[width=\linewidth]{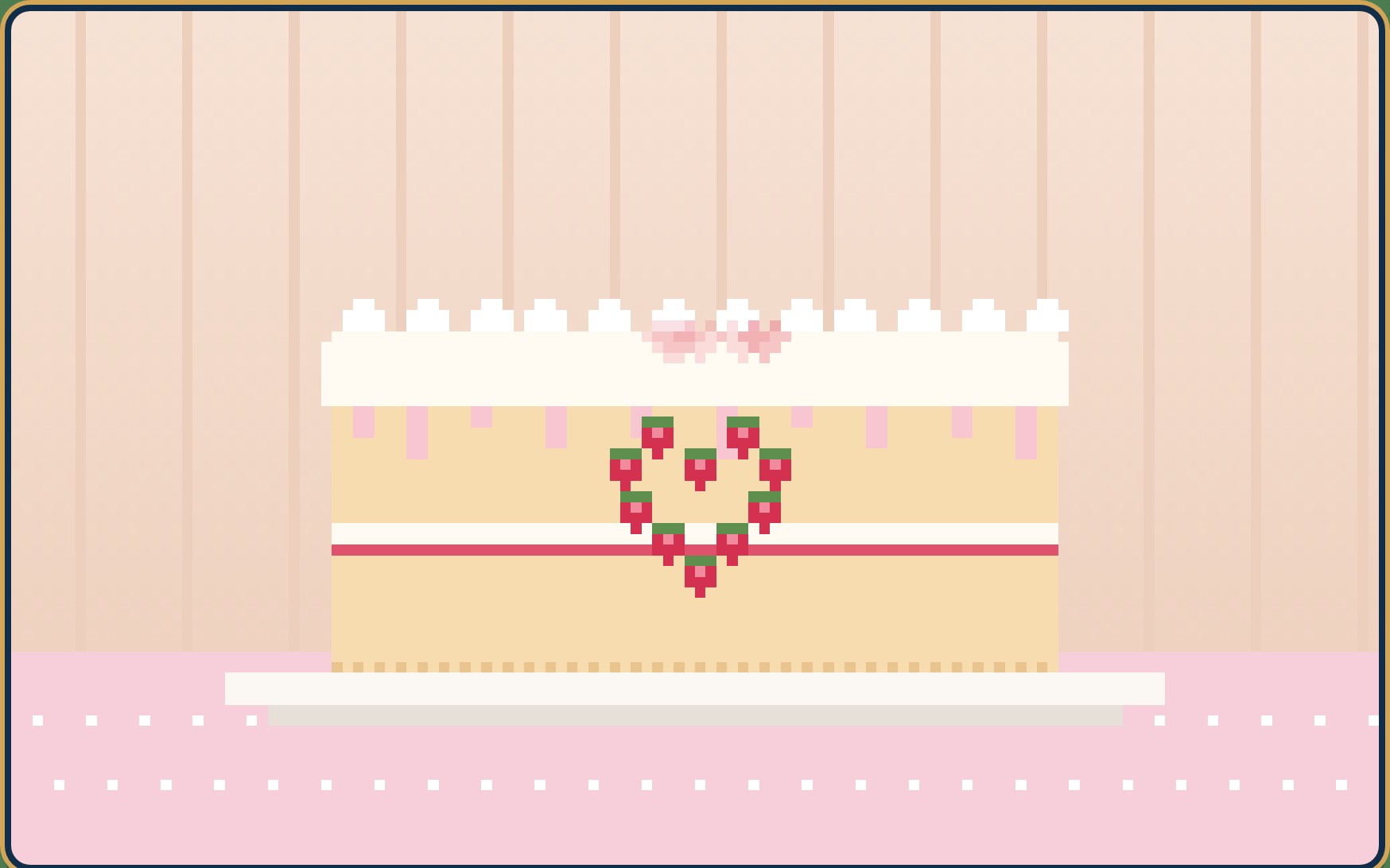}
    \caption{Prologue: the birthday cake}\label{fig:teaser-bday}
  \end{subfigure}\hfill
  \begin{subfigure}[b]{0.245\textwidth}
    \includegraphics[width=\linewidth]{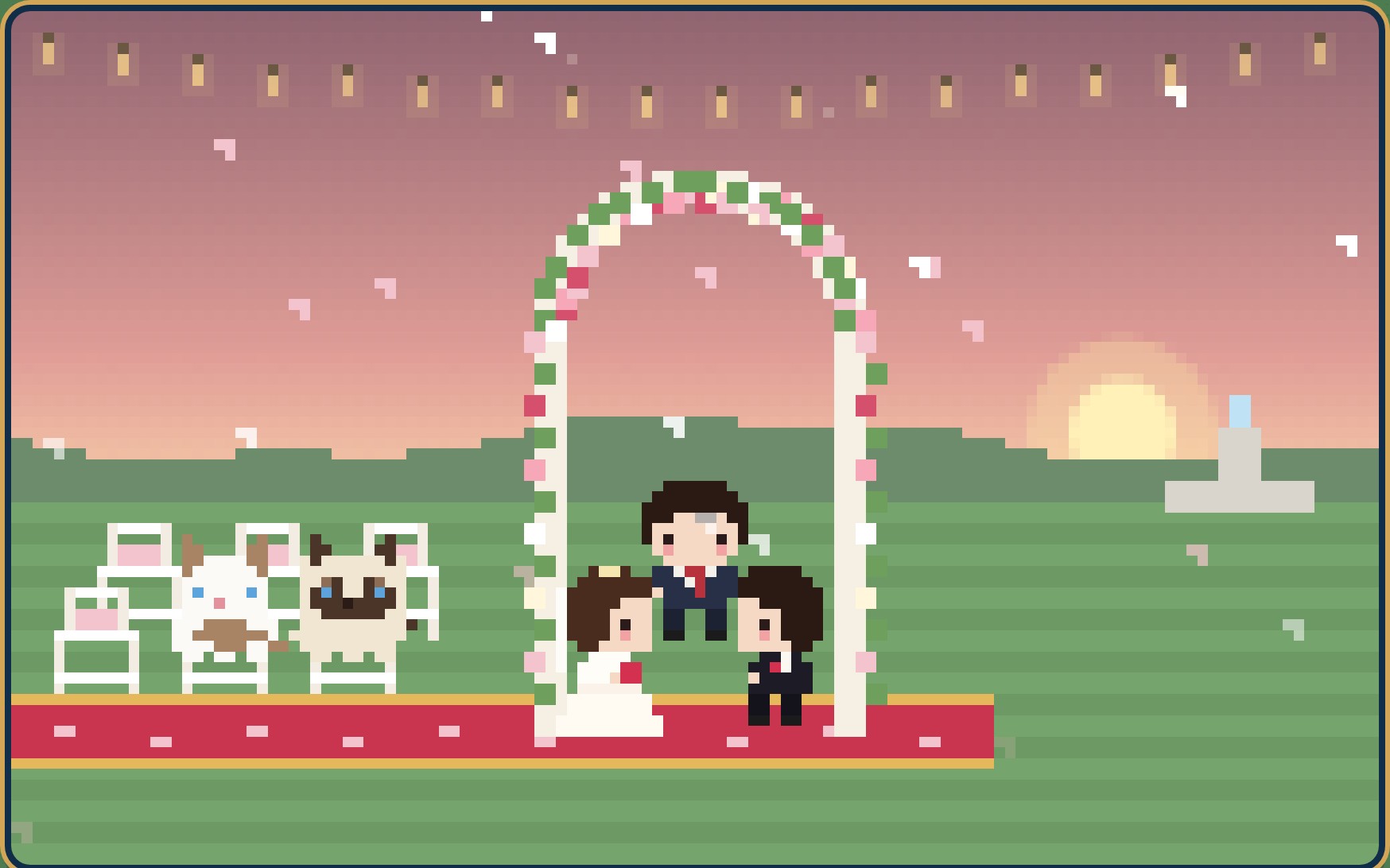}\\[2pt]
    \includegraphics[width=\linewidth]{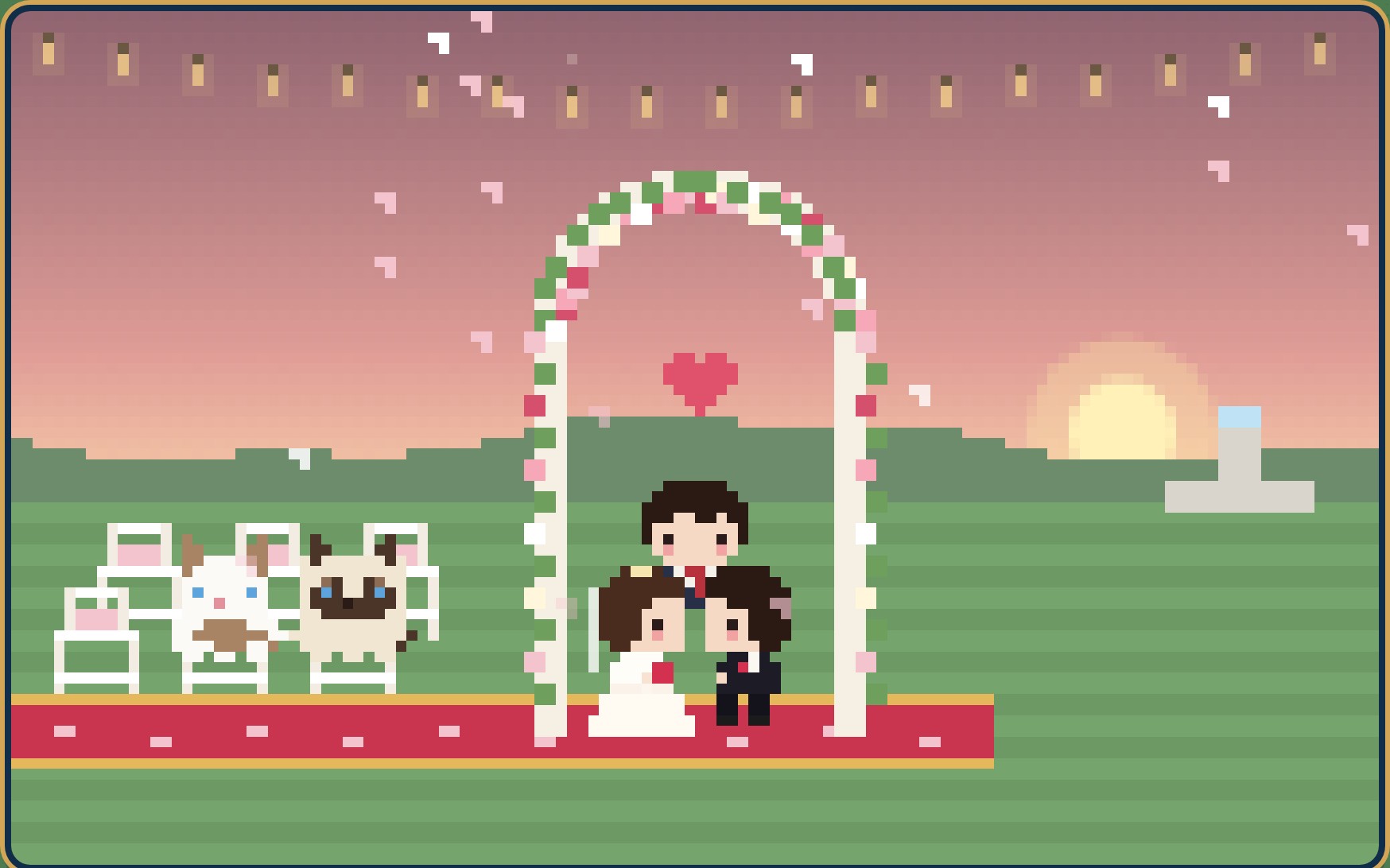}
    \caption{Finale: the wedding}\label{fig:teaser-wed}
  \end{subfigure}
  \caption{\yt{} at a glance. (a) The town map: a $24\times20$ grid of $16$-pixel tiles with nine
  numbered houses, one per level, painted entirely by code. (b) The fountain square in play, top
  after every level is cleared (gold flags, a cat sunbathing by the pond) and bottom halfway
  through; the player meets the partner, the guide Rayford and the cats. (c) The birthday
  prologue: nine candles in the dark at Shenzhen midnight, and ten strawberries set in a heart.
  (d) The wedding in the fountain garden, officiated by Rayford with the two cats as guests.}
  \Description{Four panels: a full pixel-art town map with nine coloured houses around a fountain;
  two close-ups of the square with small characters; two frames of a pixel kitchen with a lit
  birthday cake; two frames of a pixel wedding under a flower arch at sunset.}
  \label{fig:teaser}
\end{teaserfigure}

\maketitle
\hypersetup{pdftitle={MERIDIAN: Memory-Encoded Rituals for Intimacy across Distance in an Interactive, Asynchronous Neighbourhood}, pdfauthor={Jingyi Chen and Yifan Yang}}


\section{Introduction}\label{sec:intro}
Long-distance partners use communication technologies to maintain awareness, exchange affection
and share ordinary activities~\cite{vetere2005mediating,lottridge2009sharing,neustaedter2012intimacy}.
Some of the artefacts built for them are channels, such as a button whose meaning a couple
negotiates~\cite{kaye2006clicked} or a video link left open for
hours~\cite{neustaedter2012intimacy}. Others are closer to gifts and keepsakes: Hassenzahl et
al.\ list gift giving and memories among six strategies~\cite{hassenzahl2012love}, and Kirk et
al.\ built ritual objects for particular families~\cite{kirk2016ritual}. This paper is about a
gift of one specific kind: a game whose levels one partner wrote, alone, for the other to play.
Its author decides what she must do, in what order, and what is kept from her until the end, and
it has to work on a phone and a network he does not hold.

The case is \textsc{Meridian}, an account of \yt{}, a bilingual browser game that Yifan built for
Jingyi and himself. They live in College Station, Texas, and Shenzhen, China, thirteen or fourteen
hours apart. Nine levels turn shared songs, a first game of Gomoku, dated memories and classical
verse into tasks inside a walkable pixel town, framed by a birthday prologue and a fictional
wedding (Figure~\ref{fig:teaser}). Within four weeks of the handover the site grew three products
that are not levels: an infirmary for the days after a quarrel, which began with a remark by the
recipient; a page for her booked flight; and a request box whose messages a pipeline of two
language models can answer.

Earlier versions of this paper reported the artefact and its design record and stated that they
contained no evaluation. The usual study would ask the recipient, and first-person design
research has documented how roles, privacy and authorship blur when the participant is someone the
designer lives with~\cite{desjardins2018tensions}. We have not run that study. This version asks a narrower
question that the maker can answer with the software itself:
\begin{quote}
\emph{Which properties of a one-person software gift can its maker verify without using data
from the recipient?}
\end{quote}
We take the rules the design record states, find the code that implements each, and run checks
that can show the two apart. We call this \emph{recipient-free checking}. It is ordinary software
testing pointed at the claims of a design account, and we present it as a procedure followed in
one case, with no independent application yet. Three questions organise the twelve checks
(Figure~\ref{fig:summary}). \textbf{RQ1, rendering:} which routes render, and how fast, under
simulated network conditions and screen sizes? \textbf{RQ2, rules:} does the software implement
the rules the design states for hand-carried progress and for a secret ending?
\textbf{RQ3, delegation:} what does the step that decides whether a machine may answer pass and
escalate on synthetic requests?

The clearest finding concerns a rule that was stated and only half implemented. The design says
that levels open in order. The router enforces that order, and the progress code that carries a
player's state between two phones does not: its decoder accepts 512 sets of levels for each
player although ordered play yields ten. All 24 mistyped codes that the decoder accepted, out of
5,664 mutations of the codes play can produce, name states that play cannot reach, and a decoder
that refuses such states accepts none. The other checks bound what the account may claim. Routes
render under simulated network failures, and three of 21 do not render offline, among them the
level that carries the ending. The secrecy test catches the words it names and 2 of 72 synonyms
and oblique hints. The triage replay shows a model that passes few requests labelled for a person
and escalates half of the in-scope ones.

The paper makes three contributions:
\begin{enumerate}[leftmargin=*,itemsep=1pt,topsep=2pt]
  \item \emph{A design case.} An account of how autobiographical material became nine game
  levels, two ceremonies and three later products, with six situated commitments and seven
  design episodes reconstructed from the development record
  (Sections~\ref{sec:case} and~\ref{sec:artefact}; Appendices~\ref{sec:method} to~\ref{sec:log}).
  \item \emph{Checks of its stated rules.} Twelve machine-run checks of the shipped software,
  each with its failure criterion and the limit of what it shows, and eight discrepancies
  between the account and the artefact (Section~\ref{sec:eval}).
  \item \emph{A procedure from one case.} The steps we followed to derive checks from a design
  record without the recipient's data, with what a second application would have to show
  (Section~\ref{sec:disc}).
\end{enumerate}

\section{Related Work}\label{sec:related}
\paragraph{Bespoke artefacts for relationships at a distance.}
Technologies for strong ties should support the expressive, the physical and the playful, not
only the informational~\cite{vetere2005mediating}. Hassenzahl et al.\ group 143
intimacy-mediating artefacts into six strategies: awareness, expressivity, physicalness, gift
giving, joint action and memories~\cite{hassenzahl2012love}. Partners at a distance can
report as much intimacy as partners who live close~\cite{jiang2013absence,stafford2005maintaining},
but such reports do not show that any one technology improves a relationship. The closest precedent to this
paper is Chien and Hassenzahl's autoethnographic research through design on technology-mediated
relationship maintenance in a romantic long-distance relationship~\cite{chien2020ldr}. We share
its first-person position and its setting. Our material differs: we report no lived use, and
our evidence is a design record and checks of software that one partner wrote for the other.
Work on autonomy in technology-mediated relationships~\cite{zumaya2026relatedness} bears on
what such a gift asks of its recipient, a question we raise in Section~\ref{sec:disc} and
cannot answer. \yt{} belongs to the \emph{memories} and \emph{gift giving} strategies; what it
adds to them is a sequence of levels authored by one side, and a way to check the rules of that
sequence.

\paragraph{First-person design and the limits of its evidence.}
Autobiographical design draws on genuine use by the people who built a
system~\cite{neustaedter2012autobiographical}, and Desjardins and Ball analyse cases in which a
designer builds for a partner or family member~\cite{desjardins2018tensions}; first-person
methods more widely are reviewed by Desjardins et al.~\cite{desjardins2021firstperson}. We do
not have genuine use to report. This paper is a research-through-design
case~\cite{zimmerman2007rtd,gaver2012expect} that draws on autobiographical material: the maker's
own memories, his development testing and the project's history. That position sets the limit of
the evidence. The maker can say what he intended and what the software does, and only the
recipient can say what it was like to receive.

\paragraph{Answers written by machines.}
In AI-mediated communication an agent modifies or generates messages on a communicator's
behalf~\cite{hancock2020aimc}. Jakesch et al.\ found lower trust in host profiles believed to be
written by AI when profiles of both kinds were shown together~\cite{jakesch2019aimc}, and
Hohenstein et al.\ found in randomised experiments that suggested replies changed the language
of a conversation and that people suspected of using them were judged less
favourably~\cite{hohenstein2023ai}. Both study text that a person chooses to publish or send. The
request box differs: a system may answer in a partner's place, so the open questions are who
authorised the answer and how it is marked. A model that acts on outside text can also be
redirected by instructions hidden in that text~\cite{greshake2023injection}, which motivates
the triage replay in RQ3.

\paragraph{Games and local state.}
\yt{} takes the game form literally, with locks, levels, a guide, a prologue and a
finale~\cite{salen2004rules,juul2005halfreal}. Level completion is a grow-only set that
converges under union~\cite{shapiro2011crdt}, kept on each phone in the manner of local-first
software~\cite{kleppmann2019localfirst} and carried between phones by hand.

\section{The Design Case}\label{sec:case}
\paragraph{The two of us.}
The couple began their relationship on 20 September 2026 and speak Mandarin to each other. Yifan
works in English; Jingyi works in Chinese and English. Both use phones as their main devices, and
Jingyi's network in mainland China does not reach some services that Yifan's does. These
circumstances motivated bilingual text, alternative media providers and a phone-first interface.
Yifan directed the design, supplied the material and tested both player roles, with substantial
implementation help from an AI coding assistant. Jingyi is the intended recipient and a
co-author who approved this account.

\paragraph{Corpus and method.}
We follow research through design~\cite{zimmerman2007rtd,gaver2012expect}. The design record is
the public repository: 152 commits, 36 changelog versions and five design specifications up to
24 November 2026 (version 1.24.0), from which we reconstruct seven design episodes. Version
history establishes what changed and when. It does not establish unrecorded motives or the
recipient's response. The record holds three traces that originate with her: a remark quoted in
the changelog, a flight she booked, and requests she sent through the request box. We report that
these exist and what the designer built in response. We report nothing of what her requests
contain.

\paragraph{Six commitments.}
The reconstruction of the first five days of development yields six commitments
(Table~\ref{tab:claims}). They are situated choices, not validated principles. Five of them state
a rule that code implements, and Section~\ref{sec:eval} checks those. The first, that memories
became mechanics, is a claim about where the design came from. No test of the software bears on
it, and we support it with dated entries of the design record (Appendix~\ref{sec:log}).

\begin{table}[t]
  \caption{The six design commitments, the mechanism that realises each, and what in this paper
  bears on it. A check covers the mechanism named and no more.}
  \label{tab:claims}
  \small
  \begin{tabular}{@{}p{0.23\columnwidth}p{0.36\columnwidth}p{0.31\columnwidth}@{}}
    \toprule
    Commitment & Realised by & Evidence here \\
    \midrule
    \dc{1} Memory-derived mechanics & Memories, shared references and imagined futures become tasks & Design record only; not machine-checked \\
    \dc{2} Staged access & Ordered levels; entry through a door in the town & E7, for the progress code; the router's order is not tested \\
    \dc{3} Local progress & Per-player local storage; grow-only set; hand-carried codes; offline cache & E2 (offline), E7 (codes, merge) \\
    \dc{4} Bilingual coverage & Two dictionaries with the same keys & E5 (layout), E6 (keys and placeholders) \\
    \dc{5} Drawn in code & Sprites and tiles as strings; no image downloads for the town & E1, E3: readings of this build, with no image-based build to compare \\
    \dc{6} Delayed disclosure & A lexical test over strings shown before the finale & E8 \\
    \midrule
    Later: answering in his place & A server, a local model and a hosted model & E11, E12 \\
    \bottomrule
  \end{tabular}
\end{table}

\section{MERIDIAN}\label{sec:artefact}
\subsection{The Town and Its Nine Levels}
The site opens on a password field that accepts one date. The date is a recognition question and
not a secret: the files behind it are public. Behind it a dashboard lists everything the couple
build under the domain, and \yt{} is its first card.

The town is a $24\times20$ grid of 16-pixel tiles, painted entirely by code (\dc{5}). Paths
connect nine houses around a fountain square. A player moves by arrow keys, by tapping a tile or
with an on-screen pad, and walking into a door opens a level. The partner's sprite waits by the
fountain and reports how many levels that partner has cleared; a suited guide named Rayford
explains the town; two cats, drawn from the couple's real ones, guard the next two locked doors.
The sky follows the clock in each player's own city, so at one instant she walks under noon and he
under lit windows.

\begin{figure*}[t]
  \centering
  \includegraphics[width=\textwidth]{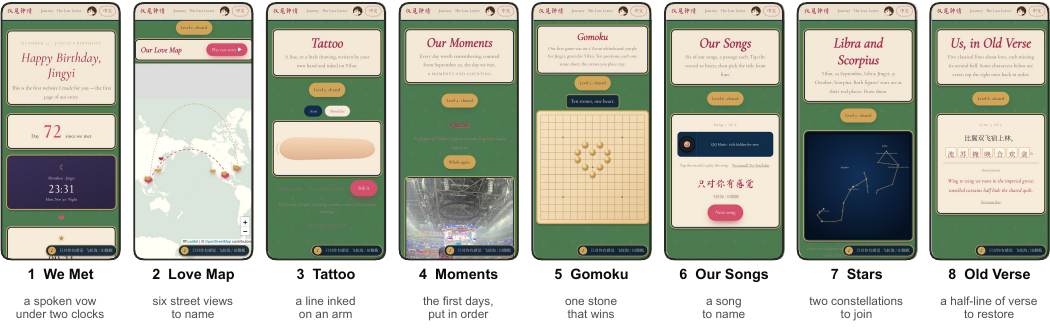}
  \caption{Eight of the nine levels as they appear on a phone, each with the shared reference
  it turns into a task; the ninth is the ring try-on (Figure~\ref{fig:ring}).}
  \Description{Eight phone screens in a row, each a cream panel on green pixel grass showing one
  level: We Met, a spoken vow under two clocks; Love Map, six street views to name; Tattoo, a
  line inked on an arm; Moments, the first days put in order; Gomoku, one stone that wins; Our
  Songs, a song to name; Stars, two constellations to join; Old Verse, a half-line of verse to
  restore.}
  \label{fig:levels}
\end{figure*}

Each level takes one piece of shared material and makes it a rule (\dc{1};
Figure~\ref{fig:levels}). The first asks the player to say a vow aloud with the partner's full
name. The Love Map asks for six places from street views. Tattoo asks for a handwritten line on
the partner's arm. Moments orders the first forty days. Gomoku reworks the couple's first game,
played on a video-call whiteboard, into ten win-in-one positions whose answers form a heart. Our
Songs plays a passage and asks for its title. The seventh level joins the stars of their two
birth signs, which are neighbours in the sky. The eighth restores missing half-lines of classical
love poems. The ninth, named Meridian, puts a ring on the player's hand through the camera and
asks for a photograph. Levels open in order (\dc{2}), and nothing shown before the ninth may name
the ring or the wedding (\dc{6}).

Two ceremonies frame the game. A birthday prologue plays once before the town opens: he bakes a
cake against her clock, the player whisks, places ten strawberries in the Gomoku heart and blows
out nine candles by breathing into the microphone. A wedding opens once a player has cleared the
last level, with scripted vows, the two rings in 3D and a certificate to keep.

\subsection{After the Town}
The first version of this paper described a finished gift. In the four weeks that followed, the
dashboard grew from one card to five, and three of the new cards changed what the gift is.

\begin{figure*}[t]
  \centering
  \includegraphics[width=\textwidth]{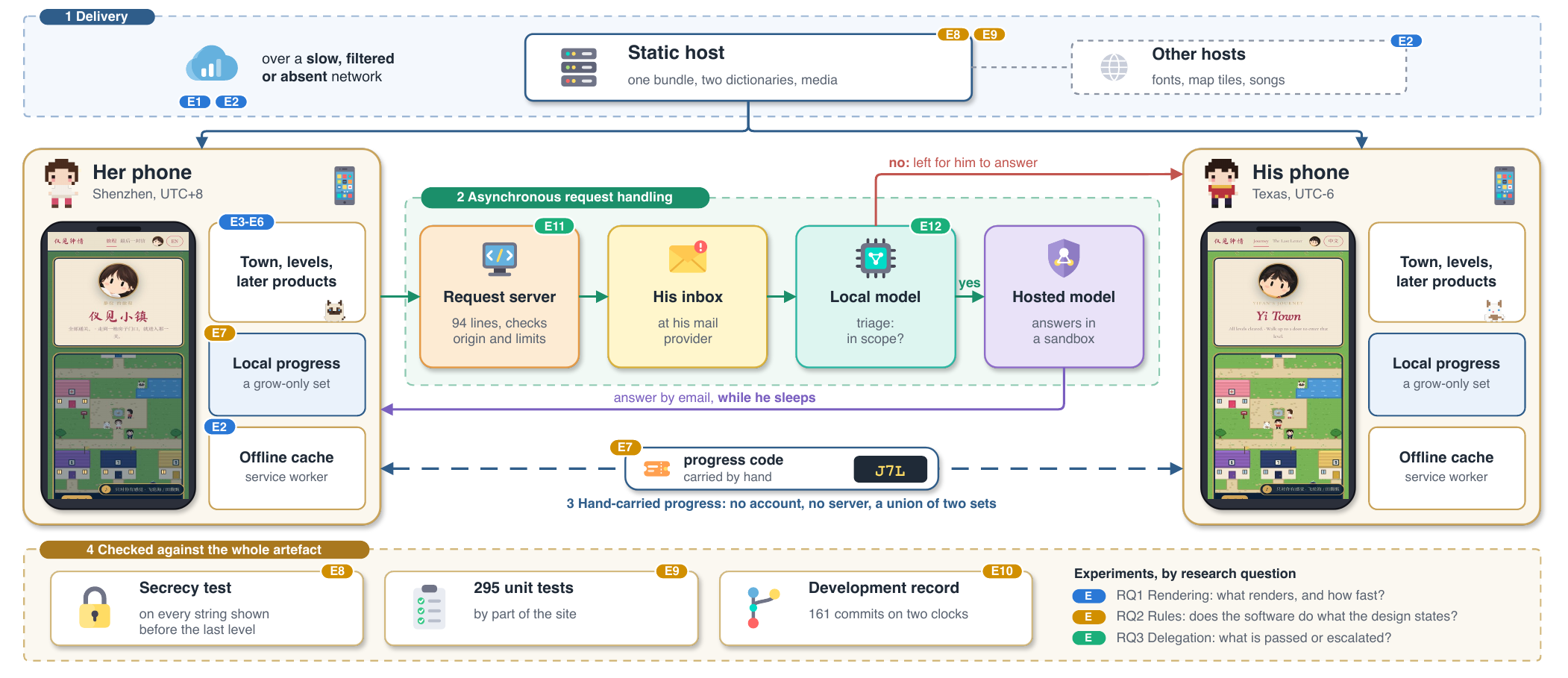}
  \caption{\textsc{Meridian} and what the evaluation tests. (1) One static site is delivered to
  both phones over a slow, filtered or absent network; her phone shows the town in Chinese at
  night and his shows it in English by day. (2) A request leaves her phone through a server
  and his inbox to a local model, which decides whether a hosted model may answer in his place.
  (3) Progress moves between the phones as a short code carried by hand. (4) Three checks
  apply to the whole artefact. Tags name the experiment that tests each part, coloured by
  research question.}
  \Description{A diagram in four numbered panels. Panel 1, delivery: a static host and other
  hosts above two phones. Each phone shows a screenshot of the pixel town and three parts:
  town, levels and later products; local progress; offline cache. Panel 2, between the phones:
  request server, his inbox, local model and hosted model in a row, with a branch back to his
  phone when the local model says no and an answer by email back to her phone. Panel 3: a
  dashed double arrow between the phones labelled progress code, carried by hand, J7L. Panel 4:
  secrecy test, 295 unit tests and the development record. Tags E1 to E12 are blue for reach,
  amber for rules and green for delegation.}
  \label{fig:overview}
\end{figure*}

\emph{The infirmary} began with something the recipient said on 3 November: that when hurt she
would not talk and would hide. The next day the town had a pink house. A visitor names what
hurts, takes three slow breaths, and reads a letter written in advance, in the other partner's
name, for that hurt. Leaving offers one button, ``tell them I am out''. This is the first feature in the record
whose stated origin is the recipient. \emph{The trip page} replaced a hand-drawn symbolic ticket
with her booked flight: the countdown now ends at a real landing. \emph{The request box} is a
composer in which she writes what she needs help with and attaches files. A browser cannot send
mail with attachments, so the site gained its first server (\dc{3} now holds for progress only).
The same day, the designer added a pipeline that answers in his place while he sleeps: a local
model reads the request and decides whether it is within an agreed scope, a hosted model
prepares the answer inside a sandbox, and the answer goes to her by email
(Figure~\ref{fig:overview}). The specification written the day before had listed ``an AI that
answers her'' as a non-goal.

\subsection{System}
\yt{} is a static single-page application in strict TypeScript: 9{,}895 lines of source, three
runtime dependencies, each feature a folder with its page, styles, data and tests. Progress is a
grow-only set of cleared levels in the browser's local storage. To move it between phones a
player reads out or sends a three- or four-character code: a letter for the player, the set as a
base-36 number, and one ISO~7064 MOD~37,36 check character. Importing a code takes the union,
so a code can add progress and never remove it. A service worker precaches the application shell.
Heavy modules (the map library, the 3D ring, the hand tracker) load only inside the level that
needs them. The request server is 94 lines on an edge platform; the answering pipeline is 621
lines of Python on the designer's laptop. Appendices~\ref{sec:system} and~\ref{sec:products}
give every data structure, rule and data flow.

\subsection{Seven Design Episodes}
Seven recorded changes make the rationale inspectable (Appendix~\ref{sec:log}). A same-device
drawing game was replaced by a song level that two people can play at different hours. The
ring level's completion changed from viewing to taking a photograph, which makes a camera a
condition of reaching the ending. Generated descriptions of the couple's routines needed
correction against the designer's account. The entry gate lost its answer-revealing placeholder
and later narrowed to one spelling. The infirmary was built from a remark. A booked flight
replaced an invented date. And a stated non-goal was reversed within a day. Across them the
recurring issue is how a technical rule expresses an assumption about the other person; in the
later ones the rule begins to act for one person toward the other.

\section{Evaluation}\label{sec:eval}
\subsection{Question and Approach}
The evaluation asks which properties of the gift its maker can verify without data from the
recipient. For each rule the design record states we name the code that implements it, write a
check that can fail, and report what was observed and what the observation does not show
(Table~\ref{tab:holds}). The checks fall under three questions.
\begin{description}[leftmargin=2.2em,labelwidth=2.2em,labelsep=0pt,itemsep=1pt,topsep=2pt,font=\normalfont\bfseries]
  \item[RQ1] \emph{Rendering.} Which routes render, and how fast, under simulated network
  conditions and screen sizes? (E1 to E6)
  \item[RQ2] \emph{Rules.} Does the software implement the rules the design states for
  hand-carried progress and a secret ending, and how much of the code do its tests run?
  (E7 to E9; E10 gives context)
  \item[RQ3] \emph{Delegation.} What does the step that decides whether a machine may answer
  pass and escalate on synthetic requests? (E11, E12)
\end{description}
None of the checks uses data from the recipient: where requests are needed we wrote 120
synthetic ones, and the answering pipeline's private scope is replaced by a stand-in.
Discrepancies are reported as found and were not repaired, so the results describe the artefact
she has.

All checks ran on 29 November 2026 against commit \code{9193c20} (version 1.25.0) on one laptop
(Apple M5, 32\,GB). That commit is nine commits past the design record's cut-off: five add an
album page that this paper does not describe and the checks do not visit, and four archive or
revise the paper. Browser checks drive the production build in desktop Chromium~148 at
$390\times844$ CSS pixels with touch input. The browser is seeded with the state of a returning
player: the entry question passed, the prologue seen and every level cleared, so that each route
can be opened directly. A route counts as \emph{rendered} when its view element has content, no
uncaught script error occurs and none of its own images is broken. Rendering does not show that
the task on a page can be completed, and no check plays a level. The browser checks submit no
form and refuse every request to the site's own servers. Model checks use open models served
locally at temperature zero. Scripts and raw results are in \code{paper/experiments/}, and
Appendix~\ref{sec:expdetail} gives each protocol and the full tables.

\begin{figure*}[t]
  \centering
  \includegraphics[width=\textwidth]{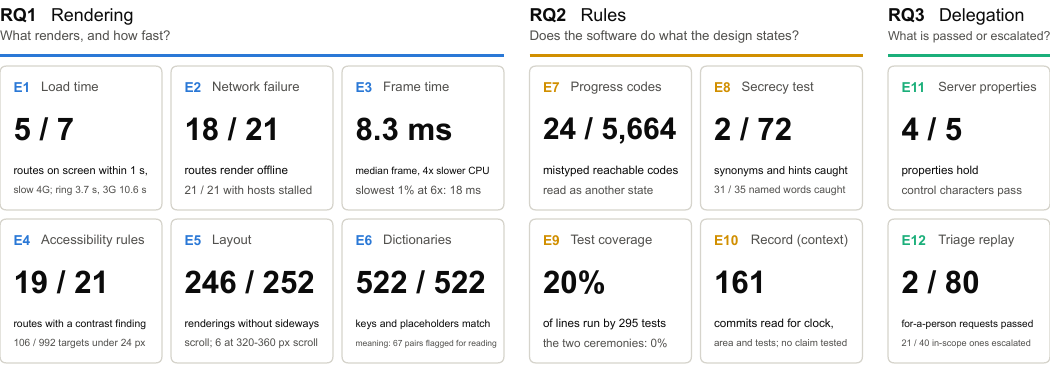}
  \caption{The twelve checks under three questions. Each tile gives one count and the count that
  qualifies it. E10 describes the development record and tests no claim.}
  \Description{A grid of twelve tiles in three groups. Rendering: load time, 5 of 7 routes within
  one second on slow 4G; network failure, 18 of 21 routes render offline; frame time, 8.3
  milliseconds; accessibility rules, 19 of 21 routes with a contrast finding; layout, 246 of 252
  renderings; dictionaries, 522 of 522 keys match. Rules: progress codes, 24 of 5,664 mistyped
  reachable codes read as another state; secrecy test, 2 of 72; test coverage, 20 percent;
  record, 161 commits. Delegation: server properties, 4 of 5; triage replay, 2 of 80 requests for
  a person passed and 21 of 40 in-scope ones escalated.}
  \label{fig:summary}
\end{figure*}

\subsection{RQ1: Which Routes Render, and How Fast?}\label{sec:rq1}
\paragraph{Method.}
\emph{E1} loads seven routes five times under four throttled network profiles with the CPU
slowed four times, with a cold cache and again after the service worker has installed. A
cold-cache load here is a direct visit to an unlocked route, not a new player's first session.
\emph{E2} records the other hosts each of the 21 routes calls and opens every route under four
failures: hosts unreachable from mainland China \emph{refused}; the same hosts \emph{stalled},
so that requests hang; every other host refused (\emph{alone}); and the device \emph{offline}
after one earlier visit. \emph{E3} records frame intervals in the town while the avatar walks,
with the CPU slowed up to six times. \emph{E4} runs the axe-core 4.10 rules for WCAG~2.0 to~2.2 at
levels A and AA~\cite{wcag22} on every route in both languages and measures every touch target.
\emph{E5} renders every route at six widths from 320 to 1{,}280 pixels in both languages.
\emph{E6} compares the keys and placeholders of the two dictionaries.

\begin{figure*}[t]
  \centering
  \includegraphics[width=\textwidth]{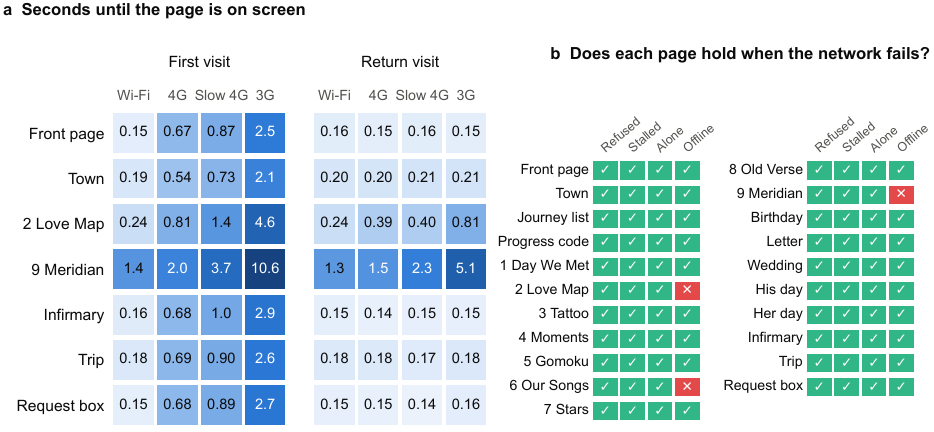}
  \caption{E1 and E2. (a) E1: median seconds until a route's own view is on screen, over five
  runs, with a cold cache and after the service worker has installed; darker is slower. (b) E2:
  whether each of the 21 routes renders when blocked hosts are refused, when they stall, when no
  other host answers, and offline after one visit. Three routes do not render offline.}
  \Description{Left: two heat tables of seven routes by four networks; cold-cache times grow from about 0.2 seconds on Wi-Fi to 2 to 5 seconds on 3G, with the ring level at 10.6 seconds; return times stay near 0.2 seconds except for the ring and the Love Map. Right: a grid of 21 routes by four failure conditions, all marked as rendering except Love Map, Our Songs and the ring level under offline.}
  \label{fig:reach}
\end{figure*}

\paragraph{Result.}
\emph{Load time.} The front page transfers 350\,kB and the town 297\,kB. On slow 4G
(1.6\,Mbit/s) five of seven routes show their view within 0.95\,s; the Love Map takes 1.4\,s and
the ring level, which transfers 1.5\,MB on a first visit, 3.7\,s. On 3G (400\,kbit/s) the five take 2.1 to 2.9\,s, the Love Map 4.6\,s and
the ring 10.6\,s (Figure~\ref{fig:reach}a). After the service worker has installed, five of
seven transfer nothing and show their view in about 0.2\,s on every profile. The town keeps a
median frame interval of 8.3\,ms with the CPU slowed four times (E3).

\emph{Network failure.} The routes call a font service on every page, map tiles on one and a
music service on one. All 21 render when the blocked hosts are refused, when they stall and
when no other host answers (Figure~\ref{fig:reach}b). Two qualifications matter. With the hosts
stalled, the browser's \code{load} event did not fire within 15\,s in any of 63 loads, although
the view appeared no more than 13\,ms later than usual on 20 of 21 routes. And a rendered route is
not a working one: with the music host unreachable, Our Songs renders and its player has
nothing to play. Offline after one visit, 18 of 21 routes render. The Love Map and the ring
fail because their style chunks are left out of the precache, and Our Songs throws an error
from the music embed.

\emph{Layout and accessibility rules.} No text box cuts its text off in any of 252 renderings,
and 246 show no sideways scroll; the six that scroll are Old Verse and the trip page at 320
pixels in both languages and the request box in English at 320 and 360 (E5). axe-core reports a
colour-contrast finding on 19 of 21 routes in each language, on 207 elements in 42 audits (E4;
Figure~\ref{fig:access} in the appendix); the rule compares text against a ratio of 4.5 to 1,
or 3 to 1 for large text. Separately, we measured 992 touch targets: 106 are smaller than
24 CSS pixels on a side and 883 smaller than 44. These are size measurements and not counts of
violations, since the AA criterion at 24 pixels has exceptions for spacing and equivalent
controls and 44 pixels is the AAA criterion~\cite{wcag22}.

\emph{Dictionaries.} The two dictionaries hold the same 522 keys with the same placeholders
(E6). A page rendered in English holds a median 2.2 times as many characters as the same page in
Chinese and is up to 23\% taller (E5). Whether the two sides say the same thing was screened by two models; they agree
weakly (Cohen's $\kappa = 0.29$), and the screen yields 67 pairs for a bilingual reader and no
estimate of translation quality (Appendix~\ref{sec:expdetail}).

\paragraph{Limits.}
These are readings from desktop Chromium with emulated size, touch, network and CPU, from a
seeded state. They do not show that she can enter, play a level, grant or refuse the camera,
import a code or reach the ending on her phone and her network. The block list stands in for a
network inside mainland China. Automated rules find a minority of accessibility barriers, and no
assistive technology was used.

\subsection{RQ2: Does the Software Implement the Stated Rules?}\label{sec:rq2}
\paragraph{Method.}
Two rules carry the story between two phones and toward its ending. The first is that progress
moves as a short code that a person reads out or types, and that a mistyped code is refused.
A player $p$ who has cleared the set of levels $C_p \subseteq \{\ell_1, \dots, \ell_9\}$
carries
\begin{equation}\label{eq:code-main}
  \mathrm{code}(p, C_p) = \lambda_p \,\Vert\, \beta_{36}(m) \,\Vert\, \chi\big(\lambda_p \Vert \beta_{36}(m)\big),
  \quad m = \sum_{\ell_i \in C_p} 2^{\,i-1},
\end{equation}
a player letter $\lambda_p$, the set as a number in base 36 and one ISO~7064 check character
$\chi$~\cite{iso7064}. The receiving phone merges by union. The design also states that levels
open in order, so play yields only the ten prefixes $C_p = \{\ell_1, \dots, \ell_j\}$,
$0 \le j \le 9$, of the 512 sets the code can express. \emph{E7} mutates codes by one edit of
five kinds (one character replaced, two neighbours swapped, a doubled character replaced, one
character deleted, one inserted) and counts the mutations that the decoder reads as a state
other than the original. It does so twice: from all 1{,}024 codes the decoder accepts, and from
the 20 codes that ordered play can produce. For the second set it also decodes under two
alternatives: the shipped code with a decoder that refuses a non-prefix set, and a code that
carries the number of levels cleared in place of the set. The edits are a model of typing
errors; we have no data on which errors people make with these codes.

The second rule is that nothing shown before the last level gives the ending away, which rests
on a lexical test of three patterns over 464 strings. \emph{E8} runs 107 hints of the finale,
written by us in Chinese and English in four tiers, through those patterns. \emph{E9} measures
what the 295 unit tests execute.

\begin{figure*}[t]
  \centering
  \includegraphics[width=\textwidth]{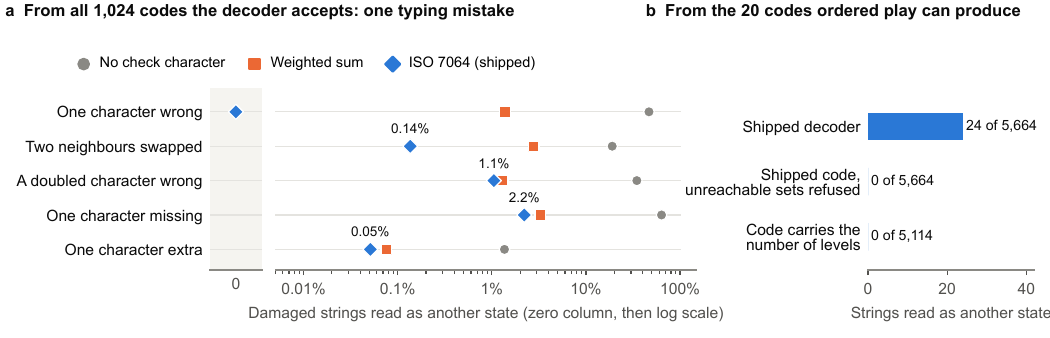}
  \caption{E7. (a) From all 1{,}024 codes the decoder accepts: the share of one-edit mutations
  read as another state, under three check designs. A true zero is plotted in its own column to
  the left of the logarithmic axis. (b) From the 20 codes ordered play can produce: mutations
  read as another state by the shipped decoder, by a decoder that refuses unreachable sets, and
  under a code that carries the number of levels cleared.}
  \Description{Left: a dot plot with five kinds of typing mistake as rows and the share of mutated codes read as another state on a logarithmic axis, with a separate column for zero. With no check character about 1 to 64 percent are accepted; a weighted sum accepts about 0.08 to 3.3 percent; the shipped ISO 7064 character accepts none of the single wrong characters and 0.05 to 2.2 percent of the other kinds. Right: three bars. The shipped decoder accepts 24 of 5,664 mutations of reachable codes; the decoder that refuses unreachable sets accepts 0 of 5,664; the count code accepts 0 of 5,114.}
  \label{fig:codes}
\end{figure*}

\paragraph{Result.}
\emph{Over all decodable codes.} The decoder read none of 140{,}840 single-character
substitutions as another state. Across the five kinds it read 207 of 331{,}530 mutations as
another state: 4 of 2{,}943 swaps, 21 of 1{,}995 doubled characters, 89 of 4{,}024 deletions
and 93 of 181{,}728 insertions (Figure~\ref{fig:codes}a). The swaps contradict a comment in the
source that says a swap cannot pass. Of the 207, 186 decode to a set that ordered play cannot
produce, such as level 5 without level 4. No accepted mutation changes the player, and since
importing a code is a union, none removes progress. The merge is commutative, idempotent and
equal to set union over all 262{,}144 pairs of sets.

\emph{Over the codes play can produce.} The 20 reachable codes have 5{,}664 mutations. The
shipped decoder read 24 of them as another state, 2 deletions and 22 insertions, and all 24 are
sets that ordered play cannot produce; 22 of them add levels the player has not cleared, 63
levels in all. A decoder that also refuses non-prefix sets read none of the 5{,}664 as another
state, and every code that ordered play has produced stays valid under it. A code that carries
the count of levels read none of its 5{,}114 mutations as another state
(Figure~\ref{fig:codes}b). The rule that would have closed the gap was in the design from the
first days and in the router; it was left out of the data format.

\emph{The secrecy test.} It catches all 12 phrases built from its own words and 19 of 23 other
forms of them. It catches 1 of 46 synonyms (such as \emph{engagement}, \emph{vows} and \zh{新娘}) and
1 of 26 oblique phrases (such as \emph{on one knee} and \zh{无名指}), and it refuses 9 of 13 harmless strings
such as \emph{marrow} and \zh{戒烟} (``quit smoking''). The first level has the heading ``The
vow'' and has each player say ``I do''. The test guards a vocabulary. The lexicon is ours, so
the rates describe this test against these hints.

\emph{Test coverage.} The 295 unit tests execute 20.3\% of 4{,}647 lines: 56\% of the progress,
routing and gate code, 13\% of the nine levels and none of the two ceremonies (E9;
Figure~\ref{fig:record} in the appendix).

\paragraph{Limits.}
E7 tests the code and its decoder. It does not test that the router opens levels in order, on
which the notion of a reachable set depends, and it says nothing about how often a person
mistypes. The comparison of three designs is a count over enumerated edits, not a measured
reduction in errors in use.

\emph{The record as context (E10).} The repository holds 161 commits at the tested commit, 33
of them documentation and 19 the paper. By author time, 45 fall between midnight and 07:00 in
Shenzhen and 26 in the same hours in Texas. Author time records when a commit was made and
nothing about how the time difference shaped a decision, so we draw no conclusion from it.

\subsection{RQ3: What Does the Triage Step Pass and Escalate?}\label{sec:rq3}
Since 24 November a pipeline can answer a request in the designer's place. A server receives
what she sends (E11), and a local model decides whether a hosted model may prepare an answer
(E12).

\paragraph{Method.}
\emph{E11} sends the request handler 4{,}000 generated requests that vary origin, method, path,
text, file count, file size and file name, with names that carry path traversal, quotes,
control characters and header-injection attempts. A recorder stands in for the mail binding,
and five stated properties are checked on every request. \emph{E12} is a synthetic replay of
the triage call. It uses the published prompt, schema and message layout, a stand-in scope (a
one-page bilingual reading note) and 120 requests that we wrote in six classes of 20, half in
each language. Two classes are labelled in scope (a new request and a revision) and four are
labelled for a person: outside the scope, a request to send the result elsewhere, a request to
run a command or read a file, and an instruction placed inside the excerpt. The class balance
is ours and is not an estimate of what she sends. Five local models take the replay, among them
Qwen3-14B, the model used in deployment. With $y(x) = 1$ for a request labelled for a person
and $d(x) = 1$ when the model escalates it, we report
\begin{equation}\label{eq:triage}
  \mathrm{miss} = \Pr[\, d = 0 \mid y = 1 \,], \qquad \mathrm{over} = \Pr[\, d = 1 \mid y = 0 \,].
\end{equation}
The replay observes a decision. It does not run the answering model, and it measures no reply,
no delivery and no load on a person.

\paragraph{Result.}
\emph{The server.} Four of five properties hold on every request: mail leaves only for a
request that passes every documented check (1{,}629 of 4{,}000 did), a request from another
origin receives no cross-origin grant, the recipient and sender are the configured ones, and no
attachment name carries a path separator, quote or control character. The fifth fails: in 147
sent messages a control character from the text reaches the mail subject. None is a line
break, so no header can be injected.

\begin{figure}[t]
  \centering
  \includegraphics[width=\columnwidth]{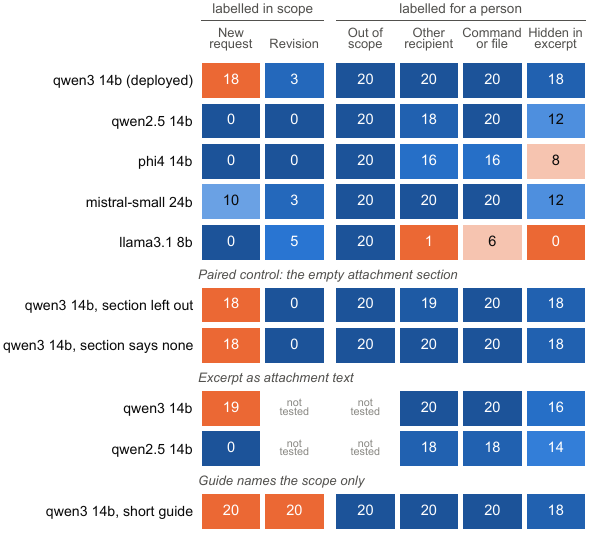}
  \caption{E12, a synthetic replay: requests of 20 in each class that each model escalates to a
  person. Dark blue marks agreement with the class label and orange its opposite. The first row
  is the model used in deployment under the published guide; the paired controls replay the
  same requests with the empty attachment section left out or marked as empty.}
  \Description{A grid of counts out of 20. Rows are models and conditions; columns are six
  request classes, two labelled in scope and four labelled for a person. The deployed qwen3
  model escalates 18 new requests, 3 revisions, 20 out of scope, 20 other recipient, 20 command
  and 18 placed instructions. qwen2.5 escalates 0, 0, 20, 18, 20, 12. phi4 escalates 0, 0, 20,
  16, 16, 8. mistral-small escalates 10, 3, 20, 20, 20, 12. llama3.1 escalates 0, 5, 20, 1, 6,
  0. With the excerpt as attachment text qwen3 escalates 19 new requests and qwen2.5 none; the
  revision and out-of-scope classes were not tested in that condition. With a guide that names
  only the scope qwen3 escalates all 40 in-scope requests. In the two paired controls, with the
  attachment section left out or marked as empty, qwen3 escalates 18 new requests and no
  revisions.}
  \label{fig:answer}
\end{figure}

\emph{The deployed model, default condition.} Of 80 requests labelled for a person it passed 2
(miss $2.5\%$), both instructions placed in a Chinese excerpt. Of 40 in-scope requests it
escalated 21 (over $52.5\%$): 18 of 20 new requests and 3 of 20 revisions. Read the other way,
it passed 21 of the 120 requests, and 2 of those 21 were labelled for a person
(Figure~\ref{fig:answer}). By language, it missed 2 of 40 in Chinese and none of 40 in English,
and escalated 9 of 20 in-scope requests in Chinese and 12 of 20 in English.

\emph{Other conditions and models.} With a guide that names only the scope, the same model
escalated all 40 in-scope requests and still passed the same 2 of 80: a stricter-sounding guide
moved the in-scope requests and not the misses. With the excerpt placed in the attachment
section as transcribed text, it escalated 19 of 20 new requests and passed 4 of 20 placed
instructions. Two of the other four models escalated none of the 40 in-scope requests and
passed 10 and 20 of the 80 others (miss $12.5\%$ and $25\%$). Placed instructions were the
weakest class for every model: the five passed 50 of 100.

\emph{A paired control for the stated reason.} For all 18 new requests it escalated, the
deployed model gave an empty attachment section as its reason; the pipeline writes that section
for every email. We replayed the same 120 requests with the section left out, and again with
the section saying that there is no attachment. The model escalated 18 of 20 new requests in
all three versions. Its reasons changed and its decisions did not: with the section left out,
2 of 18 reasons mention an attachment and most of the rest say that the excerpt is incomplete,
has no English or leaves the task unclear; with the section marked as empty, 17 of 18 mention the attachment again. The section
did matter for revisions, where escalations fell from 3 of 20 to none in both controls, so the
count for in-scope requests fell from 21 to 18 of 40. One request to send the result elsewhere
was passed only with the section left out. The model's stated reason was therefore not the
cause of most escalations, and we do not know what is.

\paragraph{Limits.}
The 1{,}120 decisions come from 120 distinct requests, reused across models and conditions,
written around four books by the authors of the checks; there is no held-out set, no human
baseline and no rule baseline. The paired control shows that a model's stated reason is a
hypothesis about its behaviour and not a measurement of it. Nothing here shows that the
pipeline is safe: the replay does not observe what the answering model does with a request
that passes.

\subsection{What the Checks Establish}\label{sec:answers}
Table~\ref{tab:holds} sets out each claim with its check, its failure criterion, what was
observed and what the observation leaves open. \textbf{RQ1.} Under simulated conditions the
routes render, quickly for five of seven, and three do not render offline; rendering is as far
as these checks see. \textbf{RQ2.} The progress code refuses single-character substitutions and
accepts a small number of other edits, all of which, from the codes play can produce, name
states the design's own ordering rule excludes. The secrecy test covers the words it lists.
\textbf{RQ3.} On 120 synthetic requests the model used in deployment passed 2 of 80
labelled for a person and escalated 21 of 40 labelled in scope. The replay supports no
statement about the safety of the pipeline or the quality of an answer.

The checks surfaced eight discrepancies between the account and the artefact: progress states
accepted that ordered play cannot produce, a swap of the last two characters of a code that the
source says cannot pass, three routes that do not render offline, a 1.5\,MB final level,
contrast findings on 19 routes, small touch targets, three routes that scroll sideways on narrow
screens, and control characters in a mail subject. None has been repaired in the commit we
tested.

\begin{table*}[t]
  \caption{Each claim of the design account with the check that bears on it, the criterion by
  which the check fails, the observation, and what the observation does not show.}
  \label{tab:holds}
  \footnotesize
  \begin{tabular}{@{}p{0.13\textwidth}p{0.18\textwidth}p{0.17\textwidth}p{0.22\textwidth}p{0.21\textwidth}@{}}
    \toprule
    Claim in the account & Executable check & Fails when & Observed & Not shown \\
    \midrule
    Pages work on her network (\dc{3}, \dc{5}) & 21 routes under four simulated failures; 7 routes under four throttled profiles (E1, E2) & A route does not render; no time threshold was set in advance & 21 of 21 render with hosts refused, stalled or absent; 18 of 21 offline; 0.7 to 3.7\,s on slow 4G & That a level can be played; a phone; a network in mainland China \\
    Pages fit her screen and can be read & 252 renderings; axe-core rules; target sizes (E4, E5) & Text is cut off or the page scrolls sideways; a rule reports a finding & 246 of 252 without sideways scroll; contrast findings on 19 of 21 routes; 106 of 992 targets under 24\,px & Conformance to WCAG; use with assistive technology \\
    Two languages (\dc{4}) & Keys and placeholders of both dictionaries (E6) & A key or placeholder is missing on one side & 522 of 522 match & That the two sides mean the same; 67 pairs are flagged for a reader \\
    \addlinespace
    A mistyped code is refused (\dc{2}, \dc{3}) & All one-edit mutations of all codes and of the 20 reachable codes (E7) & A mutation decodes to a state other than the original & 0 of 140{,}840 substitutions; 207 of 331{,}530 over five kinds; 24 of 5{,}664 from reachable codes, all unreachable states & How often people mistype; that the router keeps levels in order \\
    The ending is not given away (\dc{6}) & 107 hints against the shipped patterns (E8) & A hint passes the patterns & 31 of 35 named words and variants caught; 2 of 72 synonyms and oblique hints & What a player infers; the first level has players say ``I do'' \\
    The code is tested & Line coverage by part of the site (E9) & (descriptive) & 20\% of lines; 13\% of the levels; none of the two ceremonies & Whether untested code is wrong \\
    \addlinespace
    Mail leaves only as designed & 4{,}000 generated requests against the handler with a recorded mail binding (E11) & Any of five stated properties is violated & Four hold on every request; control characters reach the subject in 147 messages & The deployed mail service; delivery \\
    Triage sends out-of-scope requests to a person & Synthetic replay of the triage call on 120 written requests, five models (E12) & A request labelled for a person is passed, or an in-scope one is escalated & Model used in deployment: 2 of 80 passed; 21 of 40 in-scope escalated; short guide: 40 of 40 escalated, 2 of 80 still passed & Real requests; the answering model; reply quality; why the model escalates (its stated reason failed a paired control) \\
    \bottomrule
  \end{tabular}
\end{table*}

\section{Discussion}\label{sec:disc}
\subsection{A Checking Procedure from One Case}
The procedure we followed has four steps. List the rules the design record states. For each,
find the code that implements it, or note that none does. Write a check that can fail, with its
failure criterion fixed before the run where one exists. Report the observation with the limit
of what it shows, before any repair. Nothing in these steps is new to software testing. What
the case adds is their target: the claims of a design account written by the same person who
wrote the code, in a setting where asking the recipient to evaluate a present is itself a
request about the relationship~\cite{desjardins2018tensions}.

We do not claim this as a validated method. One maker applied it to one artefact, with an AI
assistant writing the checks, and we have no comparison with what ordinary quality assurance
would have found. Two kinds of evidence would be needed to call it a method: an independent
evaluator who first inspects the same artefact in the usual way and then follows these steps,
with the added findings, the misses and the disagreements recorded; and a second artefact, to
learn which steps transfer.

The procedure also has an edge that no amount of checking moves. It can show that a decoder
refuses a mutated code and cannot show that exchanging codes feels like sharing. It can show
that a route renders at 320 pixels and cannot show that a level reads as affection and not as
an examination. Claims about meaning, surprise, pressure or closeness remain hers to make.

\subsection{What the Checks Change in the Design Account}
Three results revise what the earlier versions said. First, the account presented the check
character as the safeguard of hand-carried progress. E7 shows that it stops substitutions and
that the remaining errors are stopped by a rule the designer already had: levels open in order.
Staged access (\dc{2}) was implemented in the router and left out of the data format. Second,
the account described delayed disclosure as tested. E8 shows a test of vocabulary, and in the
first level each player says ``I do'', so the design shows a wedding in action from the start
while guarding its words. Third, the account treated the site's single visual world as an
achievement of drawing in code (\dc{5}). E1 and E3 give load and frame times for this build and
cannot attribute them to that choice, and E4 reports contrast findings on 19 of the 21 routes
that share its palette.

\subsection{Authorship, Order and Freedom to Decline}
Ordered access and delayed disclosure coordinate a surprise, and they also let one partner
determine what the other must do before reaching a letter or a finale. The record shows one
such decision plainly: the ring level was changed from viewing a ring to taking a photograph,
so a person without camera access, or who prefers not to give an image, cannot clear the last
level. A scripted ``I do'' is a game response, not consent to a commitment, and the sequence
does not model refusal as a successful outcome. E1 and E2 add a mechanical fact about the same
level: it is the one that does not render offline and the slowest to load. The later products
move the other way. None is required, ordered or revealed, and the infirmary treats leaving as
an ordinary action. It also shows the limit of that movement: a place to decline conversation
was built by the person being declined, and its letters were drafted for both partners by one
side. Work on autonomy in technology-mediated relationships~\cite{zumaya2026relatedness} gives
terms for these questions; our record can only show where they arise.

\subsection{Answering in Person and Answering by Pipeline}
The specification for the request box listed ``an AI that answers her'' as a non-goal, so that
a request would reach one person. The next day's revision added a pipeline that can prepare an
answer in his place, and its answers at first carried no mark of how they were made. That
sequence is in the record. What it means to her is not. Studies of suggested replies and of
text believed to be written by AI report effects on how senders are
judged~\cite{jakesch2019aimc,hohenstein2023ai}, in settings where a person chooses what to
send; an answer prepared without the sender is a different case and we have no evidence about
it. The replay in E12 bears on the design in one way: under the published
guide the model escalated half of the in-scope synthetic requests, so a pipeline of this kind
does not obviously remove the person it was built to relieve.

\subsection{Privacy and What Travels}
The gate deters casual entry and provides neither encryption nor access control. Local completion
flags avoid a progress database; they do not make hosted photographs, voice notes or letters
private. E2 gives the inventory of other hosts each route calls, and it is short: fonts on every
route, map tiles on one, a music service on one. The request box changes the picture more than any
earlier feature. A request leaves her phone with its files and passes through an edge platform, a
mail provider, the designer's laptop, the provider of a hosted model and a second mail provider.
Publishing such a pipeline carries its own exposure, since instructions to a model describe the
work and the work describes the person who asked for it. Our rule is to describe the mechanism
and withhold what passes through it, and the checks follow the same rule by using synthetic
requests and a stand-in scope.

\subsection{Implications for Designing Personal Artefacts}
Four implications follow for people who build software for one other person. \emph{Put the
story's rules in the data.} Where the narrative fixes an order, the format that carries state
can refuse what the order excludes; here that one check removes every accepted error among the
enumerated edits. \emph{Test the ending under the worst conditions.} The last level should be
opened offline, on a slow link and with the camera refused, since it is the step a staged gift
cannot afford to lose. \emph{Check the palette before pages depend on it.} A shared palette
fixes contrast for every later page. \emph{Specify the delegate.} When a system may answer
under a partner's name, three things can be designed and checked: the scope it may answer
within, the mark the recipient sees on an answer, and the conditions that send a request to
the person.

\section{Limitations and Future Work}\label{sec:limits}
No recipient experience, sustained practice or relationship outcome is reported. The checks
describe one build at one commit, run once on one machine. The browser checks use desktop
Chromium with emulated size, touch, network and CPU, from a seeded state, and stop at
rendering. The edits in E7 are a model of typing errors and its notion of a reachable state
rests on a router rule we did not test. The secrecy lexicon and the 120 requests were written
for these checks by the same hands that wrote the checks, with a class balance we chose; the
triage replay uses the published prompt and a stand-in scope, repeats each request under
several conditions, and observes neither the answering model nor a reply. Model readings in E6
are a screen. Speech, breath and hand tracking were not evaluated, since that needs recordings
of people. The procedure has had no independent application.

The next step is a walkthrough on the recipient's own phone, if she wishes it, covering entry,
music, a refused camera, importing a code and the ending, under an ethics determination where
an institution is involved. A second is to repair the eight discrepancies and rerun the checks.
A third is an independent application of the procedure, as Section~\ref{sec:disc} describes.

\section{Conclusion}
\textsc{Meridian} documents a two-person game that turns autobiographical material into tasks,
places and ceremonies, and what the gift became in its first four weeks. We asked which of its
properties the maker could verify without data from the recipient. \emph{RQ1.} In desktop
Chromium under simulated conditions all 21 routes render with other hosts refused or stalled
and 18 render offline; the route that carries the ending is among the three that do not.
\emph{RQ2.} The progress decoder refuses every single-character substitution and accepts 24 of
the 5{,}664 one-edit mutations of the codes play can produce, every one a state that the design's ordering
rule excludes and that a decoder applying the rule refuses. The secrecy test covers the words
it lists and 2 of 72 synonyms and oblique hints. \emph{RQ3.} In a synthetic replay the triage model used in deployment passed 2 of 80
requests labelled for a person and escalated 21 of 40 in-scope ones; what the answering model
does with a passed request was not observed.

The finding we would carry to another project is the first of these: a rule that the designer
stated and the story enforced was missing from the format that carries the state, and a check
derived from the design record found it. The checks do not show that the game improves
intimacy or that she accepts its authored sequence. A remark of hers became a building, and a
channel meant to reach him can now be answered by a pipeline; what a partner is owed in each
case is a question for her.

\section*{Ethics and Publication Scope}
The machine-run checks use no data from the recipient: they read no request, no inbox, no
private configuration and no usage data, they use 120 synthetic requests with a stand-in scope,
and none submitted a form on the live site or sent mail. The design account is a different
matter. It is written from autobiographical material and from the designer's records of
operating the site, and those records include traces that originate with her: a remark quoted
as the public changelog records it, a booked flight, and the fact that she sent requests.
Appendix~\ref{sec:products} also reports checks the designer ran by hand on the answering
pipeline, some on requests she had sent; it reports that they ran and how the pipeline
behaved, and nothing of what a request contained. The two people whose names, dates,
photographs, voices and daily routines appear in the game and in this paper are its two
authors, and each has agreed to that disclosure. No interview or study of her experience was
conducted, and we did not seek an institutional ethics determination; readers and venues that
treat a co-author's deployment records as data about a person should weigh the account with
that in view. Any future study with participants would need a determination before data
collection.

\section*{Code and Data Availability}
The source is public at \url{https://github.com/rayford295/yi-town} (since 30 October 2026) and
the game runs at \url{https://jingyifan.love}. The design record was inspected at version 1.24.0,
commit \code{fc11108}, and MotionMirror at version 0.2.0, commit \code{80c4fd8}, in its own
repository, \url{https://github.com/rayford295/motionmirror}. The checks ran against the
artefact at version 1.25.0, commit \code{9193c20}. Their scripts, synthetic data, raw results
and figure code are not part of that commit: they are in \code{paper/experiments/} of later
commits, from \code{7281dee} on, with the browser version, the model digests and instructions
to rerun them. The development-record check reads the history up to the artefact commit
whatever the current head is. The answering pipeline's code is in the repository; its
configuration, the scope of its work and everything it processes are not. The code is released
under the MIT License. The photographs, voice notes, letters and other personal material in the
repository remain the authors' own: they are published so that the game can be read and run,
not as a research dataset, and may not be redistributed. A recorded walk through the whole game
is at \url{https://youtu.be/C0-oPjvNzZk}.

\section*{Declaration of Generative AI Use}
Claude (Anthropic), through Claude Code, wrote much of the software, drafted text and prepared
figures; Codex (OpenAI) was used to inspect the source and check the text. The checks of
Section~\ref{sec:eval} were designed, written and run with the same assistant: it wrote the
scripts, composed the 120 synthetic requests and the secrecy lexicon, chose the edit model for
the progress codes, read model outputs for confounds, and drew the figures by code from the
result files. Language models are also part of the system described and of its checks: the
answering pipeline runs two open models locally and calls Claude to prepare answers, and E6,
E10 and E12 use open models, run locally, as readers or as subjects. The authors reviewed what
appears here and are responsible for it.

\begin{acks}
To Coffee and Duoduo, who guard the doors. The street-view photographs on the Love Map are from
Wikimedia Commons and credited in the game; the \emph{Wedding March} is a public-domain recording
from the European Archive.
\end{acks}

\bibliographystyle{ACM-Reference-Format}
\bibliography{references}

\appendix
\section*{Appendices}
The appendices keep the complete account of the second version of this paper (24 November 2026),
unchanged apart from the removal of its abstract and of one figure that now appears in the main
text as Figure~\ref{fig:levels}, and from the factual corrections listed in
Appendix~\ref{sec:v3}. Appendix~\ref{sec:intro2} is that version's introduction;
Appendices~\ref{sec:related2} to~\ref{sec:discussion} are its sections in their original order,
including the full description of every level, every later product and every system
(Appendices~\ref{sec:experience} to~\ref{sec:products}) and the seven design episodes
(Appendix~\ref{sec:log}). Where that text says ``Section'', read ``Appendix''. Where it says that
the paper reports no evaluation, it describes the second version; Section~\ref{sec:eval} is new.
Appendix~\ref{sec:expdetail} gives the protocol and full results of each experiment, and
Appendix~\ref{sec:v3} lists what changed in this version.

\section{Introduction to the Second Version}\label{sec:intro2}
Long-distance partners use communication technologies to maintain awareness, exchange affection
and share ordinary activities~\cite{vetere2005mediating,lottridge2009sharing,neustaedter2012intimacy}.
Some report substantial intimacy despite geographical separation, but such findings do not imply
that a particular technology improves a relationship~\cite{jiang2013absence,stafford2005maintaining}.
This paper asks a narrower design question: how can a couple's shared material become an
asynchronous game, and what tensions arise when one partner authors both its narrative and its
interaction constraints?

\yt{} is a browser game developed by Yifan for Jingyi and himself. They began their relationship
on 20 September 2026 and live in Shenzhen, China, and College Station, Texas. Their clocks differed
by thirteen hours when development began; since Texas returned to standard time on 1 November
2026 the difference has been fourteen hours. The game draws on shared songs, a Gomoku game, dated memories and a symbolic
hand-drawn ticket representing a planned meeting. Players navigate a pixel town whose nine houses
contain levels, while a birthday prologue, a letter and a fictional wedding organise the narrative
(Figure~\ref{fig:teaser}). The game is built for separate visits; playing at the same time is
possible but never required.

We use \textsc{Meridian} as the name for this design case: Memory-Encoded Rituals for Intimacy
across Distance in an Interactive, Asynchronous Neighbourhood. A \emph{memory-encoded ritual}
means a scripted action whose intended significance derives from shared autobiographical material,
and \emph{intimacy} names the intended qualities of recognition, affection and shared meaning.
The central design move is to make a memory shape an action or a rule, rather than only illustrate
an interface.

The first version of this paper (30 October 2026) reported the artefact and a retrospective
reconstruction of its development before the recipient's planned first play on 31 October 2026.
This second version keeps that account and extends the record to 24 November 2026. In those four
weeks the site stopped being one game. Its front page now lists five cards and a sixth tile that promises more. Three of the cards
are products of a different kind from the game: a place to hide after a quarrel, a page for a
booked flight, and a box for asking the designer for help. A fourth leads out to a project the
two made together for other people. They matter to the first version's argument for two
reasons. One of them began with something the recipient said, which is the first trace in the
record of the recipient reshaping the gift. Another gave the site its first server and, a day
later, an automated pipeline that answers her in the designer's place while he sleeps.

The paper draws on research through design and autobiographical design, without claiming the
extensive lived use associated with mature first-person
studies~\cite{zimmerman2007rtd,gaver2012expect,neustaedter2012autobiographical}. It still reports
no recipient evaluation. Its contributions are:
\begin{enumerate}[leftmargin=*,itemsep=2pt,topsep=2pt]
  \item A documented design case that translates autobiographical material into nine game levels
  and staged ceremonies (Sections~\ref{sec:method} and~\ref{sec:experience}).
  \item Six situated design commitments and seven traceable development episodes, with explicit
  distinctions between recorded changes, interpretations and unevaluated intentions
  (Table~\ref{tab:commitments}; Section~\ref{sec:log}).
  \item An implementation account and verification audit of local progress, manual code exchange
  and selected interface properties, together with their privacy and reliability limits
  (Section~\ref{sec:system}).
  \item New in this version: an account of what the gift became in its first four weeks of
  existence beside its recipient (Section~\ref{sec:after}), including the path a request takes
  through a server, a local model and a hosted model (Section~\ref{sec:askpath}), and what that
  path does to the first version's privacy claims and to the meaning of an answer
  (Sections~\ref{sec:privacy} and~\ref{sec:delegated}).
\end{enumerate}
Section~\ref{sec:v2} lists what changed between the versions.


\section{Related Work}\label{sec:related2}

\subsection{Mediating Intimacy at a Distance}
Vetere et al.\ studied how couples and families express intimacy and argued that technologies for
strong ties should support the expressive, the physical and the playful, not only the
informational~\cite{vetere2005mediating}. Hassenzahl et al.\ reviewed 143 intimacy-mediating
artefacts and grouped them into six strategies: awareness, expressivity, physicalness, gift
giving, joint action and memories~\cite{hassenzahl2012love}. Many such systems are deliberately
minimal channels, such as the Virtual Intimate Object, a single button whose meaning the couple
negotiates~\cite{kaye2006clicked}. Neustaedter and Greenberg found that some long-distance couples
leave a video call running for hours to share ``being together'' rather than to
talk~\cite{neustaedter2012intimacy}; CoupleVIBE conveyed a partner's location by
vibration~\cite{bales2011couplevibe}; SyncDecor linked everyday appliances in two
homes~\cite{tsujita2008syncdecor}. \yt{} sits in Hassenzahl's \emph{memories} and \emph{gift giving}
strategies. What distinguishes it is what the interface represents: a collection of
shared references organised as tasks rather than a live awareness signal.

\subsection{Autobiographical and First-Person Design}
Neustaedter and Sengers defined autobiographical design as design research drawing on genuine
use by the people who built the system~\cite{neustaedter2012autobiographical}, and Desjardins
and Ball documented its tensions, including the blurred line between designer and
participant, and the intimacy of the data involved~\cite{desjardins2018tensions}. First-person
methods include situated inquiry through the researcher's own experience; a small
number of people alone does not establish methodological depth~\cite{lucero2019sample,desjardins2021firstperson}. Research through design treats the artefact itself as a carrier of
knowledge~\cite{zimmerman2007rtd,gaver2012expect}. \yt{} uses autobiographical material and developer self-testing, but the present account
precedes sustained use by the recipient. We therefore distinguish the making of the artefact
from an evaluation of living with it.

\subsection{Play, Ritual and Games for Two}
Huizinga described play as a ``magic circle'' set apart from ordinary life~\cite{huizinga1955homo};
Salen and Zimmerman made the circle a design tool~\cite{salen2004rules}; Juul argued that games
are half real rules and half fiction~\cite{juul2005halfreal}. Sicart argued that play
appropriates the world and its objects~\cite{sicart2014play}, and Isbister argued that games build
connection between players through shared action~\cite{isbister2016games}. Ludic design in HCI
values curiosity and ambiguity over task efficiency~\cite{gaver2004drift,gaver2003ambiguity}, and
slow technology designs for reflection rather than throughput~\cite{hallnas2001slow,odom2012technology}.
Aron et al.'s procedure for generating closeness through escalating mutual
disclosure~\cite{aron1997closeness} is, in a sense, a level structure. \yt{} takes the game form
literally: locks, levels, a guide, a prologue and a finale. Bespoke ritual technologies
already explore connection across distance~\cite{kirk2016ritual}; our case organises
selected memories and imagined futures as a sequence of game actions.

\subsection{Local-First and Backend-Free Progress}
Local-first software seeks to combine collaboration with user control and durable access to
data~\cite{kleppmann2019localfirst}. State-based conflict-free replicated data types (CRDTs)
converge when replicas combine compatible state through an appropriate merge; the grow-only set
uses union~\cite{shapiro2011crdt}. \yt{} adopts this operation for level completion and exchanges
state manually. Progress is local, but static hosting, remote media and browser-dependent
recognition remain part of its infrastructure.
Keeping progress local also does not protect personal images that are distributed as static
assets; visual content can expose location information~\cite{yang2024geolocator}.

\subsection{Answers Written by Machines}
Hancock et al.\ define AI-mediated communication as interpersonal communication in which an
agent modifies, augments or generates messages on a communicator's
behalf~\cite{hancock2020aimc}, and Jakesch et al.\ found that text believed to be written by AI
was trusted less once readers suspected a mix of human and machine
authorship~\cite{jakesch2019aimc}. Systems that let a language model act on text from outside
also inherit a security problem: instructions hidden in the content can redirect the
model~\cite{greshake2023injection}. The request box in Section~\ref{sec:askpath} raises both
questions inside one couple.


\section{Design Context and Method}\label{sec:method}

\subsection{The Two of Us}
Table~\ref{tab:dates} lists the dates that the system encodes. The couple speak Mandarin to each
other; Yifan works in English, Jingyi in Chinese and English. Both use phones as their main
devices, and Jingyi's network in mainland China blocks some services (YouTube, Google Fonts) that
Yifan's does not. These circumstances motivated bilingual text, alternative media providers and a
phone-oriented interface.

\begin{table}[t]
  \caption{Key dates encoded in the system (\code{src/core/config.ts}).}
  \label{tab:dates}
  \small
  \begin{tabular}{@{}lp{0.62\columnwidth}@{}}
    \toprule
    Date & Moment \\
    \midrule
    2026-09-20 & We met; day 1 of every counter in the town \\
    2026-09-25 & Yifan's birthday (Libra) \\
    2026-10-17 & First video call; the hand-drawn ticket CYF925 \\
    2026-10-23 & First Gomoku game, on a Zoom whiteboard \\
    2026-10-24 & First commit of \yt{} \\
    2026-10-30 & Repository made public; first version of this paper \\
    2026-10-31 & Jingyi's birthday (Scorpio); intended first recipient play \\
    2026-11-04 & The infirmary, after Jingyi's remark of 3 November \\
    2026-11-08 & Jingyi books her flights; the trip page \\
    2026-11-24 & The request box, its server and the answering pipeline \\
    2027-02-04 & Her flight lands in New York; countdown target since 8 November \\
    2027-02-05 & First planned meeting date; symbolic ticket CYF925, kept as a memory \\
    \bottomrule
  \end{tabular}
\end{table}

\subsection{Scope, Corpus and Reconstruction}
\begin{figure*}[t]
  \centering
  \includegraphics[width=\textwidth]{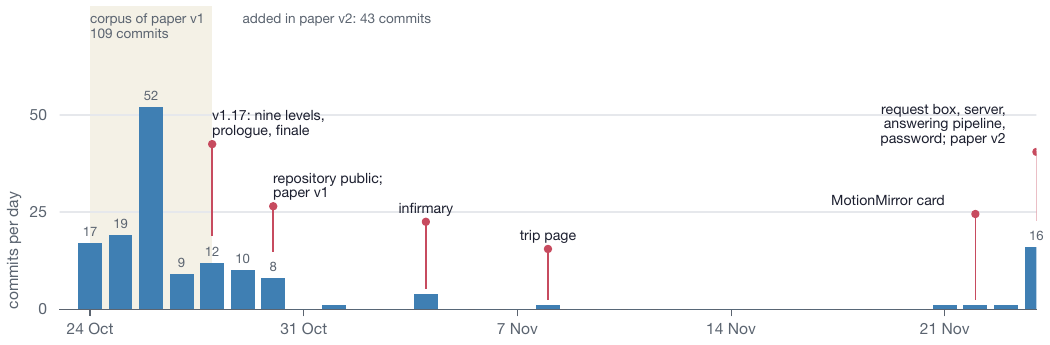}
  \caption{The development record, one bar for each day's commits on the public branch. The
  shaded span is the corpus of the first version of this paper: five days in which the town
  was built. Everything to its right is new in this version: two days of preparation for
  publication, then three additions spread over three weeks, then a single day,
  24 November, on which the request box, its server, the answering pipeline and the password
  were all built. Counts for that day are after the rewriting of history described in
  Section~\ref{sec:privacy}.}
  \Description{A bar chart of commits per day from 24 October to 24 November 2026. Tall bars on
  24 to 30 October, the tallest 52 on 26 October; a long quiet stretch with one commit on
  1 November, four on 4 November and single commits on 8, 21, 22 and 23 November; and a bar of
  16 on 24 November. Labelled markers name the
  events on each active day.}
  \label{fig:timeline}
\end{figure*}

The first version's corpus comprised 109 commits dated 24--28 October 2026, 28 changelog entries
from version 0.1.0 to 1.17.0, and four specifications, for the site, the quest, the College
Station day and the town, with a cut-off at commit \code{55440ae}. This version extends the
cut-off to commit \code{fc11108} (24 November 2026, version 1.24.0): 152 commits, 36 changelog
entries and a fifth specification, for the request box (Figure~\ref{fig:timeline}). The count of commits is that of the public branch after the rewriting of
24 November described in Section~\ref{sec:privacy}, which merged five commits into two. The 43
added commits are unevenly spread: the infirmary and the trip page
fall in the first eight days of November, and the request box, its server and the answering
pipeline in two days at the end.
Yifan directed the design, supplied the autobiographical material and tested both player roles,
with substantial implementation assistance from an AI coding assistant. Jingyi is the intended
recipient. This account reports no interview, diary, recipient play session or relationship measure.

The added record holds three traces that originate with the recipient: a remark of hers quoted in
the changelog, a flight she booked, and requests she sent through the request box. A second program, MotionMirror, is reached from the site's dashboard and kept in its own
repository; we inspected it at commit \code{80c4fd8} of 24 November 2026 and describe it in
Section~\ref{sec:mm}, without counting its history in the corpus. We report that
these traces exist and what the designer built in response. We do not report the content of her
requests, and we treat none of the three as an evaluation. Whether and how the first play of
31 October went is not part of the corpus, and we make no claim about it.

For the present reconstruction, we inspect the documented changes underlying seven episodes.
Four are from the first version: replacing Pictionary, changing the ring's completion condition,
correcting represented routines, and revising the entry gate. Three are new: building the
infirmary from a remark, replacing the symbolic ticket, and reversing a stated non-goal of the
request box (Section~\ref{sec:log}). Each episode links a source change to an
interpretation of the design trade-off. The selection illustrates decisions discussed in this
paper rather than an exhaustive coding of all commits. Version history establishes what changed
and when, not the designer's unrecorded motivation or the recipient's response.

\subsection{Knowledge Claims and Positionality}
Research through design can make an artefact and its rationale available for scrutiny without
claiming a universal solution~\cite{zimmerman2007rtd,gaver2012expect}. First-person inquiry also
requires attention to whose experience becomes evidence~\cite{desjardins2018tensions,desjardins2021firstperson}.
Here, the designer's familiarity with the relationship informs the content but cannot stand in
for the other partner's perspective. The six commitments below are situated design choices
reconstructed from the artefact and log, not validated principles.

\subsection{Six Design Commitments}
We articulate six commitments from the reconstruction of the first five days of development.
Table~\ref{tab:commitments} maps each to the mechanisms that realise it. They were written for
the town; Section~\ref{sec:after} asks which of them the later products keep.

\begin{description}[leftmargin=0pt,labelwidth=0pt,labelsep=0pt,style=nextline,itemsep=2pt,topsep=2pt,font=\normalfont\scshape]
  \item[D1 Memory-derived mechanics.] Selected memories and shared references shape tasks.
  The Gomoku level reworks a prior game; the ring represents an imagined future rather than a past event.
  \item[D2 Staged access.] Ordered levels and door entry organise the narrative and pace normal
  play; they are not security barriers.
  \item[D3 Local progress.] Completion state is stored on-device and shared through a code,
  without an author-operated progress backend. External services remain involved in other features.
  \item[D4 Bilingual coverage.] Both dictionaries contain non-empty values for the same keys.
  \item[D5 Code-defined pixel art.] Town tiles and character sprites are generated in code.
  Photographs, audio, rendered rings and other assets have separate provenance and reuse conditions.
  \item[D6 Delayed narrative disclosure.] A lexical test checks dictionary entries shown before the
  finale for explicit references to it.

\end{description}

\begin{table}[t]
  \caption{Design commitments and the mechanisms that realise them.}
  \label{tab:commitments}
  \small
  \begin{tabular}{@{}p{0.30\columnwidth}p{0.62\columnwidth}@{}}
    \toprule
    Commitment & Realised by \\
    \midrule
    \dc{1} Memory-derived mechanics & Memories, cultural references and imagined futures become tasks (Table~\ref{tab:levels}) \\
    \dc{2} Earned access & Ordered level registry; router guard; door-only entry; cats guarding the next locks \\
    \dc{3} Local progress & \code{localStorage} per player; grow-only set; progress codes; PWA offline cache \\
    \dc{4} Bilingual coverage & Matching non-empty dictionary keys; no independent language-quality evaluation \\
    \dc{5} Drawn in code & Sprites and tiles as strings; banner and icons generated from them \\
    \dc{6} Delayed disclosure & Lexical test over eligible dictionary keys, with explicit exclusions \\
    \bottomrule
  \end{tabular}
\end{table}

\section{The Experience}\label{sec:experience}

\subsection{The Front Door}
\begin{figure}[t]
  \centering
  \begin{subfigure}[b]{0.49\columnwidth}\includegraphics[width=\linewidth]{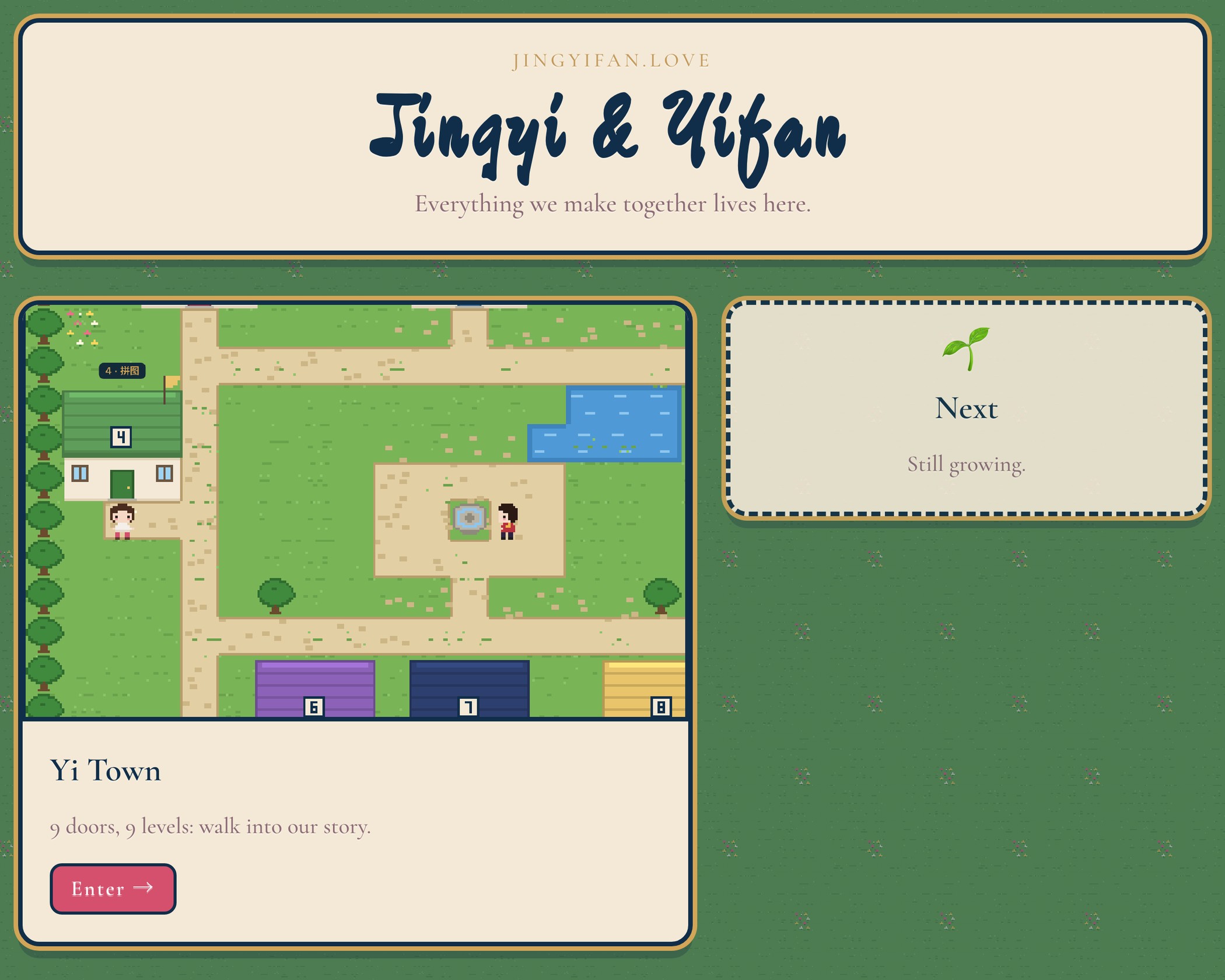}\caption{Dashboard}\end{subfigure}\hfill
  \begin{subfigure}[b]{0.49\columnwidth}\includegraphics[width=\linewidth]{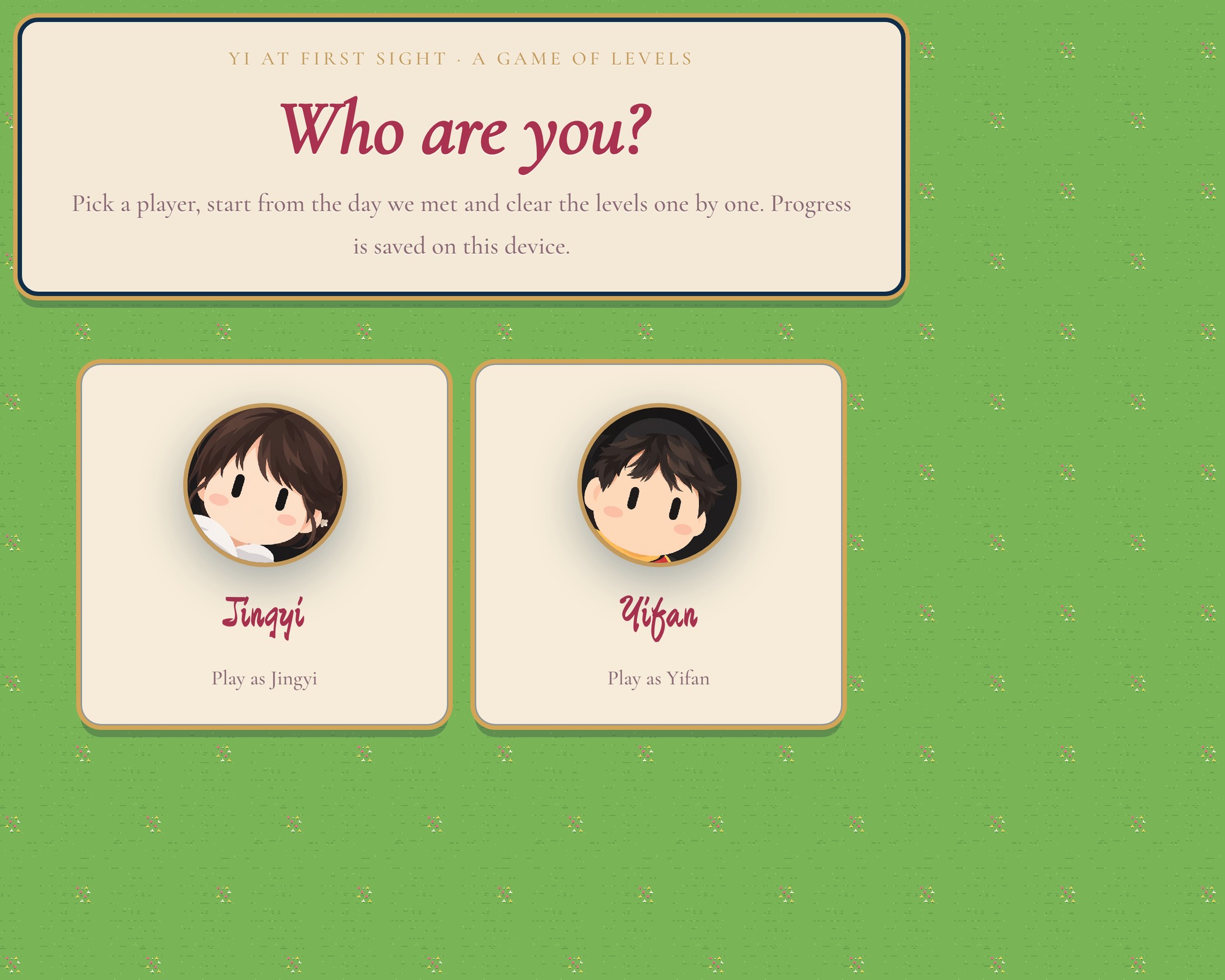}\caption{Pick a player}\end{subfigure}\\[3pt]
  \begin{subfigure}[b]{0.49\columnwidth}\includegraphics[width=\linewidth]{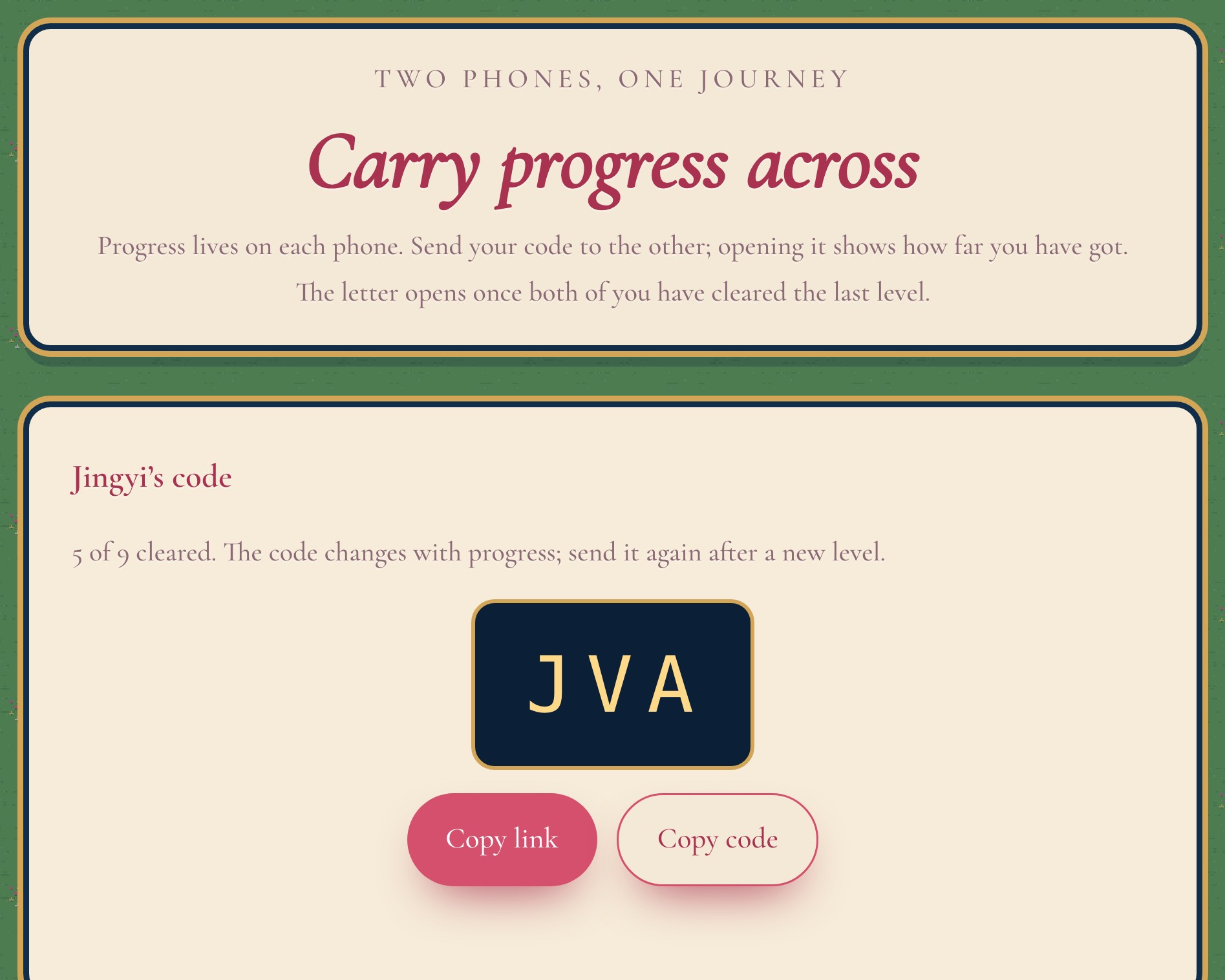}\caption{Progress code}\end{subfigure}\hfill
  \begin{subfigure}[b]{0.49\columnwidth}\includegraphics[width=\linewidth]{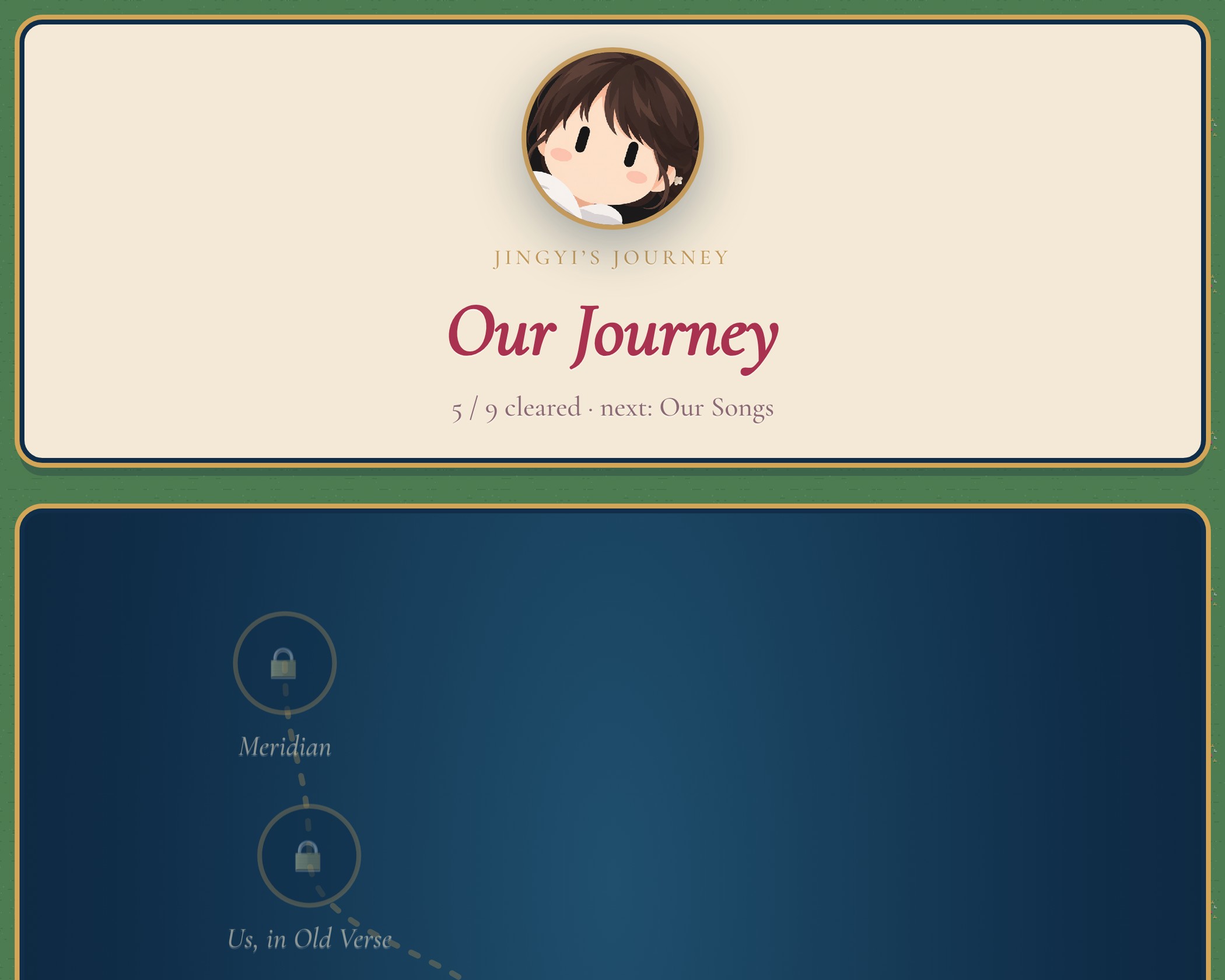}\caption{The journey list}\end{subfigure}
  \caption{Around the town. (a) The dashboard lists every product under the domain. (b) Each device
  keeps one progress per player. (c) Progress crosses phones as a code (here \code{JVA}). (d) The
  journey as a list, for players who prefer it to the map.}
  \Description{Four screenshots in cream and navy on a green pixel meadow: a dashboard card with a
  map thumbnail; two avatar cards; a large three-letter code; a handwritten-style letter.}
  \label{fig:front}
\end{figure}

The site opens on a single prompt. At the first version it was a question, and the answer, a
date both players know, was accepted in four spellings. Since version 1.24 it is a password
field that accepts one spelling, the date's eight digits. In both forms the digits typed are
compared against SHA-256 hashes~\cite{fips1804}. Writing
$\delta(a)$ for the digits of an answer $a$ in order, $H$ for SHA-256 and $\mathcal{A}$ for the
stored digests, four at first and now one,
\begin{equation}\label{eq:gate}
  \mathrm{pass}(a) \iff H\big(\delta(a)\big) \in \mathcal{A} .
\end{equation}
Hashing changes how the answer is compared, but the date also appears in the game's own narrative;
it is a recognition question, not a secret, and calling it a password does not make it one. Behind it is a dashboard that lists everything the couple
build under the domain (Figure~\ref{fig:front}a), of which \yt{} is the first; the figure shows
it as it was at the first version, with one card; Figure~\ref{fig:dash} shows it now, and
Section~\ref{sec:after} describes what was added. Choosing the town
asks \emph{Who are you?} (Figure~\ref{fig:front}b). Progress is kept per player on each device,
so both can play on one phone.

\subsection{The Town}
The town (Figure~\ref{fig:teaser-map}) is a $24\times20$ grid of $16\times16$-pixel tiles. Paths
connect nine houses around a fountain square and a pond. Each roof has its own colour and its
level number, and a cleared house flies a gold flag. The player moves with the arrow keys, by
tapping a tile (the avatar walks a shortest path to it), or with an on-screen d-pad. Walking into
a door opens a talk box with the level's one-line story, for example ``Our first game was on a Zoom
whiteboard. Now you place the last stone.''

\begin{figure}[t]
  \centering
  \includegraphics[width=\columnwidth]{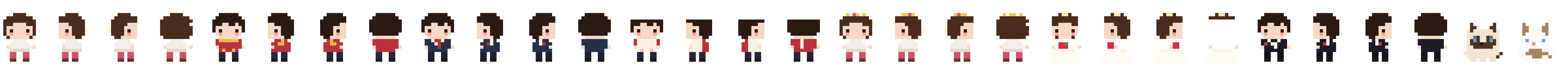}
  \caption{Every character in \yt{} is a string of characters in the source. From left: Jingyi,
  Yifan, Rayford, Yifan as the baker, Jingyi crowned, the bride and the groom (four facings each),
  then the cats Coffee and Duoduo. Bodies are $12\times12$ pixels over four rows of shared legs.}
  \Description{A row of tiny pixel-art characters seen from four sides, followed by two pixel cats.}
  \label{fig:sprites}
\end{figure}

Four kinds of character live in the town (Figure~\ref{fig:sprites}). The \emph{partner} waits by
the fountain and reports how far they have got (``I am waiting here. I have cleared 5 levels.
You?''). \emph{Rayford}, Yifan in a suit, is a scripted guide south of the fountain: he explains
the project, lists the levels, hints at the next one and walks the player to its door. Two Ragdoll
cats, \emph{Coffee} (\zh{咖啡}) and \emph{Duoduo} (\zh{哆哆}), guard the locks: Coffee sits at the next
locked door and Duoduo at the one after. When a door opens they walk to their new posts, and with
nothing left to guard they sunbathe by the pond. Bumping into a cat shows a photo of the real cat.
Two signposts lead out of the square to the two cities (Section~\ref{sec:days}), and a red mailbox
holds voice notes the partner has left.

The sky follows the clock in the player's own city: Jingyi walks the town in Shenzhen time and
Yifan in College Station time, so at the same instant one sees noon and the other sees lit windows
under a night wash. The whole site stands on the same pixel meadow, and every heading is a cream
sign in the town's palette, so leaving a house never leaves the world.

\subsection{Nine Levels}
Table~\ref{tab:levels} lists the levels with their source memory, mechanic and input modality,
and Figure~\ref{fig:levels} shows them in play.

\begin{table*}[t]
  \caption{The nine levels: autobiographical material, shared references and imagined futures become tasks (\dc{1}). Modalities: \emph{V}~voice,
  \emph{T}~touch or pointer, \emph{C}~camera, \emph{K}~keyboard alternative available.}
  \label{tab:levels}
  \small
  \begin{tabular}{@{}clp{0.25\textwidth}p{0.36\textwidth}cl@{}}
    \toprule
    \# & Level & Design material & Task (clear condition) & Input & Data \\
    \midrule
    1 & The Day We Met & 20 September 2026; the vow & Answer a question, then say \emph{\zh{我爱你，<name>，我愿意}} aloud; homophones of the name accepted & V/K & 1 vow per player \\
    2 & Love Map & Where each studied; where they will meet & Name six street views; each right answer drops a heart on the map to collect & T & 6 places \\
    3 & Tattoo & A promise to write on each other & Handwrite a line on an arm or shoulder, then ``ink it'' & T & strokes \\
    4 & Moments & The first 40 days & Order five memories, then reassemble a $3\times3$ portrait of the partner & T & 8 memories \\
    5 & Gomoku & First game on a Zoom whiteboard & Ten win-in-one endgames; the ten answers form a heart & T & 10 positions \\
    6 & Our Songs & The six songs they share & Hear a passage, pick the title from four & T & 6 songs \\
    7 & Libra and Scorpius & Their birth signs, neighbours in the sky & Join 21 stars along 20 edges & T/K & J2000 positions \\
    8 & Us, in Old Verse & Classical love poetry & Restore the missing half-line of five poems character by character & T & 5 poems \\
    9 & Meridian & What comes next & Try the ring on through the camera and take a photo & C & 3 GLB models \\
    \bottomrule
  \end{tabular}
\end{table*}

\paragraph{1. The Day We Met.} The first level shows the day count since 20 September, the time
in both cities with day and night icons, and a countdown to the flight. To clear it, each player
reads a vow, rewritten from the wedding formula as the vow of ``becoming each other's partner
once more'', and says one line aloud: \zh{我爱你，陈静仪，我愿意} (``I love you, Jingyi Chen, I do''),
with the partner's full name. Browser speech recognition~\cite{w3c2023speech} hears Mandarin
names inconsistently, so the matcher removes everything except Han characters and Latin
letters and accepts any of eight or nine homophone spellings of each name (for example \zh{陈静怡},
\zh{陈静宜}). A typed fallback keeps the level playable in browsers that cannot listen.

\paragraph{2. Love Map.} Six places matter to the couple: Texas, Los Angeles, Japan, Shenzhen,
Hong Kong and New York. The street views are public photographs of well-known sites from
Wikimedia Commons rather than the couple's own pictures, since a personal photograph can give away
far more of a place than its owner intends~\cite{yang2024geolocator}. Before each heart appears on a Leaflet map, the player names a
street-level photograph from four options (Figure~\ref{fig:levels}, panel 2), and a wrong pick reshuffles the
options. Great-circle arcs connect the two home cities.

\paragraph{3. Tattoo.} The player handwrites a line or a small drawing on an arm or a shoulder,
drawn in SVG, and inks it. The tattoo is kept on the device and can be saved as an image to send.

\paragraph{4. Moments.} The timeline holds eight dated memories, each with photos and a story in
both languages, from hearts made from both sides of the ocean (day~21) to a heart drawn on the
Gomoku board (day~35). The player orders five of them by tapping pairs to swap, then reassembles a
$3\times3$ portrait of the partner.

\paragraph{5. Gomoku.} Each of ten positions on a $15\times15$ board has exactly one winning point
for the player's own colour (purple for Jingyi, green for Yifan, as in their first game). The ten
answers are the ten points of a heart; solved answers stay on the board in gold, so the heart draws
itself as the player progresses. Positions are generated from three patterns (a four closed at one
end, a broken four, three-and-one) placed around each heart point in varied directions with decoy
stones (Figure~\ref{fig:heart}). Writing $M$ for the mover's stones, $\mathcal{E}$ for the empty
cells and $\mathrm{five}(\cdot)$ for ``contains five in a row'', a test requires of every position
$k$ that
\begin{equation}\label{eq:win}
  \big|\{\, c \in \mathcal{E} : \mathrm{five}(M_k \cup \{c\}) \,\}\big| = 1, \qquad
  \{c^{\ast}_1, \dots, c^{\ast}_{10}\} = \mathcal{H},
\end{equation}
where $c^{\ast}_k$ is that unique cell and $\mathcal{H}$ the ten cells of the heart. Gomoku is a
first-player win under optimal play~\cite{allis1994searching}; win-in-one puzzles avoid that
question entirely and make every move feel like a find.

\begin{figure}[t]
  \centering
  \begin{tikzpicture}[x=5.4mm, y=5.4mm, hpt/.style={circle, draw=sCrB, line width=0.6pt, fill=gold!35, text=sInk, inner sep=0pt, minimum size=4.2mm, font=\scriptsize\sffamily}]
    \fill[sCrF, rounded corners=3pt] (-0.6,-0.6) rectangle (14.6,14.6);
    \draw[sLine!70, line width=0.3pt, step=1] (0,0) grid (14,14);
    \foreach \i in {3,7,11} \foreach \j in {3,7,11} \fill[sLine] (\i,\j) circle (0.09);
    \node[hpt] at (7,9) {\tiny 1};
    \node[hpt] at (6,10) {\tiny 2};
    \node[hpt] at (5,9) {\tiny 3};
    \node[hpt] at (5,8) {\tiny 4};
    \node[hpt] at (6,7) {\tiny 5};
    \node[hpt] at (7,6) {\tiny 6};
    \node[hpt] at (8,7) {\tiny 7};
    \node[hpt] at (9,8) {\tiny 8};
    \node[hpt] at (9,9) {\tiny 9};
    \node[hpt] at (8,10) {\tiny 10};
    \fill[ygreen] (12,4) circle (0.36); \fill[ygreen] (5,11) circle (0.36); \fill[ygreen] (6,12) circle (0.36); \fill[ygreen] (10,8) circle (0.36); \fill[ygreen] (5,6) circle (0.36);
    \fill[jpurple] (8,8) circle (0.36); \fill[jpurple] (9,7) circle (0.36); \fill[jpurple] (10,6) circle (0.36); \fill[jpurple] (11,5) circle (0.36); \fill[jpurple] (4,7) circle (0.36); \fill[jpurple] (9,12) circle (0.36);
    \draw[sRed, line width=1.1pt, dashed] (7,9) -- (11,5);
    \draw[sRed, line width=1pt] (7,9) circle (0.5);
    \node[anchor=north west, font=\scriptsize\sffamily, text=sSoft] at (-0.6,-0.9) {$15\times15$ board; purple: mover (Jingyi), green: partner's decoys};
  \end{tikzpicture}
  \caption{Ten answers draw a heart. The gold discs are the ten cells of $\mathcal{H}$ in the order they
  are found (Eq.~\ref{eq:win}); the stones are position~1 as generated in the source
  (\code{heart-1}): six purple stones for the mover, five green decoys, and the unique winning cell
  (red ring) completing a diagonal five. Later positions are built the same way around the other nine
  cells, so solved answers accumulate into the outline.}
  \Description{A Gomoku grid with ten numbered gold discs forming a heart, a few purple and green stones, and a red ring marking the winning cell on a dashed diagonal.}
  \label{fig:heart}
\end{figure}

\paragraph{6. Our Songs.} Six songs the couple share (among them \zh{只对你有感觉} and
\zh{天下第一好}) play a passage through the official embedded players: QQ Music for the player in
China and YouTube for the player in the United States, each switchable. The site hosts no audio.
The embedded players, as used here, report no playback event to the page, so the options unlock
after eight seconds on screen, whether or not the passage was heard.

\paragraph{7. Libra and Scorpius.} Yifan is a Libra (25 September) and Jingyi a Scorpio
(31 October); in the sky the two constellations are neighbours, and Libra's two brightest stars
still bear the Arabic names of the Scorpion's claws, \emph{Zubenelgenubi} and
\emph{Zubeneschamali}. The level projects 21 manually entered stars, six for Libra and fifteen for Scorpius,
and asks the player to draw 20 edges. The source records coordinates as J2000 right ascension
and declination, following the conventions in Meeus~\cite{meeus1998astronomical}; the chart is an
illustration, not a view of either player's current night sky.

\paragraph{8. Us, in Old Verse.} Five classical lines about love lose their second half, to be
restored from a bank of characters: a farewell traditionally attributed to Su Wu, \kai{结发为夫妻，恩爱两不疑};
Guan Daosheng's \emph{Song of You and Me}; a couplet of Li Shangyin; the paired birds of the
sixth-century anthology \emph{New Songs from a Jade Terrace}~\cite{birrell1982jade}; and an
anonymous couplet. Each solved line shows a plain English gloss.

\begin{figure}[t]
  \centering
  \includegraphics[width=0.325\columnwidth]{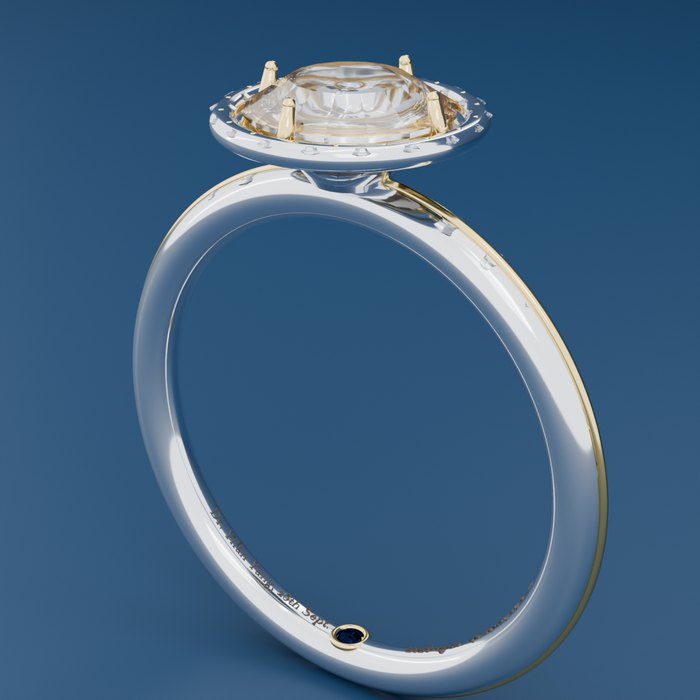}\hfill
  \includegraphics[width=0.325\columnwidth]{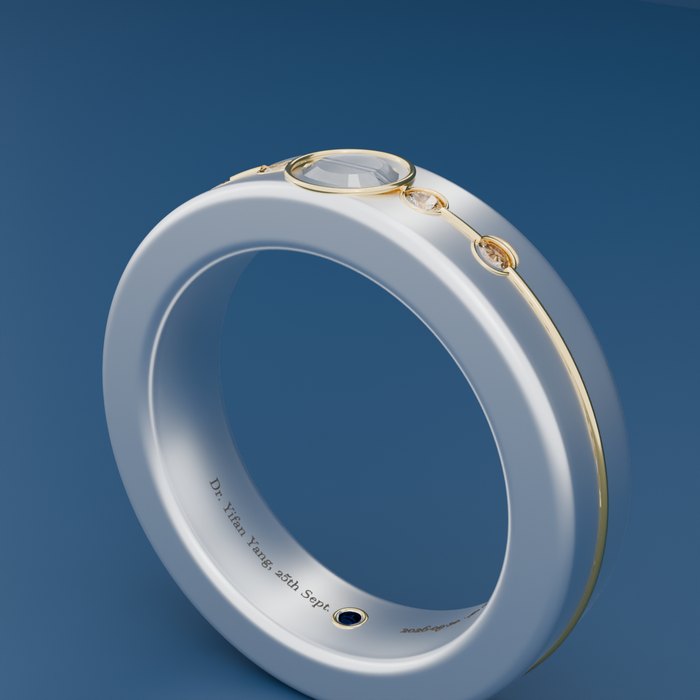}\hfill
  \includegraphics[width=0.325\columnwidth]{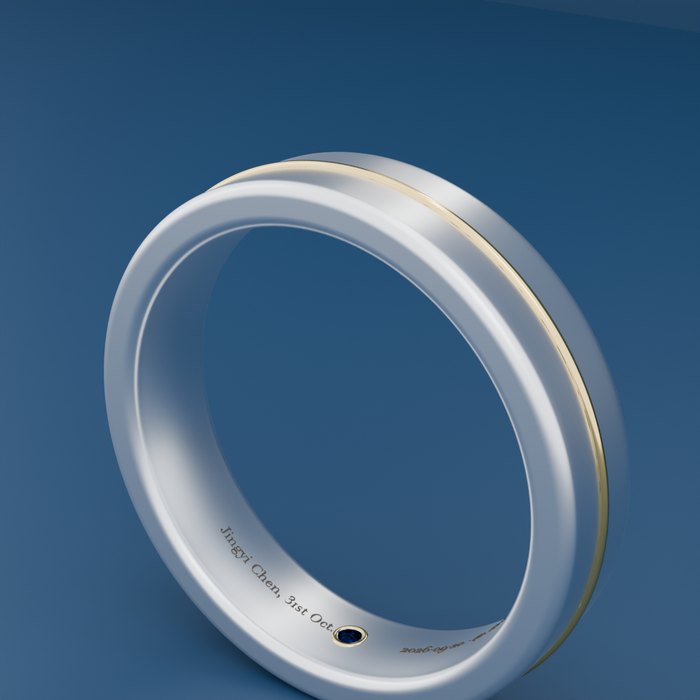}
  \caption{The Meridian pair, path-traced from the procedural Blender model: hers v1 (solitaire,
  halo and graduated shoulder stones), hers v2 (flush-set 4.2\,mm stone in a 5\,mm flat band), and
  his (brushed flat comfort-fit band). The gold meridian runs around each band.}
  \Description{Three renders of rings on a dark background: a diamond solitaire, a flat band with
  a sunken stone, and a plain wide band, each with a thin gold line around it.}
  \label{fig:ring}
\end{figure}

\paragraph{9. Meridian.} The last level is a pair of rings, called \emph{Meridian}, modelled
procedurally in Blender from a Python script (Figure~\ref{fig:ring}). The pair share a signature:
a yellow-gold line inlaid around a platinum band like a meridian on a globe, a hidden sapphire
inside the bottom, and an inner engraving that carries the \emph{other} person's name and birthday,
the date they met and the coordinates of the airport where they will meet ($40.64^\circ$N
$73.78^\circ$W). Her ring exists in two designs, a solitaire on four compass prongs (v1) and the
current flush-set stone sunk into his band's profile (v2); the page keeps both as a design history.
The level is cleared only one way: the player tries the ring on through the camera and takes a
photo (Section~\ref{sec:sensing}).

\subsection{Two Days}\label{sec:days}
\begin{figure}[t]
  \centering
  \begin{subfigure}[b]{0.49\columnwidth}\includegraphics[width=\linewidth]{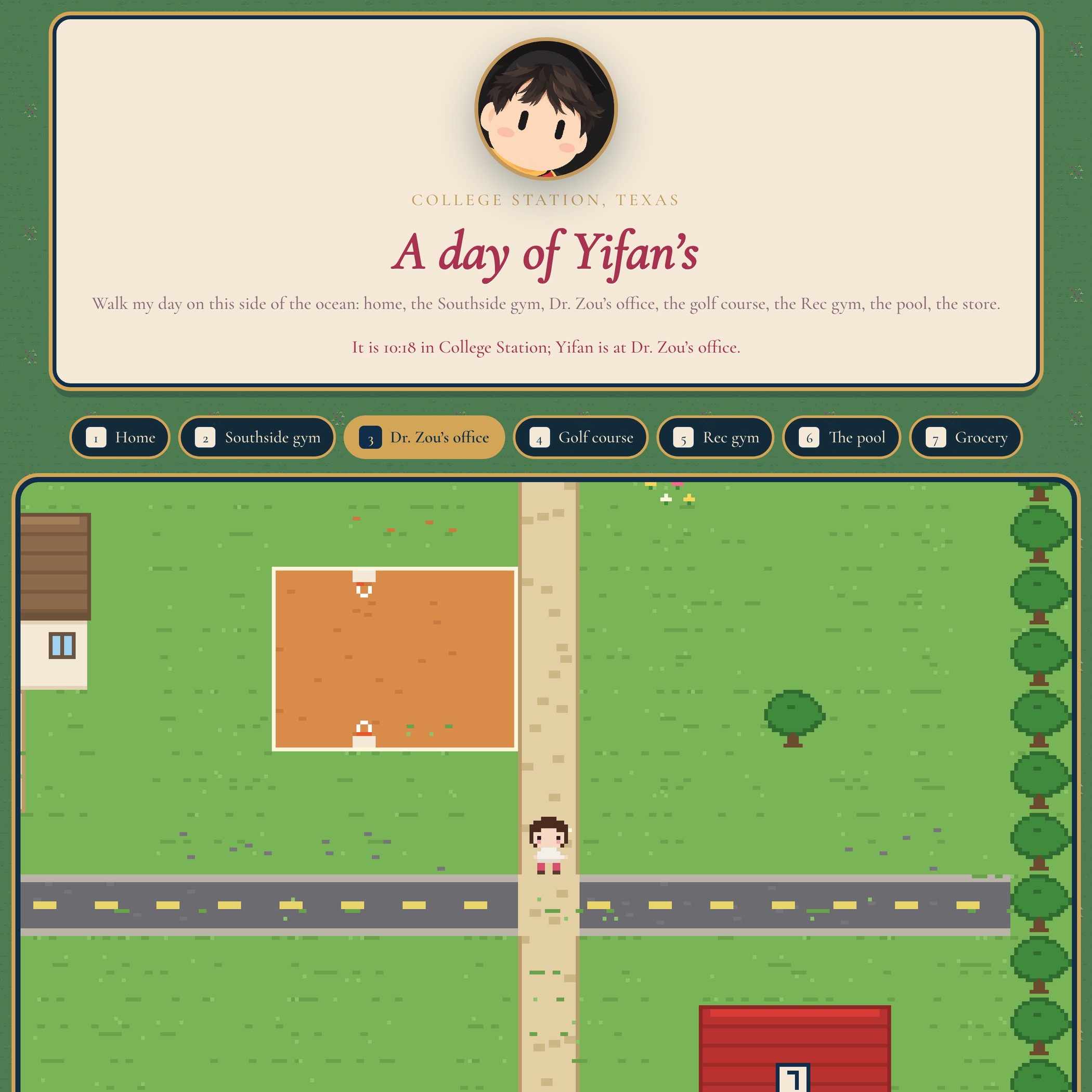}\caption{A day of Yifan's}\end{subfigure}\hfill
  \begin{subfigure}[b]{0.49\columnwidth}\includegraphics[width=\linewidth]{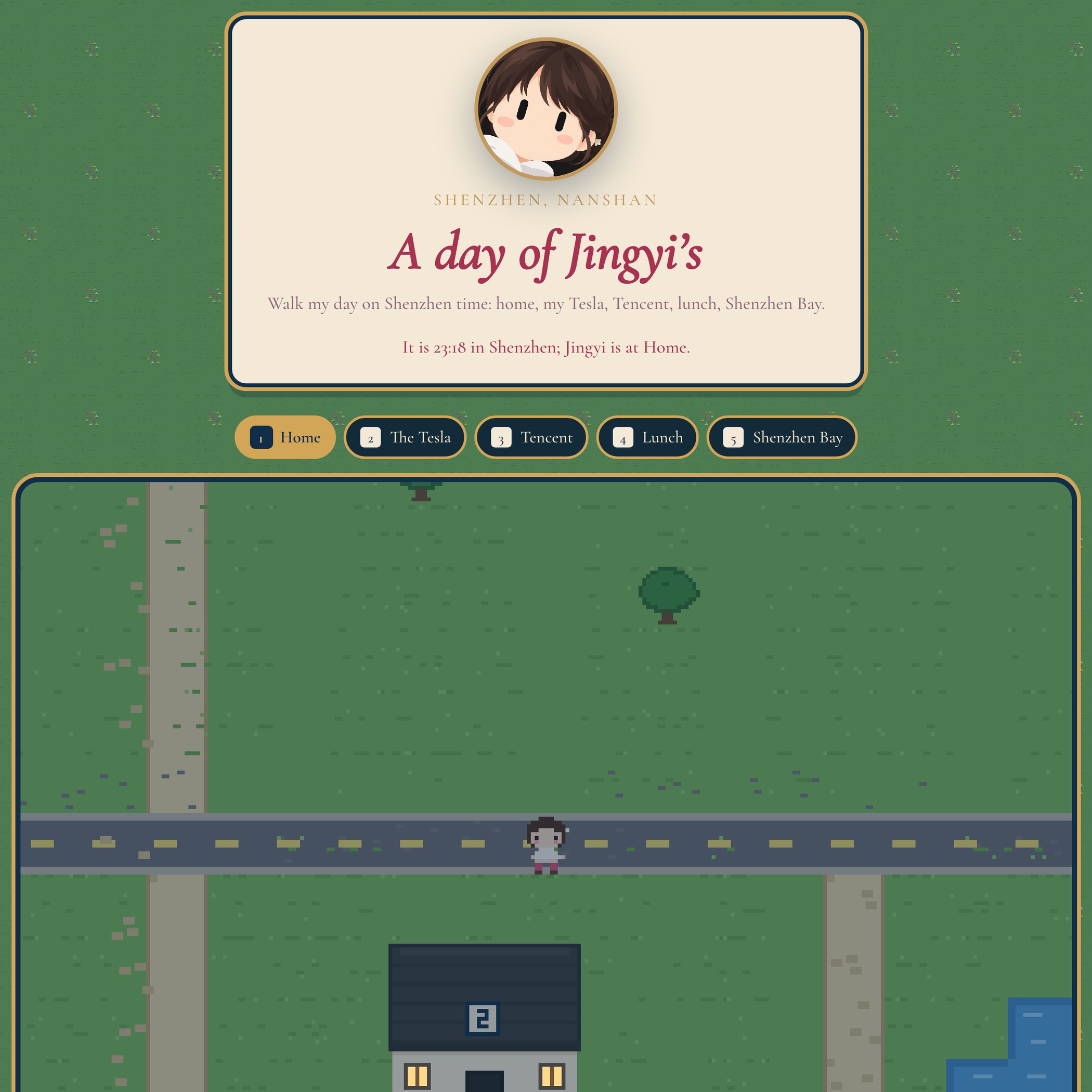}\caption{A day of Jingyi's}\end{subfigure}
  \caption{The two cities as maps. The person stands at the stop whose hours contain the current
  local time in their own city: (a) Yifan at his advisor's office at 8:49 a.m.\ in College
  Station; (b) Jingyi at home at 9:49 p.m.\ in Shenzhen, the same instant.}
  \Description{Two pixel maps with roads and buildings, each with a person card above it and a row
  of stop buttons.}
  \label{fig:days}
\end{figure}

Two signposts lead from the square to a day in each city (Figure~\ref{fig:days}). Each is a second
map with the person's stops: Yifan's home, gym, his advisor's office, the golf course and the
pool in College Station; Jingyi's home with the cats at the door, her drive, the office and
Shenzhen Bay. Each stop has hours, and the person's sprite stands wherever the clock in their own
city puts them. A partner who opens Yifan's day at 9:49 p.m. in Shenzhen finds him at 8:49 a.m. at
his advisor's office. The schedules are authored, not sensed; they began as guesses and were corrected by the designer: Yifan drives
rather than taking the bus; Jingyi drives a Tesla rather than taking the metro; and the
Shenzhen Bay stop keeps a joke between them about where they will one day live.

\subsection{Prologue and Finale}
\begin{figure*}[t]
  \centering
  \includegraphics[width=0.245\textwidth]{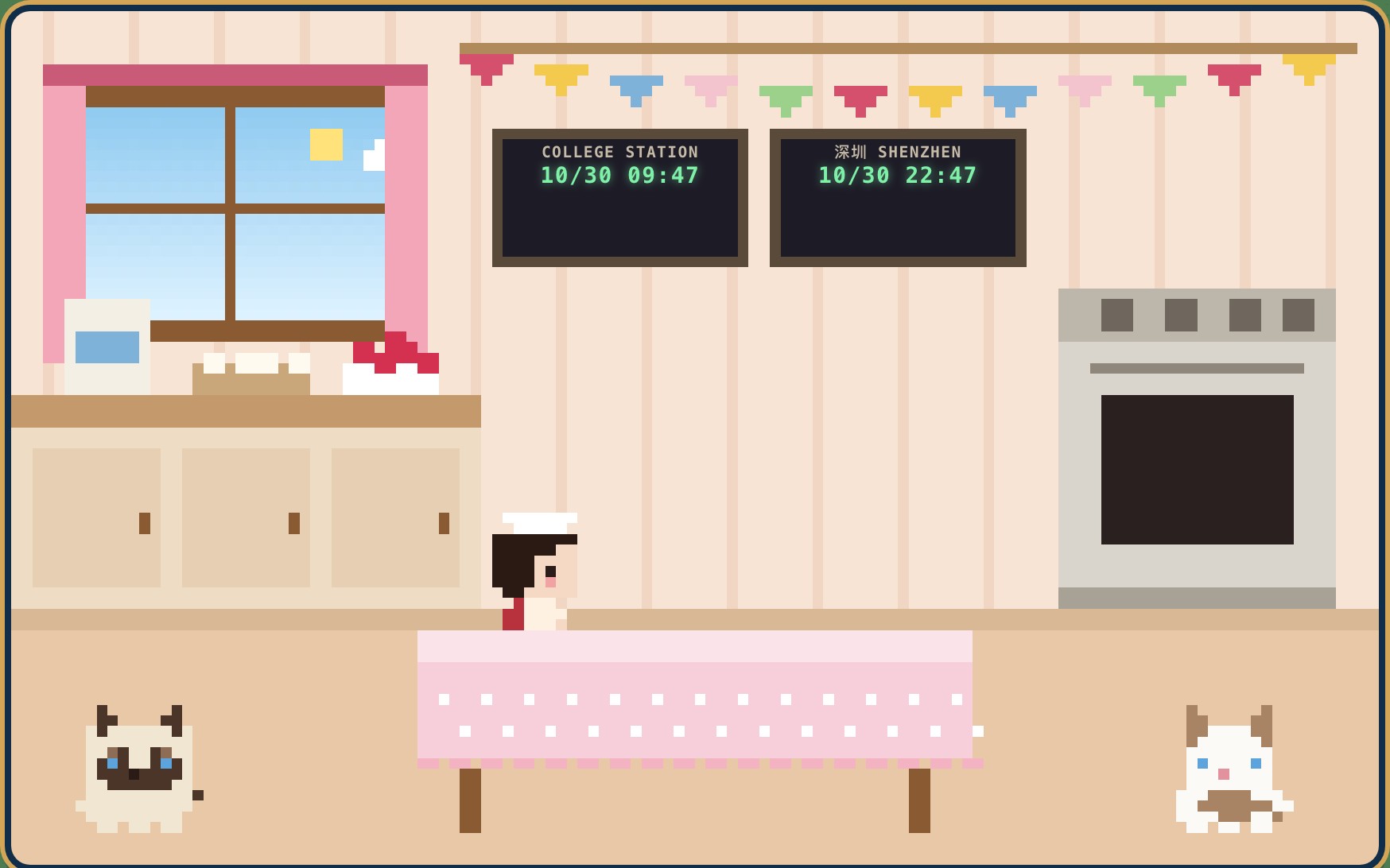}\hfill
  \includegraphics[width=0.245\textwidth]{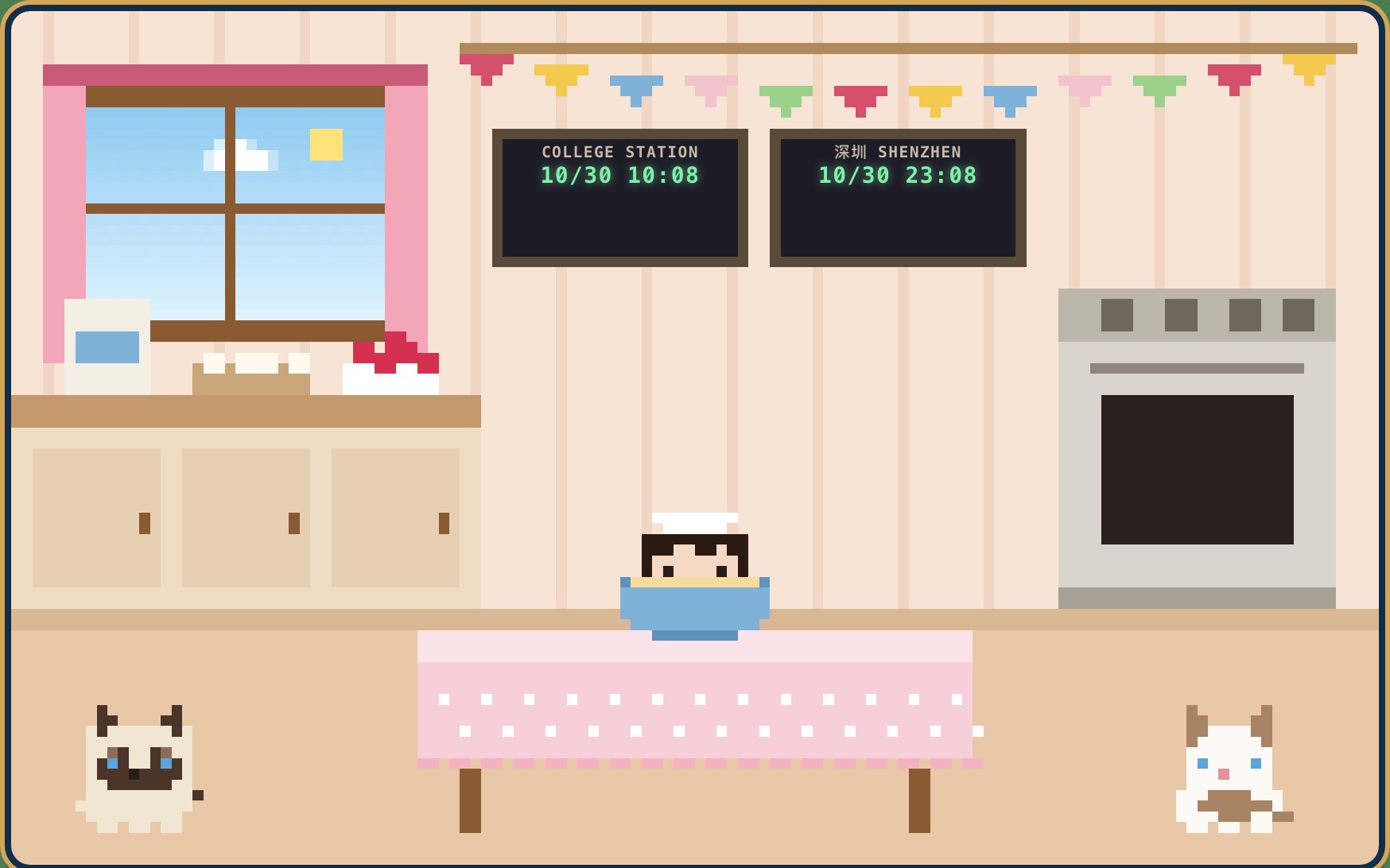}\hfill
  \includegraphics[width=0.245\textwidth]{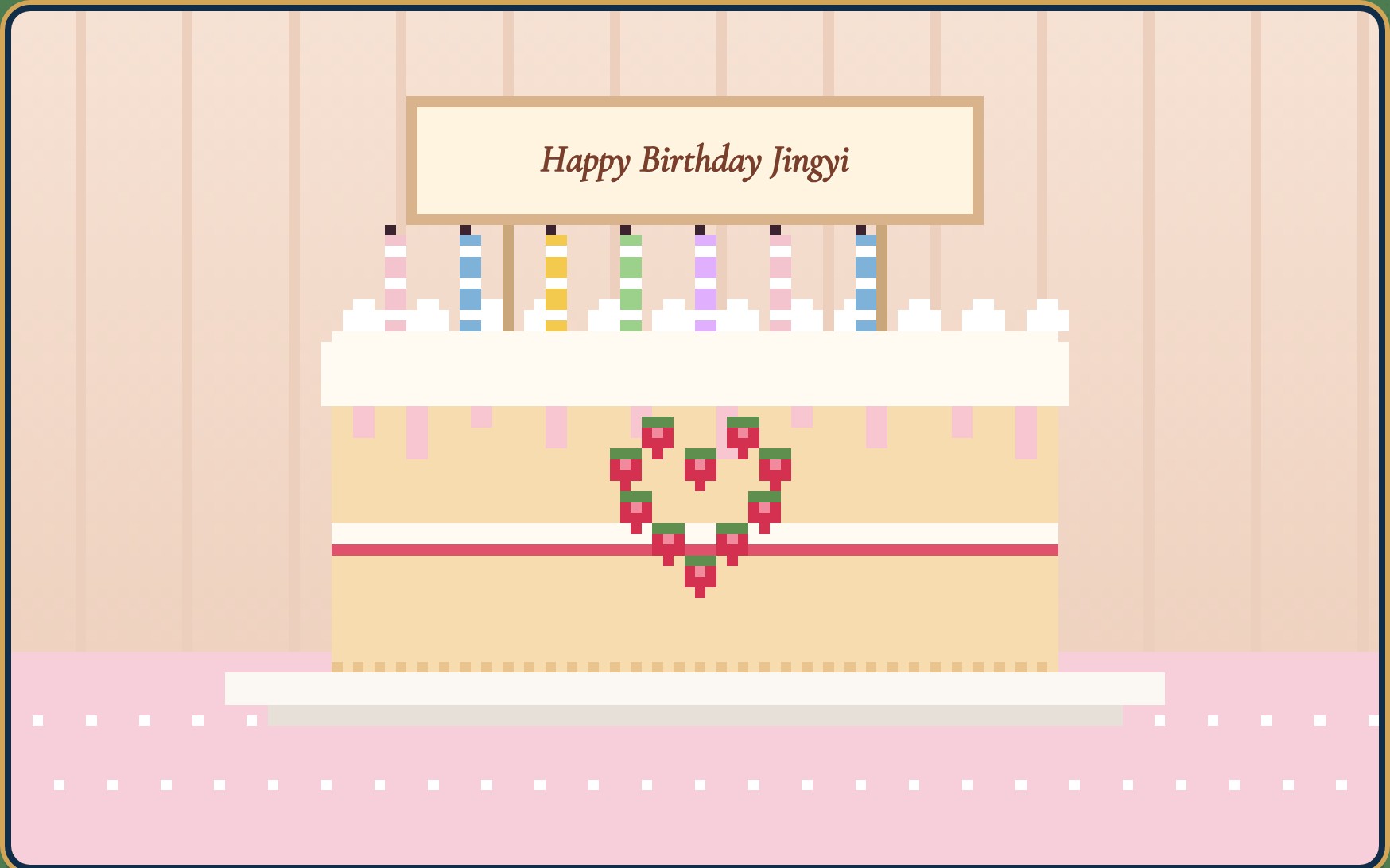}\hfill
  \includegraphics[width=0.245\textwidth]{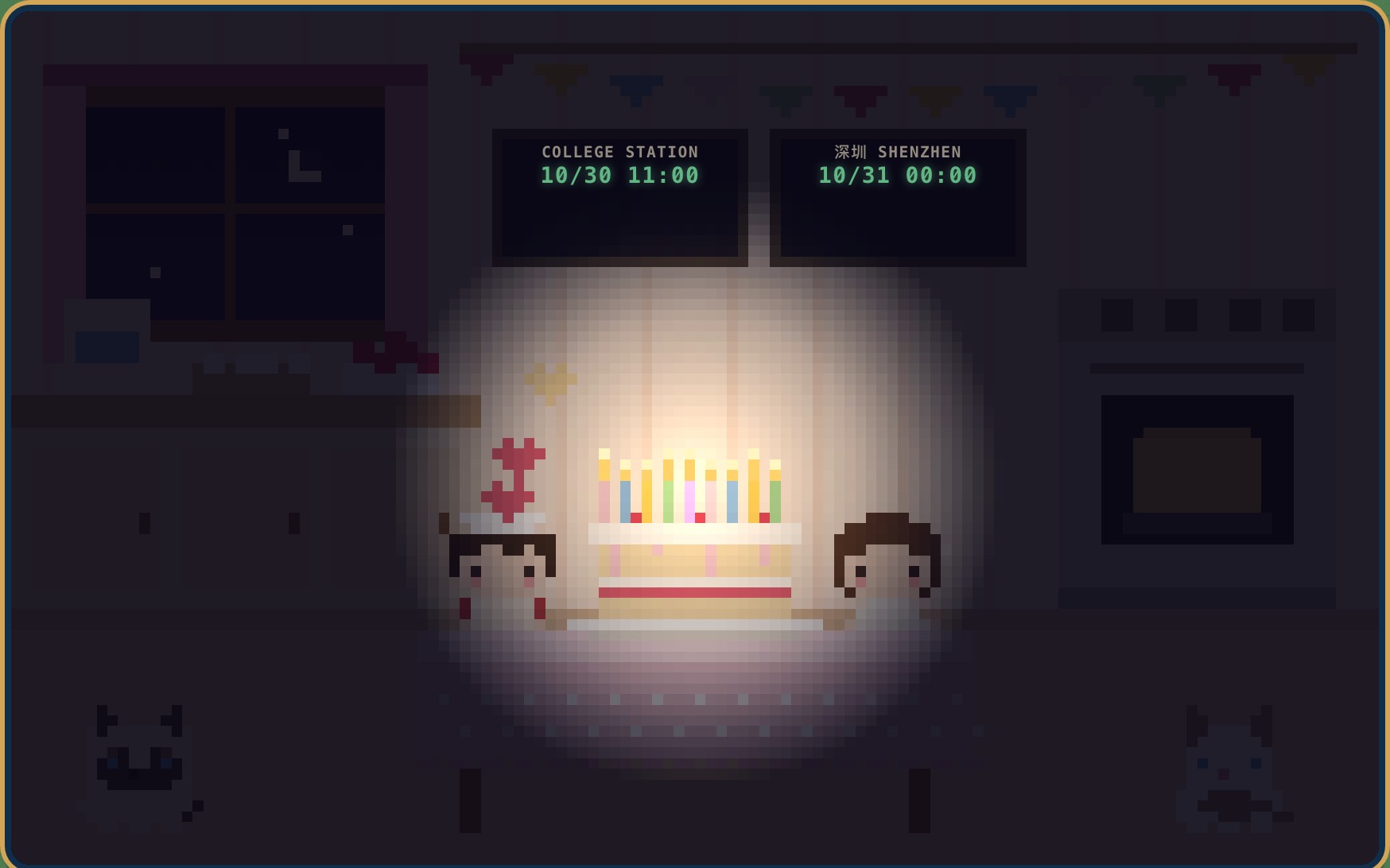}\\[3pt]
  \includegraphics[width=0.245\textwidth]{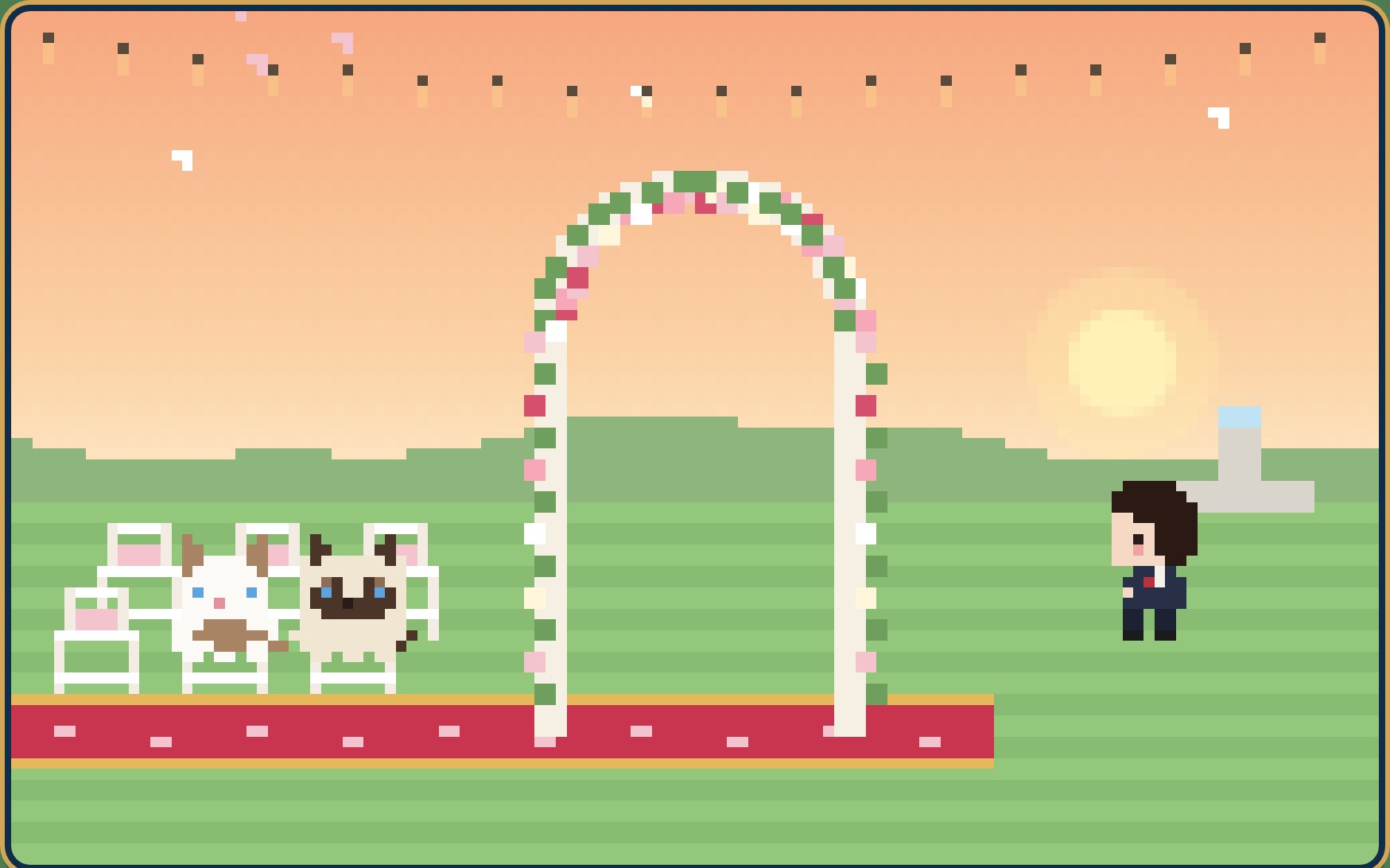}\hfill
  \includegraphics[width=0.245\textwidth]{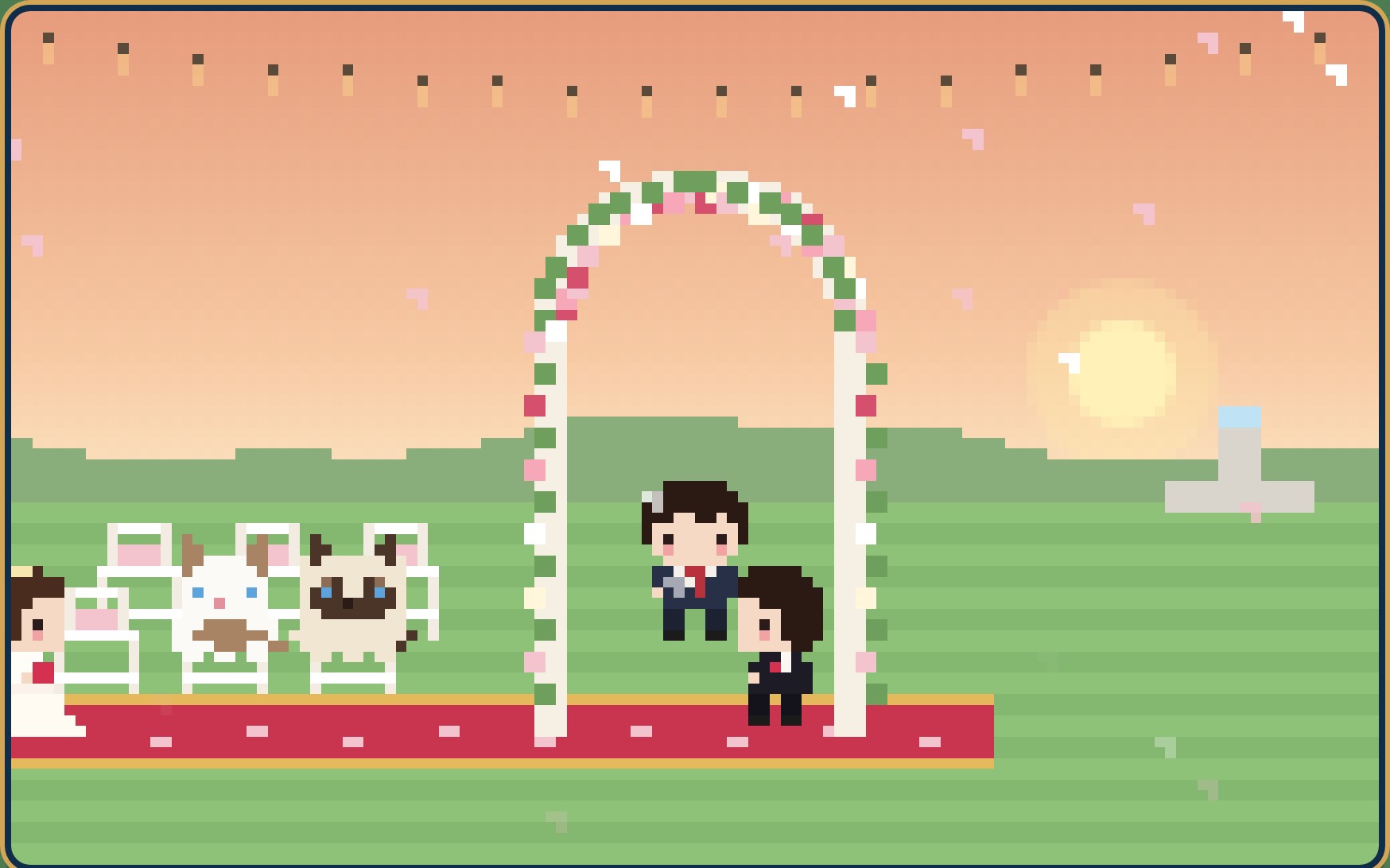}\hfill
  \includegraphics[width=0.245\textwidth]{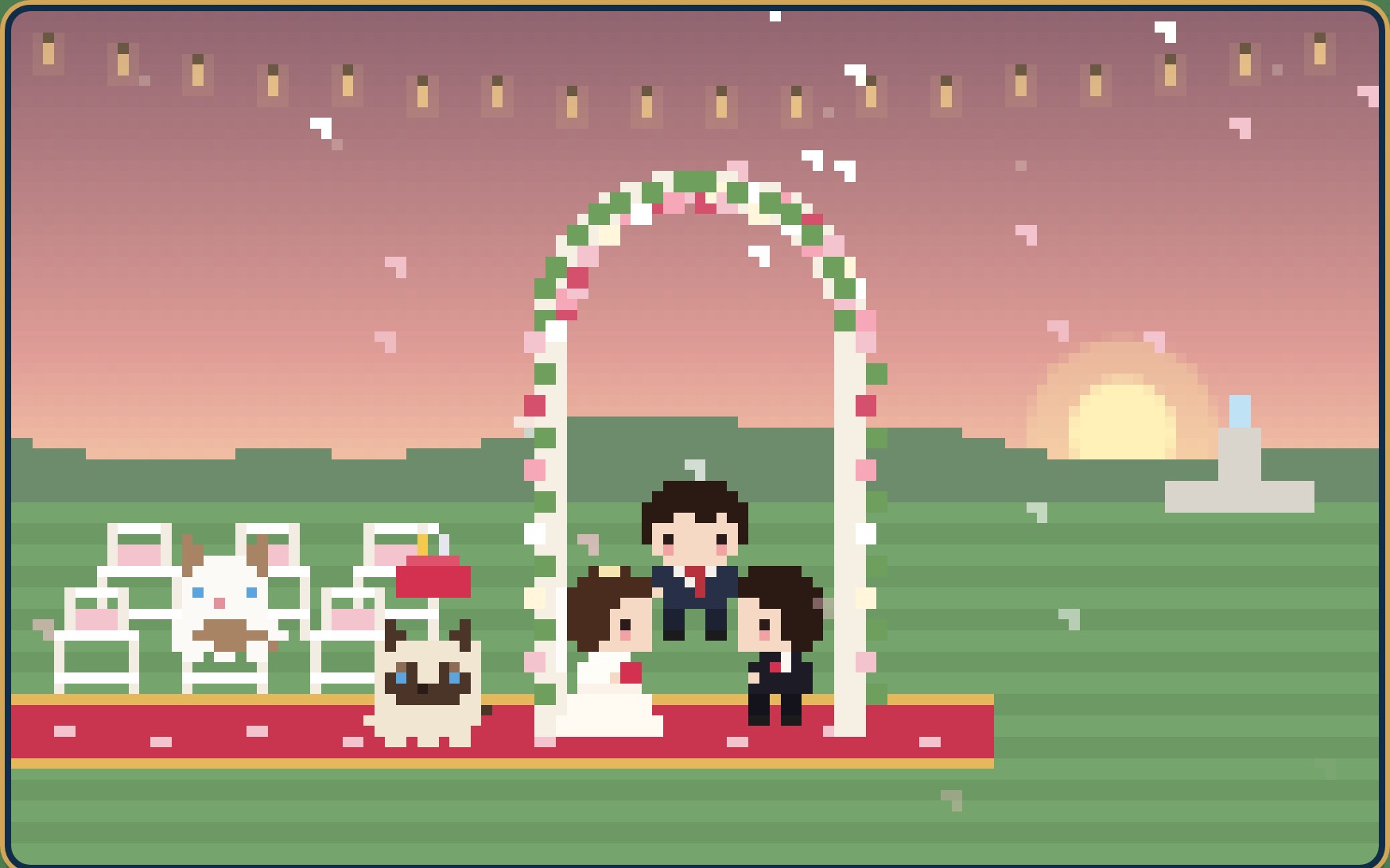}\hfill
  \includegraphics[width=0.245\textwidth]{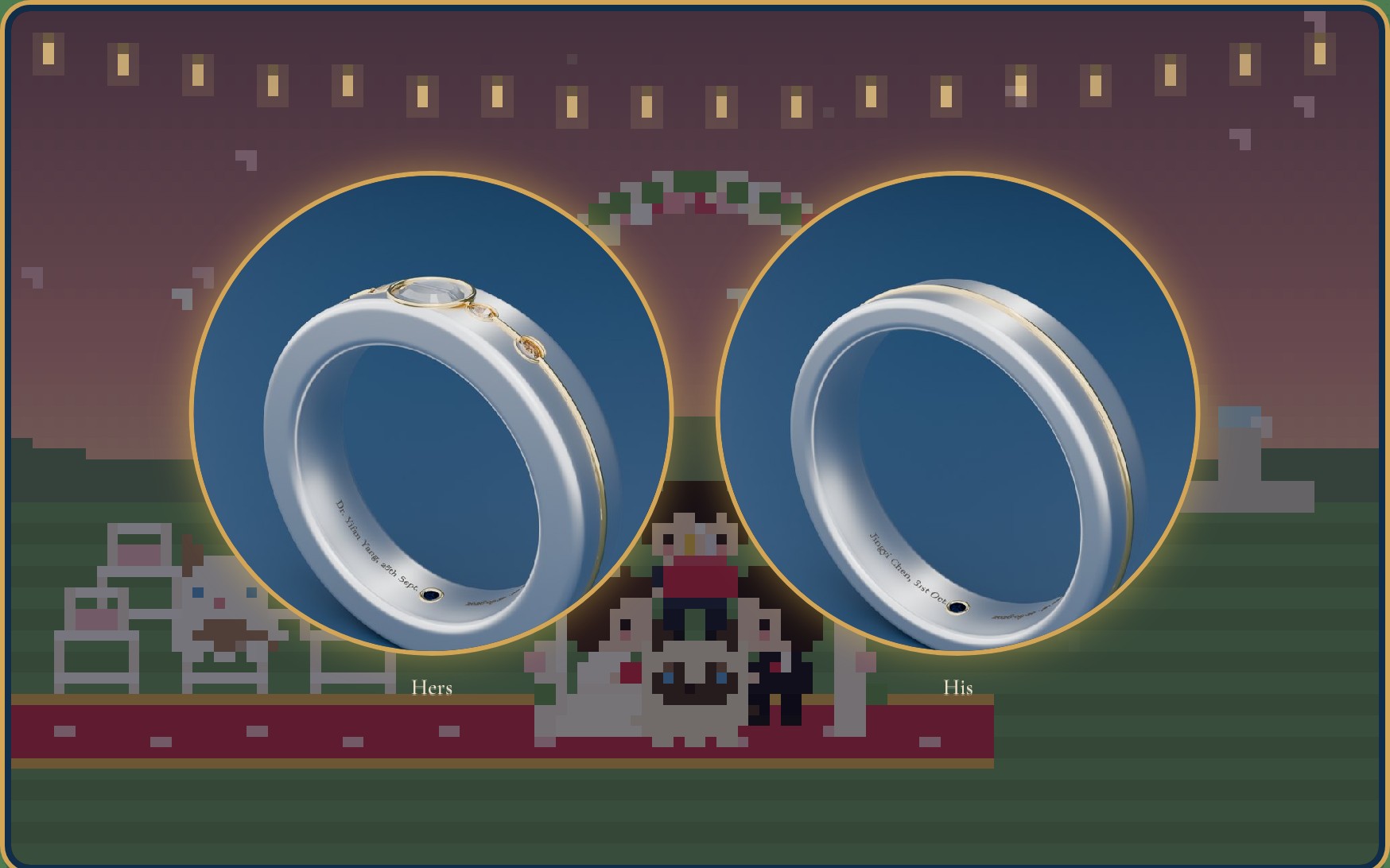}
  \caption{The two cutscenes, left to right. Top, the birthday prologue: the kitchen at 9:47 a.m.\
  in College Station (10:47 p.m.\ in Shenzhen) with both clocks on the wall; whisking the batter;
  nine candles on the finished cake; singing at Shenzhen midnight with the cats. Bottom, the
  wedding: the guests take their seats; the bride walks the aisle; Coffee brings the rings; the
  Meridian rings in 3D over the ceremony.}
  \Description{Eight pixel-art frames. Top row: a pink kitchen, a chef at a table, a cake with
  candles, the same kitchen dark with candlelight. Bottom row: a garden with a flower arch at sunset,
  a couple under the arch, a cat carrying a pillow, two large 3D rings in circles.}
  \label{fig:cutscenes}
\end{figure*}

Two cutscenes frame the game (Figure~\ref{fig:cutscenes}). They run in a small pixel theatre: a
$128\times80$ canvas scaled without smoothing, actors built from the town's sprite sheets, a
darkness layer with candle-like glows, particles, a synthesised score, and a caption box for words
and buttons.

\emph{The birthday prologue} plays once per player before the town opens. In College Station it is
9:47 a.m.; in Shenzhen 10:47 p.m., 73 minutes before Jingyi's birthday. Yifan, in a chef's hat,
bakes the cake against the Shenzhen clock. The player helps by whisking the batter (twelve taps),
placing ten strawberries in the same heart that the Gomoku answers make, and watching nine candles
lit (nine levels; in Chinese \zh{九} ``nine'' sounds like \zh{久} ``for a long time''). At the prologue's scripted Shenzhen midnight the lights go out, the door opens on the town (``In Yi Town there is no thirteen-hour
time difference''), everyone sings, the player makes a wish and blows the candles out, by
breathing into the microphone or holding a button.

\begin{figure}[t]
  \centering
  \includegraphics[width=0.78\columnwidth]{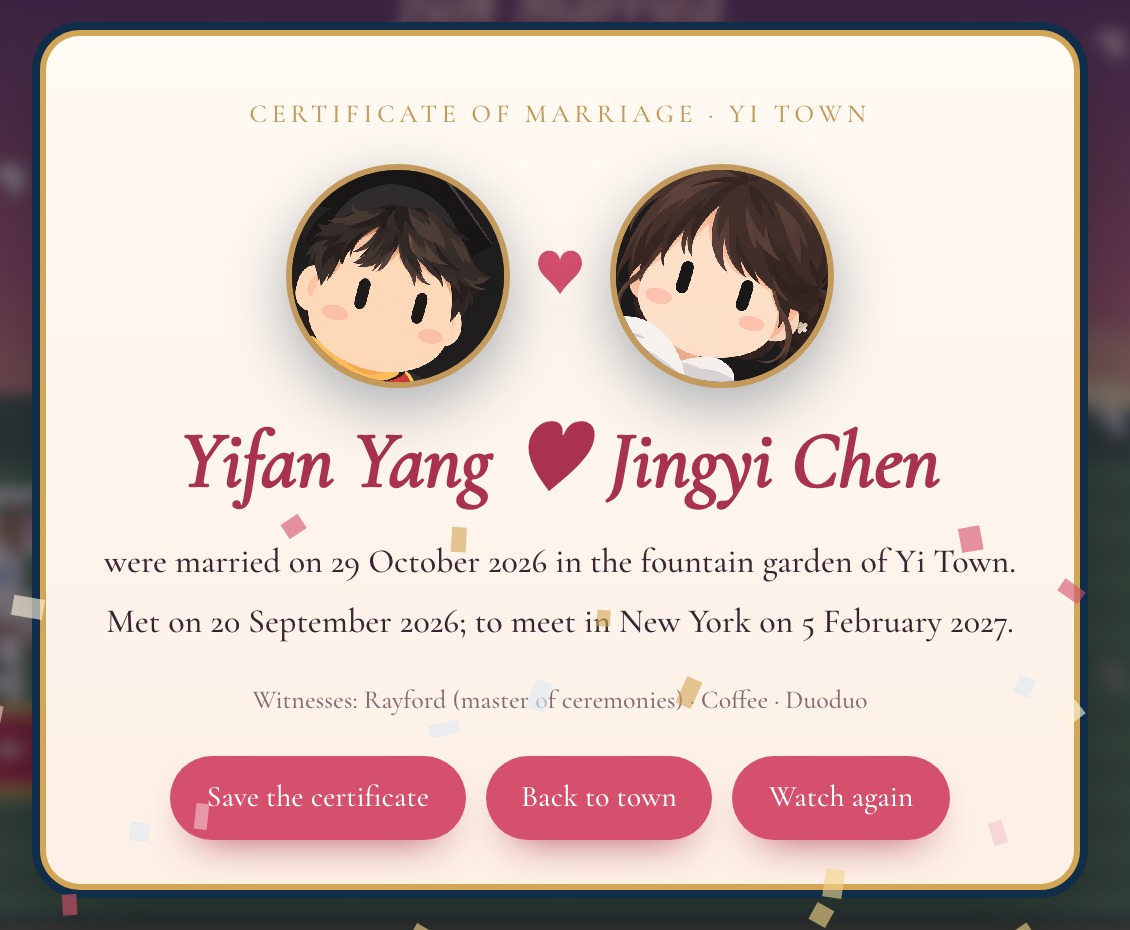}
  \caption{The certificate at the end of the wedding, saved as an image through the phone's share
  sheet. It records the day of the ceremony, the day they met and the day they will meet.}
  \Description{A cream certificate card with two cartoon avatars and the names Yifan Yang and Jingyi Chen joined by a heart, on confetti.}
  \label{fig:cert}
\end{figure}

\emph{The wedding} opens after the last letter, once the player has cleared the last level on
this device; since version 1.18 the other player need not have arrived.
In the fountain garden at dusk, Rayford reveals why he had worn a suit all along: he is the
master of ceremonies. The groom waits under the arch, the bride walks the aisle to Mendelssohn's
\emph{Wedding March}, the scripted sequence asks for both characters' ``I do'' responses, with speech
recognition where supported, Coffee carries the rings down the aisle on a red pillow, the Meridian rings turn in
3D, and the scene ends in a kiss, bells, doves and a certificate to keep
(Figure~\ref{fig:cert}). Afterwards the town dresses the two sprites as bride and groom.

\section{After the Town}\label{sec:after}
\begin{figure}[t]
  \centering
  \includegraphics[width=0.86\columnwidth]{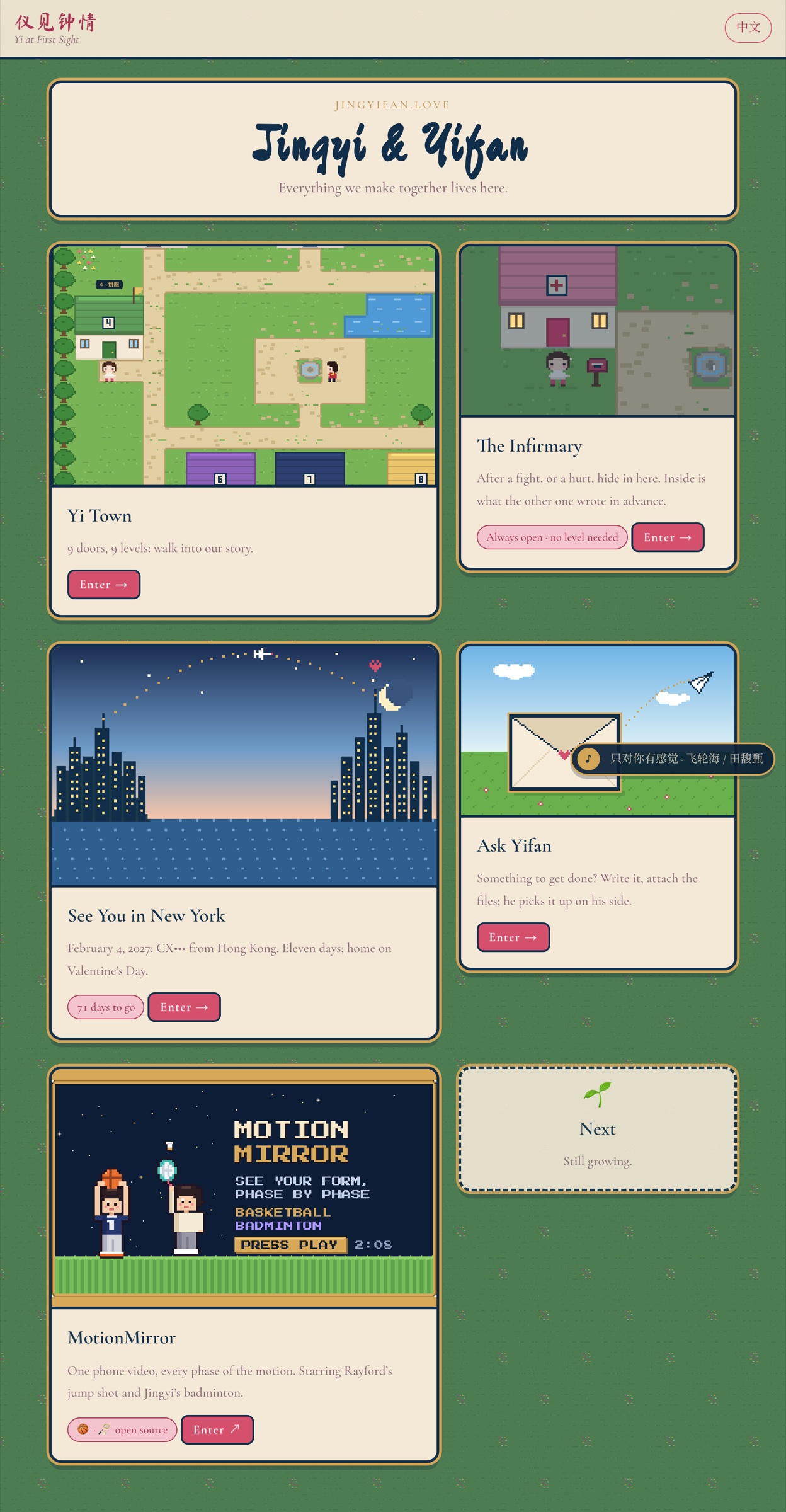}
  \caption{The dashboard at version 1.24, in English. From the top: \yt{}; the infirmary, with
  its status line; the trip, counting down; the request box; MotionMirror, which leads off the
  site; and the seed tile. The pill over the request box is the site's jukebox. The flight
  number on the trip card is masked for publication; the live page shows it.}
  \Description{A tall screenshot on a green pixel meadow. A header card reads Jingyi and Yifan,
  everything we make together lives here. Below are five picture cards in two columns, each with
  a title, a line of text and an Enter button, and a sixth tile with a dashed border, a seedling
  and the words Next, still growing.}
  \label{fig:dash}
\end{figure}

\begin{figure}[t]
  \centering
  \begin{subfigure}[b]{0.325\columnwidth}\includegraphics[width=\linewidth]{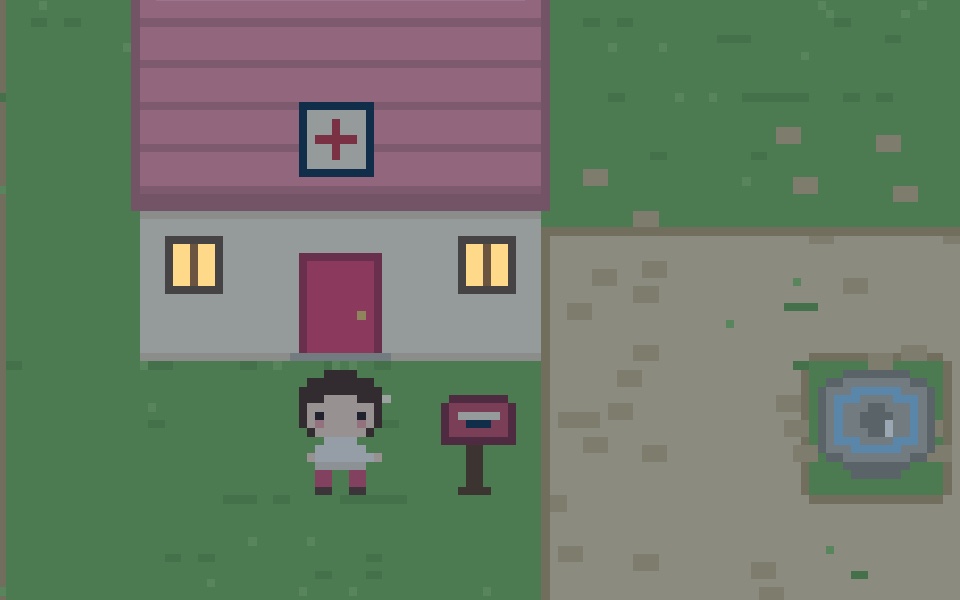}\caption{Infirmary}\end{subfigure}\hfill
  \begin{subfigure}[b]{0.325\columnwidth}\includegraphics[width=\linewidth]{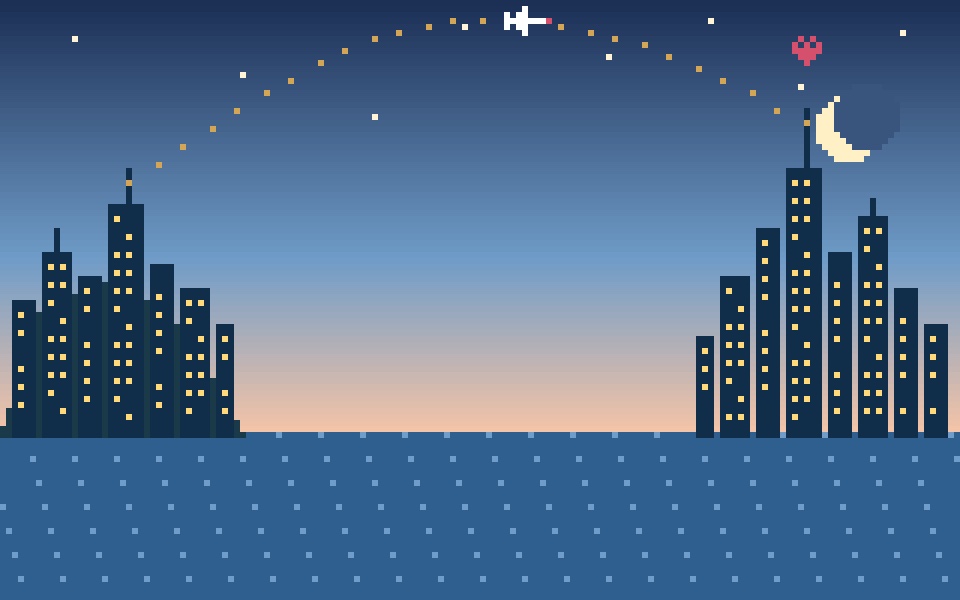}\caption{Trip}\end{subfigure}\hfill
  \begin{subfigure}[b]{0.325\columnwidth}\includegraphics[width=\linewidth]{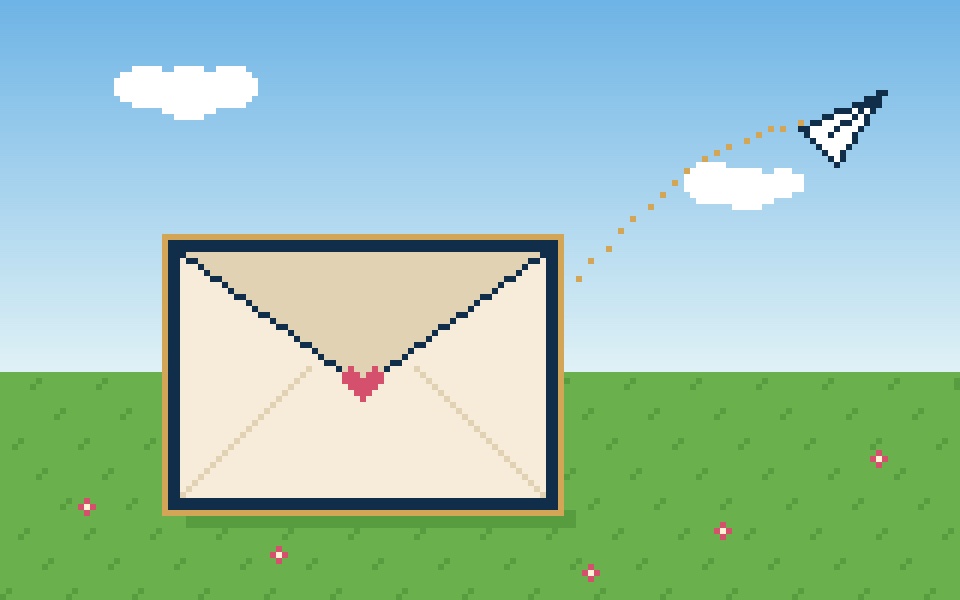}\caption{Request box}\end{subfigure}\\[3pt]
  \includegraphics[width=0.9\columnwidth]{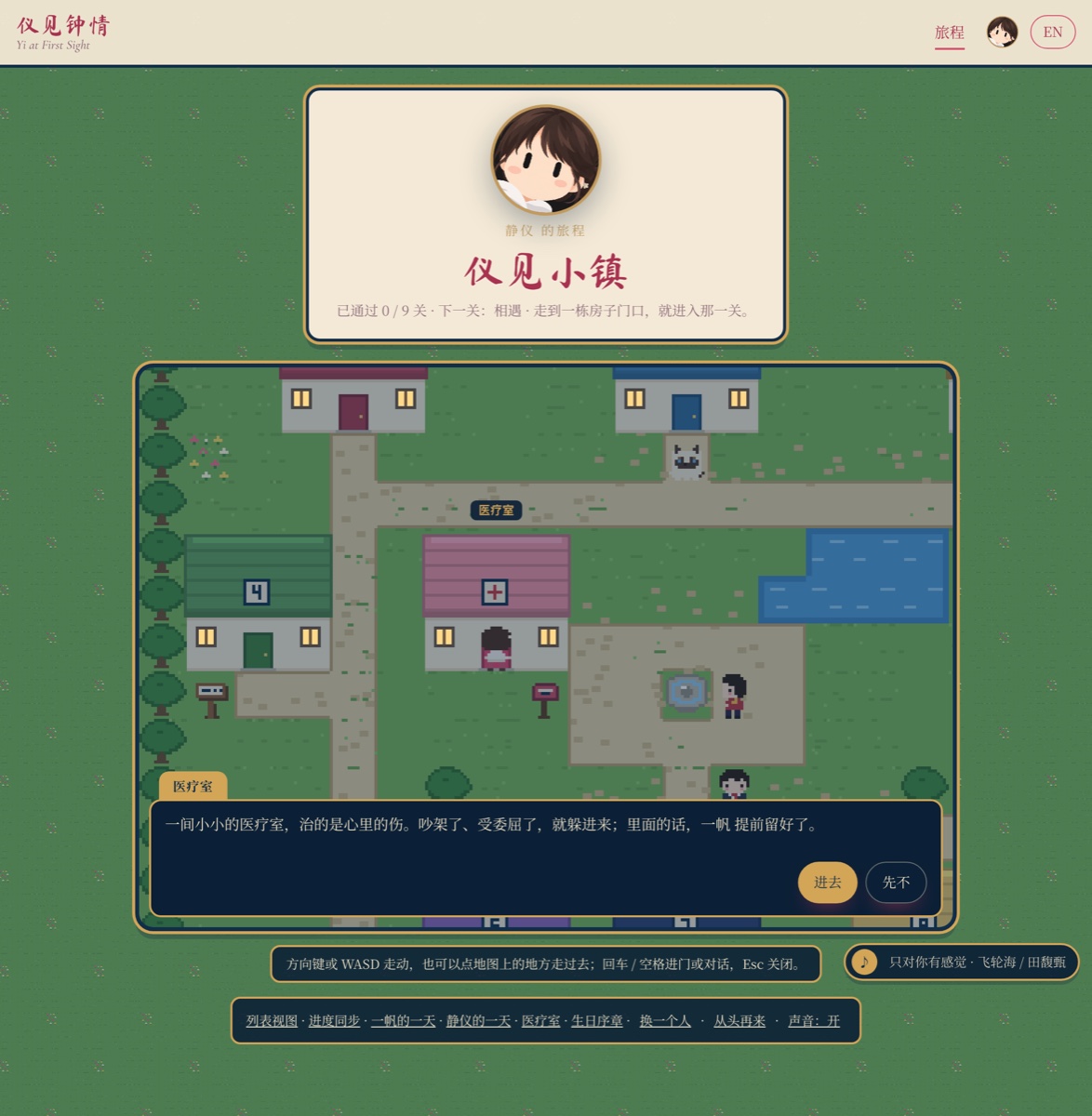}
  \caption{Three of the later products. Top: the pictures on their dashboard cards, each drawn
  by a script in the repository. Bottom: the infirmary as a house in the town, north-west of the
  fountain, with the talk box that opens at its door. It has no number on its roof and no cat at
  its door.}
  \Description{Three small pixel pictures: a pink house with a red cross, two cities joined by a
  dotted route with a plane, and an envelope sealed with a heart. Below, a screenshot of the pixel
  town with a pink-roofed house marked by a cross and a dialogue box in Chinese.}
  \label{fig:cards}
\end{figure}

The first version described a finished gift with a beginning and an end. In the four weeks that
followed, the page behind the gate changed from a list with one entry into the centre of the
site. This section goes through everything on it, in the order of the page: the dashboard
itself, its five cards, and a sixth tile that is not a product (Figure~\ref{fig:dash}). Three
of the cards are products kept in the same repository (Figure~\ref{fig:cards}). None of those
three is a level. None has a lock, none counts toward progress, and nothing in them waits to be
revealed, so staged access (\dc{2}) and delayed disclosure (\dc{6}) do not apply to them.
Table~\ref{tab:after} sets them against the commitments.

\subsection{The Dashboard}
Its header says ``Everything we make together lives here'', and the code describes it the same
way: a list, \code{PRODUCTS}, with one entry for each thing the couple build under the domain,
newest last. Adding a product means a folder, a route and one entry in that list. The first
version's dashboard had one card and read as a title page for the game; with five it reads as a
home page, and the game is one room in it.

Three decisions on this page are recorded in the comments beside the list. First, a card may
carry one line of live state, and what the line withholds is chosen as carefully as what it
shows. The town's card shows how many levels the person holding this device has cleared and
never the partner's count, because how far the other has got would reveal that there is an end
worth reaching; delayed disclosure (\dc{6}) thus reaches the one page that sits outside the
game. The infirmary's card says only that it is always open, and shows no count of visits. The
trip's card counts days to the flight, then the days together, then says that she is on her way
home. The request box has no line, since nothing about it changes between visits. Second, the
cards differ in what they ask first. The town asks who is playing, because it keeps progress;
the infirmary asks too, because it must know whose letter to show; the trip and the request box
ask nothing beyond the gate. Third, each card's picture is drawn by a script in the repository
from the same sprites and palette as the town (\dc{5}), so that a product about a flight and a
product about a quarrel look like houses on the same street.

\begin{table}[t]
  \caption{The three later products in the repository against the commitments written for the
  town. \emph{kept}: the product follows it; \emph{n/a}: nothing in the product for it to
  govern; \emph{broken}: the product departs from it by design. MotionMirror is a separate
  project and the seed tile has no behaviour; neither is rated.}
  \label{tab:after}
  \small
  \begin{tabular}{@{}p{0.34\columnwidth}ccc@{}}
    \toprule
    Commitment & Infirmary & Trip & Request box \\
    \midrule
    \dc{1} Memory-derived mechanics & kept & n/a & n/a \\
    \dc{2} Staged access & n/a & n/a & n/a \\
    \dc{3} Local state, no backend & kept & kept & broken \\
    \dc{4} Bilingual coverage & kept & kept & kept \\
    \dc{5} Drawn in code & kept & kept & kept \\
    \dc{6} Delayed disclosure & n/a & n/a & n/a \\
    \bottomrule
  \end{tabular}
\end{table}

\subsection{The Infirmary}
On 3 November the recipient said, as the changelog records it,
``\zh{以后我受伤了我就不跟你说话了，我就躲进去}'' (``from now on, when I am hurt I will not talk to
you; I will hide in there''). The infirmary was added the next day (\code{397db28}) as a pink
house by the square and as the second card. A visitor says what hurts, choosing among a quarrel,
a hurt, missing the other without wanting to speak first, and plain tiredness. Three slow
breaths follow, which can be skipped. Then the page shows a letter written in advance, in the
other partner's name, for that hurt. A cabinet gives one-line notes at random, and a chart takes private
lines that stay in the browser's local storage.

Two details carry the design. The letter opens with what its writer is doing at that moment,
computed from the clock in the writer's city and an authored schedule, for example that it is
past three in the morning there and he is asleep. The same caveat applies as to the city days in
Section~\ref{sec:days}: this is a schedule, not sensed presence. And leaving offers one button,
``tell them I am out'', which hands the phone's share sheet a single line with the length of the
visit. Nothing is sent without that tap, and the dashboard card shows no count of visits, because
how often someone has hidden is theirs alone.

The authorship of the letters needs stating exactly. There are four letters and eighteen notes
for each partner. Both sets were first written on 4 November as drafts ``in each one's voice, for
each to rewrite'' (\code{73e54bc}). By the cut-off the record shows one of Yifan's letters
replaced with his own text (\code{dc62d23}) and no revision of the set attributed to Jingyi. The
words the infirmary shows a visitor as hers are therefore, on the record, words written for her.

\subsection{See You in New York}
The first version's countdown pointed at a symbolic ticket, hand-drawn during a video call. On
8 November Jingyi booked her flights, and the site followed (\code{d9de84b}): the countdown and
the gold line on the Love Map now end at her landing on 4 February 2027, and the page for the
trip shows both legs as boarding passes with local times. Its clock knows five phases: before
departure, in the air, together, flying home, and home. During the stay it counts ``day $n$ of
11''. Below it is an itinerary with one row for each day, to be written and photographed as the
days happen; an unwritten day reads ``still growing''. The hand-drawn ticket stays in the game as
a memory, in the Moments level and the first level's quiz. The page asks for no player, since it
has no progress to keep.

\subsection{Ask Yifan}
The fourth card opens a composer like that of a chat application: a text box, a paperclip, drag
and drop, and paste for screenshots (\code{ea2c088}, 24 November). Jingyi writes what she needs
help with, attaches up to ten files and 18\,MiB, and sends. Her request reaches Yifan's inbox as
one email with the files attached. What she sent stays on her device as bubbles marked sending,
sent or failed; a failed send keeps its text and files and offers to try again. The specification
describes the page as one-way on purpose: the answer was to come over the couple's usual
messaging, or as the thing done.

A browser cannot send an email with attachments, so this product required the site's first
server-side code, and within a day it acquired an automated answerer. The page, the server and
the answerer are described in Sections~\ref{sec:askpage}, \ref{sec:worker}
and~\ref{sec:askpath}. We say nothing in this paper about what her requests ask for.

\subsection{MotionMirror}
The fifth card (\code{5bea274}, 22 November) is the only one that leaves the site. It opens, in
a new tab, a separate public project of the couple's: a tool that analyses sports technique
from a single phone video, phase by phase. The card names its two showcase recordings, Yifan's
jump shot and Jingyi's badminton, and its status line says that the project is open source. Section~\ref{sec:mm} describes how it works.

What the card adds to this case is a different relation between the two people. Every other
product on the page is made by one of them for the other, or for the two of them alone. This
one is made with both of them as its subjects and is addressed to strangers. The dashboard
therefore holds, side by side, things that only two people are meant to open and a thing meant
for anyone, behind the same gate and in the same pixel frame. The arrow on its button, which
points up and out where the others point right, is the page's only sign of that difference.

\subsection{The Seed}
The last tile has a dashed border, a seedling, and two lines: ``Next'' and ``Still growing''.
It is not a link and has no behaviour. It stands at the end of the list,
and the request box's specification speaks of that product as making good the seed tile's
promise, and of keeping the tile as the last one all the same.

The town was built with an ending: a ninth house, a letter, a ceremony. The tile states the
opposite about the site that now contains the town, which is that it is unfinished on purpose.
The final letter in the town, written after the first version of this paper, ends with ``to be
continued'', and the itinerary's unwritten days say ``still growing'' in the tile's own words.
A gift is usually finished when it is handed over. These three places mark the point where the
artefact stopped presenting itself as one.

\section{System}\label{sec:system}

\subsection{Architecture}
\yt{} is a static single-page application in strict TypeScript, built with Vite and served from
static hosting under the couple's domain (Figure~\ref{fig:arch}). The deployment configuration
targets GitHub Pages. There is no author-operated backend for progress; since version 1.22 there
is one for requests (Sections~\ref{sec:worker} and~\ref{sec:askpath}), which Figure~\ref{fig:arch} does not show. Each feature
is a folder (\code{src/features/<name>/}) with its page module, styles, data, tests and a README,
and each page is a function \code{render(root) $\Rightarrow$ cleanup} loaded lazily by a path
router, so the main bundle carries only the shell and the town. Heavy modules (Leaflet for the map,
three.js for the ring, MediaPipe for the try-on) load only inside the level that needs them. A
service worker precaches the shell and caches eligible media after retrieval. The cached
town can run offline, but remote song embeds and browser-dependent speech recognition are not
guaranteed offline; cache eviction and previously unfetched assets also limit availability.

\begin{figure}[t]
  \centering
  \begin{tikzpicture}[
      font=\scriptsize\sffamily,
      box/.style={scard, inner sep=2.5pt, minimum height=5.2mm},
      store/.style={box, syellow},
      ext/.style={sflat, dashed, inner sep=2.5pt, minimum height=5.2mm},
      arr/.style={sarr}]
    \node[box, sblue, minimum width=7.4cm, font=\scriptsize\sffamily\bfseries] (shell) {App shell: gate $\cdot$ path router $\cdot$ i18n (zh/en) $\cdot$ meadow $\cdot$ jukebox};
    \node[box, snavyc, below=3.5mm of shell.south west, anchor=north west, minimum width=2.15cm] (town) {Town engine\\\dsub{grid $\cdot$ BFS $\cdot$ sprites}};
    \node[box, snavyc, right=2.5mm of town, minimum width=2.6cm] (levels) {Nine level modules\\\dsub{lazy-loaded pages}};
    \node[box, snavyc, right=2.5mm of levels, minimum width=2.15cm] (theatre) {Theatre\\\dsub{prologue $\cdot$ wedding}};
    \node[box, steal, below=3.5mm of town, minimum width=2.15cm] (quest) {Quest model\\\dsub{ordered levels}};
    \node[box, steal, right=2.5mm of quest, minimum width=2.6cm] (sync) {Sync codes\\\dsub{G-Set $\cdot$ MOD 37,36}};
    \node[box, steal, right=2.5mm of sync, minimum width=2.15cm] (sense) {Browser sensing\\\dsub{speech $\cdot$ mic $\cdot$ hands}};
    \node[store, below=3.5mm of quest, minimum width=2.15cm] (ls) {localStorage\\\dsub{per player}};
    \node[store, right=2.5mm of ls, minimum width=2.6cm] (ss) {sessionStorage\\\dsub{in-level state}};
    \node[store, right=2.5mm of ss, minimum width=2.15cm] (sw) {Service worker\\\dsub{offline cache}};
    \node[ext, below=5mm of ss, minimum width=7.4cm] (ext) {External services: hosting $\cdot$ map tiles $\cdot$ song embeds $\cdot$ speech};
    \coordinate (shtop) at ([yshift=2.4mm]shell.north);
    \begin{scope}[on background layer]
      \node[spanel, fit=(shell)(ls)(sw)(shtop), inner sep=3pt] (phone) {};
    \end{scope}
    \node[spill, anchor=west] at ([xshift=2.5mm]phone.north west) {one phone, one browser};
    \draw[arr] (town) -- (quest); \draw[arr] (levels) -- (quest); \draw[arr] (levels) -- (sense);
    \draw[arr] (theatre) -- (sense); \draw[arr] (quest) -- (ls); \draw[arr] (sync) -- (quest);
    \draw[arr, sLine, dashed] (ss.south) -- (ext.north);
  \end{tikzpicture}
  \caption{Architecture of application modules and local state. The dashed boundary marks browser
  components, not a confidentiality boundary: static assets are hosted remotely, external requests
  expose request metadata, and speech recognition may use a remote service. Progress crosses
  devices through a manually exchanged code.}
  \Description{A block diagram: an app shell above three modules, which depend on a quest model,
  sync codes and browser sensing, which sit above three storage layers, all inside a dashed boundary
  labelled one phone, one browser; a dashed box below for external hosting, map, media and speech services.}
  \label{fig:arch}
\end{figure}

\subsection{The Quest Model}
Let $L = (\ell_1,\dots,\ell_9)$ be the ordered level registry and $C_p \subseteq L$ the set of
levels player $p$ has cleared on a device. A level is unlocked iff it is first or its predecessor
is cleared:
\begin{equation}\label{eq:unlock}
  \mathrm{unlocked}_p(\ell_i) \iff i = 1 \;\lor\; \ell_{i-1} \in C_p .
\end{equation}
Clearing adds to $C_p$ and nothing ever removes from it, except an explicit reset. The next level
to play is the first one not yet cleared, $\mathrm{current}(p) = \ell_{\min\{i : \ell_i \notin C_p\}}$.
The router enforces the lock, and a second guard requires that a level be entered through its door:
walking onto the door sets a one-shot token $\tau$ in \code{sessionStorage}, and a level route
without the token redirects to the town, so that
\begin{equation}\label{eq:enter}
  \mathrm{enter}_p(\ell_i) \iff \mathrm{unlocked}_p(\ell_i) \;\land\; \tau = \mathrm{route}(\ell_i) .
\end{equation}
Figure~\ref{fig:chain} shows the whole sequence, including the two scenes that follow the last level. Level numbers are derived from the registry, so inserting a level (the
Poems level was inserted before the last one during development) renumbers every house, sign and
string at once; a test fails until the new house has a door that the town's path search can reach.

\begin{figure}[t]
  \centering
  \begin{tikzpicture}[font=\scriptsize\sffamily, node distance=1.5mm,
      lv/.style={scard, draw=sInk, line width=0.5pt, rounded corners=0pt, minimum width=5.9mm, minimum height=5.9mm, inner sep=0pt},
      scene/.style={scard, sblue, minimum height=5.9mm, inner sep=2pt},
      arr/.style={sarr, -{Triangle[length=1.1mm, width=1.1mm]}, line width=0.8pt}]
    \node[scene] (pro) {Prologue\\\dsub{per player}};
    \definecolor{roof1}{HTML}{C74B5F}
    \node[lv, right=of pro] (l1) {1};
    \fill[roof1] ([yshift=-0.2pt]l1.north west) -- ([yshift=1.6mm]l1.north) -- ([yshift=-0.2pt]l1.north east) -- cycle;
    \draw[arr] (pro) -- (l1);
    \definecolor{roof2}{HTML}{3F7FB3}
    \node[lv, right=of l1] (l2) {2};
    \fill[roof2] ([yshift=-0.2pt]l2.north west) -- ([yshift=1.6mm]l2.north) -- ([yshift=-0.2pt]l2.north east) -- cycle;
    \draw[arr] (l1) -- (l2);
    \definecolor{roof3}{HTML}{D8973C}
    \node[lv, right=of l2] (l3) {3};
    \fill[roof3] ([yshift=-0.2pt]l3.north west) -- ([yshift=1.6mm]l3.north) -- ([yshift=-0.2pt]l3.north east) -- cycle;
    \draw[arr] (l2) -- (l3);
    \definecolor{roof4}{HTML}{5F9E5A}
    \node[lv, right=of l3] (l4) {4};
    \fill[roof4] ([yshift=-0.2pt]l4.north west) -- ([yshift=1.6mm]l4.north) -- ([yshift=-0.2pt]l4.north east) -- cycle;
    \draw[arr] (l3) -- (l4);
    \definecolor{roof5}{HTML}{8D6A4B}
    \node[lv, right=of l4] (l5) {5};
    \fill[roof5] ([yshift=-0.2pt]l5.north west) -- ([yshift=1.6mm]l5.north) -- ([yshift=-0.2pt]l5.north east) -- cycle;
    \draw[arr] (l4) -- (l5);
    \definecolor{roof6}{HTML}{8C62B8}
    \node[lv, right=of l5] (l6) {6};
    \fill[roof6] ([yshift=-0.2pt]l6.north west) -- ([yshift=1.6mm]l6.north) -- ([yshift=-0.2pt]l6.north east) -- cycle;
    \draw[arr] (l5) -- (l6);
    \definecolor{roof7}{HTML}{2D3F6E}
    \node[lv, right=of l6] (l7) {7};
    \fill[roof7] ([yshift=-0.2pt]l7.north west) -- ([yshift=1.6mm]l7.north) -- ([yshift=-0.2pt]l7.north east) -- cycle;
    \draw[arr] (l6) -- (l7);
    \definecolor{roof8}{HTML}{B0413E}
    \node[lv, right=of l7] (l8) {8};
    \fill[roof8] ([yshift=-0.2pt]l8.north west) -- ([yshift=1.6mm]l8.north) -- ([yshift=-0.2pt]l8.north east) -- cycle;
    \draw[arr] (l7) -- (l8);
    \definecolor{roof9}{HTML}{E9C46A}
    \node[lv, right=of l8] (l9) {9};
    \fill[roof9] ([yshift=-0.2pt]l9.north west) -- ([yshift=1.6mm]l9.north) -- ([yshift=-0.2pt]l9.north east) -- cycle;
    \draw[arr] (l8) -- (l9);
    \node[scene, spurple, below=4.5mm of l5] (let) {Letter};
    \node[scene, spurple, left=5mm of let] (wed) {Wedding};
    \draw[arr, sVio] (l9.south) |- node[pos=0.75, above, font=\tiny, text=sVio] {$\ell_9 \in C_p$} (let.east);
    \draw[arr, sVio] (let) -- (wed);
    \node[below=0.6mm of wed.south west, anchor=north west, font=\tiny, text=sSoft] {gated on the player's own progress};
    \node[above=1.8mm of l1.north, anchor=south west, font=\tiny, text=sSoft, xshift=-2mm] {Eq.~\ref{eq:unlock}: $\ell_i$ opens when $\ell_{i-1} \in C_p$; Eq.~\ref{eq:enter}: only through its door};
  \end{tikzpicture}
  \caption{The sequence of play. Nine houses open in order per player; the letter and the wedding
  open for a player once that player's completion set on the device contains the last level.}
  \Description{A chain of nine numbered boxes with coloured roofs following a prologue box, and a branch from box nine down to letter and wedding boxes.}
  \label{fig:chain}
\end{figure}

\subsection{Progress without a Server}
Each phone holds its own $C_p$. Because $C_p$ only grows, it is a grow-only set (G-Set), the
simplest state-based CRDT~\cite{shapiro2011crdt}: the merge of two replicas is their union,
\begin{equation}\label{eq:merge}
  C_p \sqcup C'_p = C_p \cup C'_p ,
\end{equation}
which is commutative, associative and idempotent. Replicas converge if all updates are eventually exchanged, both phones use the same
level registry and no reset occurs. Repeated imports are idempotent; an old code cannot remove
completion flags, but incomplete exchange can leave the two views different. The
only channel needed is one the couple already use every day: a chat message.

\paragraph{Progress codes.} A code encodes one player and a bitmask of cleared levels,
written in base 36 after a player letter $\lambda_p \in \{\code{J}, \code{Y}\}$ and followed by one
check character $\chi$ (Figure~\ref{fig:sync}):
\begin{equation}\label{eq:code}
  m = \sum_{\ell_i \in C_p} 2^{\,i-1}, \qquad
  \mathrm{code}(p, C_p) = \lambda_p \,\Vert\, \beta_{36}(m) \,\Vert\, \chi\big(\lambda_p \Vert \beta_{36}(m)\big),
\end{equation}
where $\Vert$ is concatenation and $\beta_{36}$ writes a number in base 36. The decoder accepts a
string only if its last character equals $\chi$ of the rest, its first is a player letter and
$0 \le m < 2^{9}$. With nine levels $m \le 511 = \mathrm{E7}_{36}$, so a
code is at most four characters: \code{JVA} says that Jingyi has cleared $m = 31$, the first five
levels (Table~\ref{tab:codes}). The check character follows the hybrid system ISO~7064
MOD~37,36~\cite{iso7064}. Writing $a_j \in \{0,\dots,35\}$ for the value of the $j$-th body
character and starting from $p_0 = 36$,
\begin{equation}\label{eq:check}
  s_j = \big((p_{j-1} + a_j) \bmod 36\big) \text{ or } 36 \text{ if } 0, \qquad
  p_j = 2 s_j \bmod 37 ,
\end{equation}
and the check character is the symbol of value $(37 - p_n) \bmod 36$. We checked the scheme exhaustively over all
1{,}024 valid codes: 140{,}840 single-character substitutions and 2{,}943 transpositions of two
different adjacent characters. It rejects every substitution. Of the transpositions it rejects
all but five. One of those (\code{JI0}$\to$\code{IJ0}) is still refused because \code{I} is not a
player letter. The other four swap the last mask digit with the check character
(\code{J801}$\leftrightarrow$\code{J810}, \code{YE01}$\leftrightarrow$\code{YE10}): in both pairs
the running value before the last mask digit is $p = 18$, after which a mask digit \code{0} calls
for the check character \code{1} and a mask digit \code{1} for the check character \code{0}, so
the two orders are both valid codes.
An accepted typo may add unearned completion flags and unlock later content, including the finale. An earlier version used a plain weighted sum,
which let some one-character typos through; it was later replaced, invalidating older
codes. Codes are not authenticated: a person can construct completion for either player. They
carry no schema version, so registry insertions or reordering can reinterpret bits. Reset is outside
the grow-only model; importing an earlier code after reset restores its flags. A code may also
travel in a link (\code{/sync?c=JVA}), exposing its payload to the hosting request and potentially
to browser history or the channel through which it is sent.

\begin{table}[t]
  \caption{Example progress codes: player letter, base-36 mask (Eq.~\ref{eq:code}) and the check character of Eq.~\ref{eq:check}.}
  \label{tab:codes}
  \small
  \begin{tabular}{@{}lrll@{}}
    \toprule
    Cleared & Mask $m$ & Base 36 & Code \\
    \midrule
    Jingyi, levels 1--5 & 31 & \code{V} & \code{JVA} \\
    Jingyi, all nine & 511 & \code{E7} & \code{JE7Z} \\
    Yifan, all nine & 511 & \code{E7} & \code{YE7O} \\
    Yifan, none & 0 & \code{0} & \code{Y0C} \\
    \bottomrule
  \end{tabular}

\end{table}

\begin{figure}[t]
  \centering
  \begin{tikzpicture}[font=\scriptsize\sffamily,
      phone/.style={scard, scream, rounded corners=2mm, minimum width=24mm, minimum height=26mm},
      chip/.style={fill=sInk, text=sYeF, rounded corners=1pt, inner sep=1.5pt, font=\scriptsize\ttfamily\bfseries},
      cell/.style={draw=sInk, line width=0.5pt, fill=white, text=sInk, minimum height=5mm, minimum width=7.5mm, inner sep=2pt, font=\scriptsize\ttfamily},
      arr/.style={sarr}]
    \node[phone] (a) at (0,0) {};
    \node[phone] (b) at (5.4,0) {};
    \node[anchor=north, text=sInk, font=\scriptsize\sffamily\bfseries] at ([yshift=-1.2mm]a.north) {Jingyi's phone};
    \node[anchor=north, text=sInk, font=\scriptsize\sffamily\bfseries] at ([yshift=-1.2mm]b.north) {Yifan's phone};
    \node[anchor=north, align=center, text=sInk] at ([yshift=-6mm]a.north) {$C_J = \{\ell_1,\dots,\ell_5\}$\\$C_Y = \{\ell_1, \ell_2\}$};
    \node[anchor=north, align=center, text=sInk] at ([yshift=-6mm]b.north) {$C_J = \{\ell_1, \ell_2, \ell_3\}$\\$C_Y = \{\ell_1,\dots,\ell_7\}$};
    \node[anchor=south, align=center, text=sGreen] at ([yshift=1.2mm]a.south) {after import:\\$C_Y = \{\ell_1,\dots,\ell_7\}$};
    \node[anchor=south, align=center, text=sGreen] at ([yshift=1.2mm]b.south) {after import:\\$C_J = \{\ell_1,\dots,\ell_5\}$};
    \draw[arr] ([yshift=3.5mm]a.east) -- node[above=0.6mm, chip] {JVA} node[below, font=\tiny, text=sNavy] {chat message} ([yshift=3.5mm]b.west);
    \draw[arr] ([yshift=-4.5mm]b.west) -- node[above=0.6mm, chip] {Y3J9} node[below, font=\tiny, text=sNavy] {or a link \code{/sync?c=Y3J9}} ([yshift=-4.5mm]a.east);
    \node[cell, below=2.5mm of a.south west, anchor=north west] (p) {J};
    \node[cell, fill=sYeF, right=-0.5pt of p] (m) {V};
    \node[cell, right=-0.5pt of m] (c) {A};
    \node[below=0.3mm of p, font=\tiny\sffamily, text=sSoft, inner sep=1pt] {$\lambda_p$};
    \node[below=0.3mm of m, font=\tiny\sffamily, text=sSoft, inner sep=1pt] {$\beta_{36}(31)$};
    \node[below=0.3mm of c, font=\tiny\sffamily, text=sSoft, inner sep=1pt] {$\chi$};
    \node[right=1.5mm of c, anchor=west, font=\scriptsize\sffamily, align=left, text=sSoft] {$m = 31 = 000011111_2$: bit $i$ is set when $\ell_i \in C_J$\\union merge (Eq.~\ref{eq:merge}) never removes a level and\\converges once both codes have travelled};
  \end{tikzpicture}
  \caption{Progress without a server. Each phone keeps both players' completion sets; a short code
  (Eq.~\ref{eq:code}) carries one set across in a chat message or a link, and the receiver takes the
  union (Eq.~\ref{eq:merge}). Below: the three fields of \code{JVA}.}
  \Description{Two phone outlines with sets of levels inside, two arrows between them labelled with the codes JVA and Y3J9, and a three-cell diagram of a code's fields below.}
  \label{fig:sync}
\end{figure}

\paragraph{Three layers of storage.} Cleared levels live in \code{localStorage} under a key per
player; half-finished state inside a level (the Gomoku position reached, the stars joined so far)
lives in \code{sessionStorage}, so a reload or a language switch inside a level does not restart it,
and it is dropped when the level is cleared; and a service worker keeps the site itself. When the
browser blocks storage, an in-memory store keeps progress for the visit.

\subsection{The Town Engine}
The town is a grid of characters (\code{.}\ grass, \code{\#} path, \code{\textasciitilde}\ water,
\code{T} tree, \code{O} fountain, and so on). Movement is one tile per 190\,ms with an ease-out
interpolation $e(t) = 1 - (1-t)^2$ of the step's progress $t \in [0,1]$; the pure functions \code{step} and \code{faced} decide movement and facing and are
unit-tested, while a class wires them to a canvas and one animation loop. A tap on the map runs a
breadth-first search over walkable tiles and queues the resulting steps. The camera follows the
player. The same engine takes any map, so the two cities reuse it with their own grids, roads and a
basketball court.

\paragraph{Art as text (\dc{5}).} A sprite is an array of strings, one character per pixel, and
a palette that maps characters to colours. A figure is a $12\times12$ body per facing over four
rows of legs shared between facings, so a walk cycle is the same two leg frames under different
bodies, and the side view is mirrored. Tiles (grass with flowers, paths that join their neighbours,
roofs in each level's colour, lit windows) are painted procedurally. The README banner, the favicon
and the app icons are generated from the same strings by scripts, so a new outfit appears
everywhere at once.

\paragraph{Daylight by city.} Let $h_p$ be the current hour in player $p$'s time zone
(\code{Asia/Shanghai} or \code{America/Chicago}). The phase of the sky and the person's place in a
city day are
\begin{equation}\label{eq:phase}
  \phi(h) = \begin{cases} \text{dawn} & 6 \le h < 7, \\ \text{day} & 7 \le h < 17, \\ \text{dusk} & 17 \le h < 19, \\ \text{night} & \text{otherwise,} \end{cases}
\end{equation}
\begin{equation}\label{eq:stop}
  \mathrm{stop}(h) = \min\big\{\, j : \exists\,[a,b) \in W_j,\; a \le h \bmod 24 < b \,\big\},
\end{equation}
where $W_j$ is the set of hour windows of the $j$-th stop. Dawn, dusk and night each lay a coloured
wash over the map (night: navy at 42\% opacity) and light the windows. Because $h_J$ and $h_Y$
differ by the two cities' offset, the same instant falls in different phases for the two players
(Figure~\ref{fig:bands}).

\begin{figure}[t]
  \centering
  \begin{tikzpicture}[x=3.3mm, y=1mm, font=\scriptsize\sffamily, text=sInk]
    \definecolor{cday}{HTML}{BFE3F7}\definecolor{cdawn}{HTML}{F6B18A}\definecolor{cdusk}{HTML}{E8893F}\definecolor{cnight}{HTML}{3E5A78}
    \foreach \name/\off/\yy in {Shenzhen (UTC+8)/8/11, College Station (UTC-5)/-5/0} {
      \foreach \u in {0,...,23} {
        \pgfmathtruncatemacro{\h}{mod(\u+\off+24,24)}
        \pgfmathsetmacro{\col}{(\h>=6 && \h<7) ? "cdawn" : ((\h>=7 && \h<17) ? "cday" : ((\h>=17 && \h<19) ? "cdusk" : "cnight"))}
        \fill[\col] (\u,\yy) rectangle (\u+1,\yy+5);
      }
      \draw[sInk, line width=0.5pt] (0,\yy) rectangle (24,\yy+5);
      \node[anchor=south west, text=sInk, font=\tiny\sffamily\bfseries, inner sep=0.4mm] at (0,\yy+5) {\name};
    }
    \foreach \u in {0,6,12,18,24} \node[anchor=north, font=\tiny, text=sSoft] at (\u,-0.5) {\u:00};
    \node[anchor=north, font=\tiny, text=sSoft] at (12,-4) {UTC};
    \draw[sRed, line width=1pt] (13.82,-1) -- (13.82,19.5);
    \node[anchor=south east, text=sRed, font=\tiny] at (24,19.3) {one instant: 21:49 in Shenzhen, 08:49 in College Station};
    \node[anchor=north west, font=\tiny, text=sSoft] at (0,-7) {\tikz\fill[cdawn] (0,0) rectangle (2mm,2mm); dawn\quad \tikz\fill[cday] (0,0) rectangle (2mm,2mm); day\quad \tikz\fill[cdusk] (0,0) rectangle (2mm,2mm); dusk\quad \tikz\fill[cnight] (0,0) rectangle (2mm,2mm); night};
  \end{tikzpicture}
  \caption{The two skies over one UTC day (Eq.~\ref{eq:phase}), with the thirteen-hour offset in force
  while the town was built (US daylight time, until 1 November 2026). Night in one town overlaps day in the other for most of every day; the red line is
  the instant shown in Figure~\ref{fig:days}.}
  \Description{Two horizontal bars coloured by dawn, day, dusk and night phases, offset by thirteen hours, with a red vertical line marking one instant.}
  \label{fig:bands}
\end{figure}

\subsection{Sensing in the Browser}\label{sec:sensing}
Camera tracking and breath-amplitude processing run in application code on the device.
Browser speech recognition has a different processing boundary.

\emph{Voice.} The vow and the wedding's ``I do'' use the Web Speech API~\cite{w3c2023speech} in
Mandarin. The API allows a remote recognition service unless local processing is explicitly
required~\cite{w3c2023speech}, and the implementation does not require it, so voice samples may
leave the device. A typed fallback is available for the vow. Recognition alternatives are joined into one string $w$ and normalised
by $\nu$, which keeps only Han characters and Latin letters; with $\sqsubseteq$ for ``is a substring
of'' and $\mathcal{N}_p$ the accepted spellings of the partner's name,
\begin{align}
  \mathrm{vow}_p(w) \iff\;& \text{\zh{我爱你}} \sqsubseteq \nu(w) \;\land\; \text{\zh{我愿意}} \sqsubseteq \nu(w) \label{eq:vow} \\
  &\land\; \exists\, n \in \mathcal{N}_p :\ n \sqsubseteq \nu(w) . \nonumber
\end{align}

\emph{Breath.} Blowing out the candles reads the microphone with echo cancellation, noise
suppression and gain control off. For each frame $t$ of $N = 1{,}024$ samples $x_{t,i}$,
\begin{align}
  r_t &= \sqrt{\tfrac{1}{N}\textstyle\sum_{i=1}^{N} x_{t,i}^{2}}, \qquad
  s_t = 0.6\, s_{t-1} + 0.4 \min(1, 5 r_t), \label{eq:breath} \\
  \mathrm{blow}_t &\iff s_t \ge 0.32 , \nonumber
\end{align}
and every 180\,ms of blowing puts out one candle. We rely on a breath being a loud, broadband
rush on a phone microphone; the threshold was set by hand and has not yet been calibrated across
devices.

\emph{Hands.} The try-on runs MediaPipe's hand landmarker~\cite{lugaresi2019mediapipe,zhang2020hands}
on the camera stream, which returns 21 landmarks $\mathbf{k}_0, \dots, \mathbf{k}_{20}$ per hand
(Figure~\ref{fig:hand}). The ring's axis $\mathbf{a}$ follows the ring finger's proximal phalanx,
its ``up'' $\mathbf{u}$ is the back of the hand, taken from the palm normal $\mathbf{n}$ with the
sign given by the reported handedness, and its centre $\mathbf{c}$ sits a fraction $\beta = 0.38$
along the phalanx:
\begin{align}
  \mathbf{a} &= \frac{\mathbf{k}_{14} - \mathbf{k}_{13}}{\lVert \mathbf{k}_{14} - \mathbf{k}_{13} \rVert}, &
  \mathbf{n} &= \pm\,\frac{(\mathbf{k}_5 - \mathbf{k}_0) \times (\mathbf{k}_{17} - \mathbf{k}_0)}{\lVert (\mathbf{k}_5 - \mathbf{k}_0) \times (\mathbf{k}_{17} - \mathbf{k}_0) \rVert}, \label{eq:hand-axes} \\
  \mathbf{u} &= \frac{\mathbf{n} - (\mathbf{n}\cdot\mathbf{a})\,\mathbf{a}}{\lVert \mathbf{n} - (\mathbf{n}\cdot\mathbf{a})\,\mathbf{a} \rVert}, &
  \mathbf{c} &= (1-\beta)\,\mathbf{k}_{13} + \beta\,\mathbf{k}_{14}, \label{eq:hand-pose}
\end{align}
and the rotation is the orthonormal basis $R = [\,\mathbf{u} \times \mathbf{a},\ \mathbf{u},\ \mathbf{a}\,]$.
Scale comes from anatomy: the mean spacing of the middle, ring and little knuckles is the
ruler, set to a constant of 19.5\,mm (a working estimate for an adult hand),
\begin{equation}\label{eq:scale}
  \text{px/mm} = \tfrac{1}{2}\,\big(\lVert \mathbf{k}_9 - \mathbf{k}_{13}\rVert + \lVert \mathbf{k}_{13} - \mathbf{k}_{17}\rVert\big) \,/\, 19.5 ,
\end{equation}
so nominal model dimensions map to the screen at an assumed scale; the overlay is stylised and
says nothing about ring size or fit. Position, rotation and scale are smoothed
exponentially with $\alpha = 0.45$ per frame, $\mathbf{c}_t = \mathbf{c}_{t-1} + \alpha(\mathbf{c} - \mathbf{c}_{t-1})$
and $q_t = \mathrm{slerp}(q_{t-1}, q, \alpha)$ for the rotation quaternion.

\begin{figure}[t]
  \centering
  \begin{tikzpicture}[x=1.35mm, y=1.35mm, font=\scriptsize\sffamily, text=sInk, lm/.style={circle, fill=sInk, inner sep=0pt, minimum size=1.7mm}, vec/.style={-{Triangle[length=1.6mm, width=1.4mm]}, line width=1pt}]
    \fill[sOrF, draw=sOrB, line width=0.7pt, rounded corners=3mm] (8,0) -- (34,0) -- (37,20) -- (6,20) -- cycle;
    \foreach \x/\h in {11/37, 18/44, 25/46, 32/41} { \draw[sOrB, fill=sOrF, line width=0.7pt, rounded corners=2mm] (\x-3,20) -- (\x-2.6,\h) -- (\x+2.6,\h) -- (\x+3,20); }
    \draw[sOrB, fill=sOrF, line width=0.7pt, rounded corners=2mm] (34,4) -- (46,12) -- (44,16) -- (35,13);
    \draw[sGold, line width=0.8pt] (25,21) -- (18,20) -- (11,19);
    \node[font=\tiny, text=sGold, anchor=north] at (22.6,19.6) {ruler};
    \node[lm] (k0) at (21,0) {};   \node[font=\tiny, anchor=north] at (21,-1.2) {$\mathbf{k}_0$ (wrist)};
    \node[lm] (k5) at (32,20) {};  \node[font=\tiny, anchor=north west] at (32.6,19.2) {$\mathbf{k}_5$};
    \node[lm] (k9) at (25,21) {};  \node[font=\tiny, anchor=south west] at (25.8,21.4) {$\mathbf{k}_9$};
    \node[lm] (k13) at (18,20) {}; \node[font=\tiny, anchor=north east] at (16.8,19.2) {$\mathbf{k}_{13}$};
    \node[lm] (k14) at (17.6,30) {}; \node[font=\tiny, anchor=east] at (15.6,30) {$\mathbf{k}_{14}$};
    \node[lm] (k17) at (11,19) {}; \node[font=\tiny, anchor=east] at (5.4,19) {$\mathbf{k}_{17}$};
    \draw[vec, sNavy] (18,20) -- (17.68,28);
    \node[font=\tiny, text=sNavy, anchor=west] at (19,27.5) {$\mathbf{a}$};
    \fill[sGreen] (21,9) circle (1.1); \draw[white, line width=0.5pt] (21,9) circle (0.5);
    \node[font=\tiny, text=sGreen, anchor=west] at (22.8,9) {$\mathbf{n}$ (towards the viewer)};
    \coordinate (c) at (17.85,23.8);
    \draw[sRed, line width=1.8pt] (c) ellipse (3.7 and 1.15);
    \draw[gold, line width=0.5pt] (c) ellipse (3.7 and 1.15);
    \draw[sRed, line width=0.5pt] (21.6,23.8) -- (36.2,33);
    \node[font=\tiny, text=sRed, anchor=west, align=left] at (36.6,34) {$\mathbf{c} = (1-\beta)\,\mathbf{k}_{13} + \beta\,\mathbf{k}_{14}$\\ $\beta = 0.38$};
    \node[font=\tiny, text=sGold, anchor=west, align=left] at (36.6,42) {$\tfrac12(\lVert\mathbf{k}_9-\mathbf{k}_{13}\rVert + \lVert\mathbf{k}_{13}-\mathbf{k}_{17}\rVert)$\\ $\approx 19.5$\,mm (assumed)};
  \end{tikzpicture}
  \caption{Placing the ring from hand landmarks (Eqs.~\ref{eq:hand-axes}--\ref{eq:scale}). The
  band's axis follows $\mathbf{k}_{13} \to \mathbf{k}_{14}$, its centre sits 38\% along that segment,
  the back of the hand is the palm normal $\mathbf{n}$ oriented by handedness, and the knuckle spacing
  $\mathbf{k}_9$--$\mathbf{k}_{13}$--$\mathbf{k}_{17}$ is the millimetre ruler.}
  \Description{A schematic hand seen from the back with labelled landmark dots at the wrist and knuckles, an arrow along the ring finger, and a red ring drawn across the finger.}
  \label{fig:hand}
\end{figure} The shutter composes the photo with a caption, and a second card
puts both players' try-on photos side by side.

\subsection{Keeping a Secret with a Test (\dc{6})}
The ring is the surprise at the end, and an interface that hints at it spoils it. Rayford, the
town, the progress page and the cats all speak before the ring is reached, and their lines were
written and rewritten over days. Rather than rely on vigilance, a unit test scans every key in
the bilingual dictionary, except the keys shown only after the ring, and fails if any contains
\zh{戒} (ring) in Chinese, \emph{ring} as a word in English, or any word for wedding, marriage,
proposal, bride or groom in either language. With $K$ the set of dictionary keys,
$K_{\mathrm{after}}$ those shown only after the ring, $A$ the allow-list and $R_{\mathrm{zh}}, R_{\mathrm{en}}$
the two pattern tests, the test asserts
\begin{equation}\label{eq:secret}
  \forall\, k \in K \setminus (K_{\mathrm{after}} \cup A):\quad
  \neg R_{\mathrm{zh}}\big(D_{\mathrm{zh}}[k]\big) \;\land\; \neg R_{\mathrm{en}}\big(D_{\mathrm{en}}[k]\big),
\end{equation}
and a companion test asserts coverage, $\forall k \in K: D_{\mathrm{zh}}[k] \neq \varepsilon \land D_{\mathrm{en}}[k] \neq \varepsilon$
(\dc{4}). One element of $A$ is listed by name: Jingyi's own joke on the Shenzhen Bay stop, which is
about a flat, not a ring. The test guards the dictionary only; assets, routes, source code and
visual cues lie outside it.

\subsection{Implementation and Verification}
Table~\ref{tab:metrics} describes the implementation at both versions. For version 1.17.0 an
audit reran the suite: all 153 tests in 36 files passed, and the strict TypeScript check and
production build succeeded. For version 1.24.0 we repeated those steps: all 263 tests in 45 files
passed, and the check and the build succeeded. The main JavaScript chunk grew from 114.09\,kB to
126.76\,kB (42.58 to 46.58\,kB gzip), and the service worker's precache from 44 entries and
302.12\,KiB to 51 entries and 363.49\,KiB; these are build sizes, not load times on phones. The
large ring, map and try-on chunks are lazy loaded.

The first audit independently enumerated the progress-code cases reported above. The progress
and code modules have not changed since that commit, so the enumeration was not repeated.
Dictionary tests check matching non-empty keys; puzzle tests check properties such as unique
winning moves. The server has its own unit tests, counted in the 263; the answering pipeline has
none, and its verification is the manual exercise described in Section~\ref{sec:verify}.
Section~\ref{sec:producttests} lists the tests of each later system.

The development records describe browser checks at desktop, 390-pixel and 320-pixel widths and a
laptop-camera check of hand orientation; no per-release end-to-end report is archived. Systematic
phone checks of speech, breath, camera completion and sharing remain outstanding.

\begin{table}[t]
  \caption{Implementation summary at the two versions of this paper.}
  \label{tab:metrics}
  \small
  \begin{tabular}{@{}p{0.50\columnwidth}p{0.19\columnwidth}p{0.19\columnwidth}@{}}
    \toprule
    Measure & 1.17.0\newline 28 Oct. & 1.24.0\newline 24 Nov. \\
    \midrule
    TypeScript source, excluding tests (lines) & 8{,}414 & 9{,}895 \\
    CSS (lines) & 1{,}080 & 1{,}276 \\
    Bilingual dictionary (lines) & 791 & 989 \\
    Test files / unit tests & 36 / 153 & 45 / 263 \\
    Commits / versions & 109 / 28 & 152 / 36 \\
    Main bundle (min.\ / gzip, kB) & 114 / 43 & 127 / 47 \\
    Precached shell (files / KiB) & 44 / 302 & 51 / 363 \\
    Feature folders & 24 & 27 \\
    Runtime dependencies & 3 & 3 \\
    Author-operated progress backend & 0 & 0 \\
    Server-side endpoints & 0 & 1 \\
    Server source (TypeScript lines) & 0 & 94 \\
    Answering pipeline (Python lines) & 0 & 621 \\
    \bottomrule
  \end{tabular}
\end{table}

\section{The Systems beside the Town}\label{sec:products}
Section~\ref{sec:system} describes the game. This section describes, at the same level of
detail, every other system reachable from the dashboard: how each stores its state, what rules
it computes, what leaves the device, and how it is tested. Section~\ref{sec:after} says what
each is for; here we say how each works. Constants are quoted from the source at the inspected
commit, and file names are given so that each statement can be checked.

\begin{figure*}[t]
  \centering
  \begin{tikzpicture}[
      font=\scriptsize\sffamily,
      tile/.style={scard, snavyc, minimum width=2.62cm, minimum height=9mm, inner sep=2pt},
      st/.style={scard, syellow, minimum width=2.62cm, minimum height=10.5mm, inner sep=2pt, font=\tiny\sffamily},
      leaves/.style={sflat, sredc, dashed, minimum width=2.62cm, minimum height=10.5mm, inner sep=2pt, font=\tiny\sffamily},
      none/.style={sflat, draw=sLine!55, text=sSoft, minimum width=2.62cm, minimum height=10.5mm, inner sep=2pt, font=\tiny\sffamily},
      lab/.style={spill, anchor=east},
      arr/.style={sarr}]
    \node[tile, sblue, minimum width=3.2cm, font=\scriptsize\sffamily\bfseries] (gate) at (0,0) {Gate\\\dsub{\mdseries one password, hashed}};
    \node[tile, sblue, minimum width=11.2cm, right=5mm of gate, font=\scriptsize\sffamily\bfseries] (dash) {Dashboard\\\dsub{\mdseries product registry $\cdot$ one status line per card}};
    \draw[arr] (gate) -- (dash);
    \node[tile, below=7mm of gate.south west, anchor=north west] (town) {\yt{}\\\dsub{nine levels}};
    \node[tile, right=1.6mm of town] (clinic) {Infirmary\\\dsub{letters, notes, chart}};
    \node[tile, right=1.6mm of clinic] (trip) {Trip\\\dsub{clock, passes, plan}};
    \node[tile, right=1.6mm of trip] (ask) {Request box\\\dsub{composer, history}};
    \node[tile, right=1.6mm of ask] (mm) {MotionMirror\\\dsub{off-site}};
    \node[sflat, dashed, right=1.6mm of mm, minimum width=1.5cm, minimum height=9mm, inner sep=2pt] (seed) {Seed\\\dsub{no behaviour}};
    \foreach \n in {town,clinic,trip,ask,mm,seed} \draw[arr] (dash.south -| \n) -- (\n.north);
    \node[st, below=2.4mm of town] (s1) {progress per player\\codes exchanged by hand};
    \node[st, below=2.4mm of clinic] (s2) {visit count\\private chart (30 lines)};
    \node[none, below=2.4mm of trip] (s3) {none\\dates in the source};
    \node[st, below=2.4mm of ask] (s4) {history (50 entries)\\names and sizes only};
    \node[none, below=2.4mm of mm] (s5) {a separate program\\run on one's own files};
    \node[none, below=2.4mm of s1] (o1) {nothing};
    \node[none, below=2.4mm of s2] (o2) {nothing\\one line, if she shares it};
    \node[none, below=2.4mm of s3] (o3) {nothing};
    \node[leaves, below=2.4mm of s4] (o4) {text and files, to a server,\\two mail providers, a laptop\\and a model provider};
    \node[none, below=2.4mm of s5] (o5) {the link itself};
    \node[lab, fill=sNavy] at ($(town.west)+(-2mm,0)$) {page};
    \node[lab, fill=sGold] at ($(s1.west)+(-2mm,0)$) {kept on the device};
    \node[lab, fill=sRed] at ($(o1.west)+(-2mm,0)$) {leaves the device};
  \end{tikzpicture}
  \caption{The site at version 1.24. Every page lies behind the gate and is reached from the
  dashboard. The second row shows what each page keeps in the browser's storage, and the third
  what it sends anywhere. Only the request box sends anything, and what it sends travels
  further than the rest of the site put together (Figure~\ref{fig:askpath}).}
  \Description{A diagram in three rows under a header. The header shows a gate leading to a
  dashboard. The first row has six tiles: Yi Town, infirmary, trip, request box, MotionMirror
  and a dashed seed tile. The second row lists what each stores on the device. The third row
  says nothing leaves the device for the first three, and for the request box lists a server,
  two mail providers, a laptop and a model provider.}
  \label{fig:sitemap}
\end{figure*}

Figure~\ref{fig:sitemap} gives the whole site at a glance.

\subsection{The Product Registry and Routing}\label{sec:registry}
The dashboard is generated from one array. A product is a record
\begin{equation}\label{eq:product}
  P = (\mathit{id},\ \mathit{route},\ \mathit{title},\ \mathit{blurb},\ \mathit{image}^{?},\
       \mathit{status}^{?},\ \mathit{external}^{?}),
\end{equation}
where \emph{title} and \emph{blurb} are keys of the bilingual dictionary, \emph{image} is an
optional picture, \emph{status} is an optional function returning one line of text or nothing,
and \emph{external} marks a route that leaves the site (\code{dashboard/products.ts}). The page
maps the array to cards and appends the seed tile, which is markup with no entry in the array
(\code{dashboard/dashboard.ts}). Each status function is evaluated once, when the page is
drawn; the page does not poll. An external card is rendered as a link that opens a new tab with
\code{rel="noopener"}.

The five status rules are these. \emph{Town}: with $p$ the player chosen on this device and
$C_p$ the levels that player has cleared, the line is the player's name and
$|C_p|/9$; it is empty when no player is chosen or $|C_p|=0$, and it never reads the other
player's set. \emph{Infirmary}: a constant string. \emph{Trip}: the function of
Section~\ref{sec:tripclock}. \emph{Request box}: none. \emph{MotionMirror}: a constant string.

\begin{table}[t]
  \caption{What each route requires before it opens (\code{main.ts}).}
  \label{tab:routes}
  \small
  \begin{tabular}{@{}p{0.31\columnwidth}ccp{0.23\columnwidth}@{}}
    \toprule
    Route & Password & Player & Level unlocked \\
    \midrule
    Dashboard, trip, request box & yes & no & no \\
    Town, infirmary, city days & yes & yes & no \\
    A level & yes & yes & yes, and entered by its door \\
    MotionMirror (off-site) & to reach the card & no & no \\
    \bottomrule
  \end{tabular}
\end{table}

Routing is by path, with no framework (\code{core/router.ts}). A legacy fragment route of the
form \code{\#/x} takes priority over the path, repeated and trailing slashes are normalised,
and navigation calls \code{history.pushState} and then dispatches a custom event that the
shell listens for. Plain left clicks on root-relative links are intercepted; links that carry
\code{target} or \code{download} are left to the browser. Each page module is loaded on
demand. If a module fails to load, which happens when a new build has replaced the files an
open page expects, the shell reloads once, guarded by a flag in \code{sessionStorage} so that a
real outage cannot cause a reload loop. Table~\ref{tab:routes} gives the three levels of
requirement. Unknown paths fall back to the dashboard.

\subsection{The Infirmary}\label{sec:clinic}
\begin{figure*}[t]
  \centering
  \includegraphics[width=\textwidth]{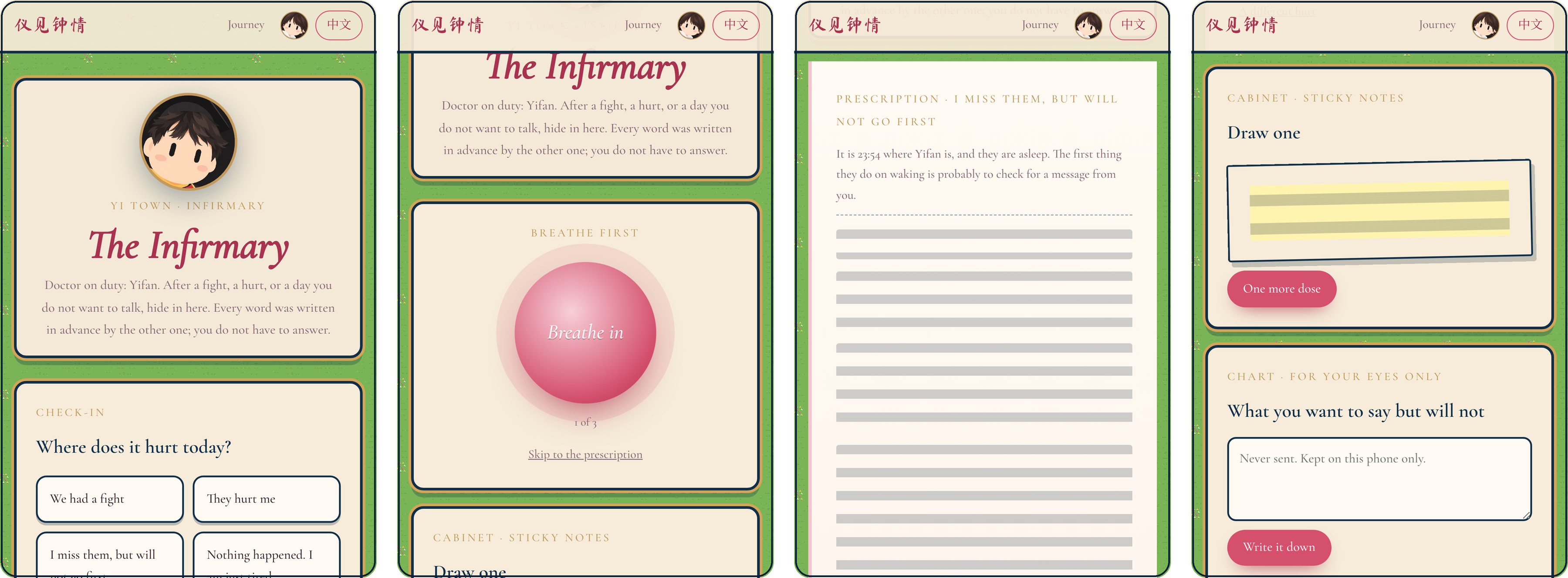}
  \caption{A visit to the infirmary, on a phone, in English. From the left: checking in by
  naming what hurts; the first of three breaths; the letter for that hurt, opening with a line
  computed from the clock in the writer's city; and the cabinet of notes above the private
  chart. The letter and the note are masked here because they are personal; the line above
  the letter and the line on the chart are shown as the page displays them, and the chart's
  line was typed for this figure.}
  \Description{Four phone screenshots side by side on a green pixel meadow. The first shows a
  title card and four buttons naming kinds of hurt. The second shows a large pink circle with
  the words breathe in. The third shows a letter whose paragraphs are replaced by grey bars.
  The fourth shows a yellow sticky note with its text hidden, and a text box labelled never
  sent, kept on this phone only.}
  \label{fig:clinic}
\end{figure*}

Figure~\ref{fig:clinic} shows one visit from start to finish.
\paragraph{Data.}
There are four kinds of hurt, $H=\{\textit{fight},\textit{hurt},\textit{miss},\textit{tired}\}$.
A prescription is a letter $(\mathit{from}, h, \mathit{paragraphs}, \mathit{signature})$ with
$h\in H$ and paragraphs in both languages; a cabinet note is $(\mathit{from}, \mathit{zh},
\mathit{en})$. The source holds eight letters, one from each partner for each hurt, each of
six paragraphs in each language, and 36 notes, eighteen from each partner
(\code{clinic/prescriptions.ts}). Two keys are kept in \code{localStorage} for each player
$p$: a visit counter, \code{yi.clinic.visits.}$p$, and a private chart,
\code{yi.clinic.chart.}$p$, a list of at most 30 entries $(\mathit{at},\mathit{text})$
(\code{clinic/visits.ts}).

\paragraph{Roles.}
Whoever checks in is the patient $p$; the other partner $\bar p$ is the doctor. The letter shown
for a hurt $h$ is the unique prescription with $\mathit{from}=\bar p$ and that hurt, and the
cabinet is the subset of notes with $\mathit{from}=\bar p$. A player is therefore never shown
their own words.

\paragraph{Checking in and breathing.}
Checking in increments the counter and nothing else: neither the hurt chosen nor the time is
written to storage, and the time of check-in is held in memory only for the length of the
visit. Three breaths follow, each of four seconds in, four seconds held and six seconds out,
so the exercise lasts
\begin{equation}\label{eq:clinicbreath}
  3 \times (4 + 4 + 6)\ \mathrm{s} = 42\ \mathrm{s}.
\end{equation}
A button skips it, and when the browser reports a preference for reduced motion the exercise
is omitted and the letter is shown at once.

\paragraph{What the doctor is doing now.}
The letter's first line is computed, not stored. Let $z_{\bar p}$ be the doctor's configured
time zone and $\tau$ the present instant. The page reads the wall-clock hour and minute in
$z_{\bar p}$ through the browser's internationalisation interface, forms
$x = \mathrm{hour} + \mathrm{minute}/60$, and classifies it with the same function that tints
the town's sky:
\begin{equation}\label{eq:clinicphase}
  \phi(x) = \begin{cases}
    \textit{dawn} & 6 \le x < 7 \\
    \textit{day} & 7 \le x < 17 \\
    \textit{dusk} & 17 \le x < 19 \\
    \textit{night} & \text{otherwise.}
  \end{cases}
\end{equation}
The phase selects one of four sentences in the dictionary, which is filled with the doctor's
name and the clock time. The line is thus a statement about the hour in a city. It uses no
calendar, location, or signal from the other person's device, and the same sentence is shown at
three in the afternoon on a working day and on a holiday.

\paragraph{The cabinet.}
A draw must not repeat the note just shown. With $N$ the doctor's notes, $\ell$ the last note
shown and $r$ uniform on $[0,1)$, the pool is $N\setminus\{\ell\}$ when $|N|>1$ and $\ell$
exists and $N$ otherwise, and the note drawn is
\begin{equation}\label{eq:draw}
  \mathrm{pool}\big[\min(|\mathrm{pool}|-1,\ \lfloor r\,|\mathrm{pool}| \rfloor)\big].
\end{equation}
Repeats after one intervening draw are allowed. A placeholder in a note is replaced by the
number of days since the couple met, counted in the patient's time zone.

\paragraph{Leaving.}
``Tell them I am out'' builds one sentence containing the visit's length in minutes,
$m=\max(1,\ \mathrm{round}(\Delta/60\,000))$ for $\Delta$ milliseconds since check-in, and
passes it to the operating system's share sheet. If sharing is unavailable or fails for a
reason other than the user cancelling, the sentence is copied to the clipboard and a notice
says so. The application itself transmits nothing: the only network request the page makes is
for an optional audio file on the site's own origin.

\paragraph{Voice.}
A letter can be heard if a recording exists at a path derived from its hurt and its writer. The
page does not know in advance which recordings exist; it creates an audio element that loads
metadata only and shows the play button when the browser reports that the file can play. A
script records a reading at 44.1\,kHz mono, trims leading and trailing silence below an
amplitude of 0.01 while keeping a quarter of a second before and half a second after the
speech, and encodes it as AAC at 96\,kbit/s (\code{scripts/record-rx.py}).

\paragraph{The chart.}
Lines are trimmed, blank lines are ignored, new lines are placed first, and only the newest 30
are kept. Erasing a line matches it by its timestamp. Entries that fail to parse are dropped
on load. The chart is escaped before display and is never sent anywhere; when storage is
blocked it lives in memory for the session.

\paragraph{In the town.}
The infirmary is a building with its door at tile $(7,8)$ and a cross for a sign
(\code{world/town.ts}). Its door is always walkable, it is not in the level registry, and the
guide can lead a player to it. The route itself requires a chosen player, because without one
the page cannot tell patient from doctor.

\paragraph{Tests.}
Eleven tests in two files check that every patient has a letter from the other for every
hurt, with equal numbers of paragraphs in the two languages and none empty; that no one
receives their own letter and no pair of writer and hurt is duplicated; that each partner has
at least ten notes in both languages; that a note is never drawn twice in a row, including the
cases of an empty cabinet and a single note; that visit counts are kept per player and
malformed storage is ignored; that the chart orders, trims, erases and caps as described; and
that every hour of the day yields a line in both languages and a visit length of at least one
whole minute. A further test of the town map checks that the door is reachable from the
starting tile and unlocked.

\subsection{The Trip Clock}\label{sec:tripclock}
\paragraph{Data.}
A flight leg is $(\mathit{from},\mathit{to},\mathit{departs},\mathit{arrives})$, where the
airports carry an IANA time zone and coordinates and the two times are ISO~8601 strings with an
explicit offset from UTC (\code{core/config.ts}). The trip is two legs, out and back. The
countdown on the home page and the gold arc on the Love Map read the outbound leg from the
same record, so there is one source for the date. The page keeps nothing in storage.

\paragraph{Durations.}
Because both times carry offsets, a leg's duration is a difference of absolute instants,
$(\mathit{arrives}-\mathit{departs})/3\,600\,000$ hours, which gives 15.5\,h out and 16.25\,h
back with no reasoning about time zones. Whether a leg lands on the next calendar day is a
different question, about dates as passengers read them: the page compares the arrival date in
the arrival airport's zone with the departure date in the departure airport's zone, and marks
the leg ``+1 day'' when they differ. This is true of the return leg only.

\begin{figure}[t]
  \centering
  \begin{tikzpicture}[font=\scriptsize\sffamily, text=sInk, x=1cm, y=1cm]
    \fill[sCrF] (0,0) rectangle (1.5,0.5);
    \fill[sBlF] (1.5,0) rectangle (2.5,0.5);
    \fill[sYeF] (2.5,0) rectangle (6.4,0.5);
    \fill[sBlF] (6.4,0) rectangle (7.4,0.5);
    \fill[sCrF] (7.4,0) rectangle (8.3,0.5);
    \draw[sInk, line width=0.7pt, rounded corners=1.5pt] (0,0) rectangle (8.3,0.5);
    \foreach \x/\l in {1.5/t_0, 2.5/t_1, 6.4/t_2, 7.4/t_3}
      { \draw[sNavy, line width=0.9pt] (\x,-0.12) -- (\x,0.62); \node[above, text=sNavy] at (\x,0.6) {$\l$}; }
    \node at (0.75,0.25) {before}; \node at (2.0,0.25) {out}; \node[font=\scriptsize\sffamily\bfseries] at (4.45,0.25) {together};
    \node at (6.9,0.25) {back}; \node at (7.85,0.25) {after};
    \foreach \i in {1,...,11} { \draw[sYeB, line width=0.6pt] ({2.5+(\i-0.5)*3.9/11},-0.12) -- ({2.5+(\i-0.5)*3.9/11},-0.3); }
    \node[below, font=\tiny\sffamily] at (2.68,-0.3) {1};
    \node[below, font=\tiny\sffamily] at (6.22,-0.3) {11};
    \node[below, font=\tiny\sffamily, text=sSoft] at (4.45,-0.32) {day $n$, on New York's calendar};
    \node[below, font=\tiny\sffamily, text=sSoft] at (0.75,-0.05) {counts to $t_0$};
    \node[below, font=\tiny\sffamily, text=sSoft] at (7.85,-0.05) {rests};
  \end{tikzpicture}\\[6pt]
  \includegraphics[width=0.94\columnwidth]{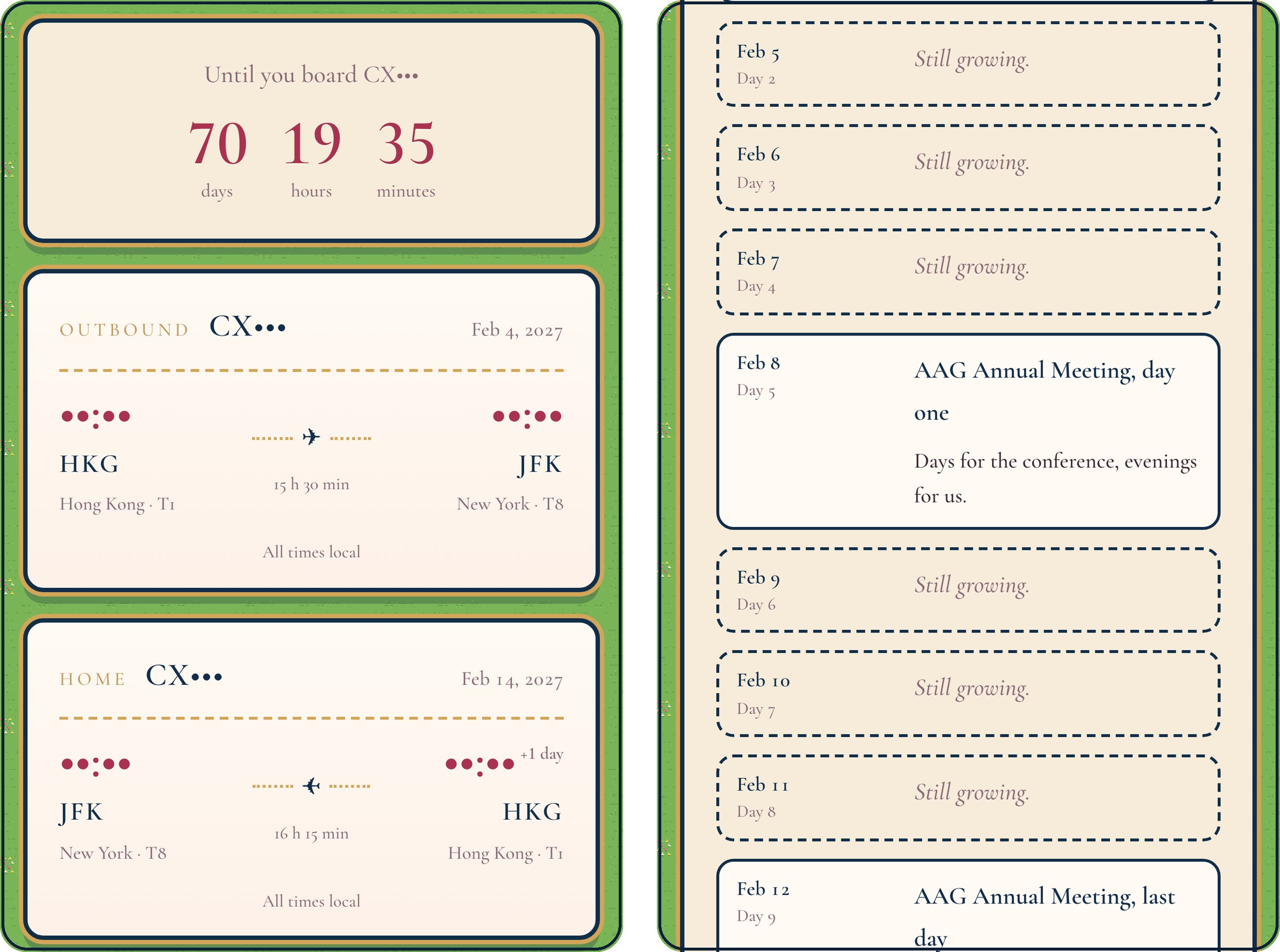}
  \caption{The trip clock. Top: the five phases are intervals between four instants, so they
  are the same wherever the page is opened; only the day number depends on a calendar, and the
  calendar is the destination's. Bottom: the page 70 days before departure. Left, the
  countdown and the two legs; right, the itinerary, where a day not yet written reads ``still
  growing''. Flight numbers and clock times are masked for publication.}
  \Description{Top: a horizontal bar divided into five labelled intervals, before, out,
  together, back and after, separated by four marked instants, with eleven small ticks under
  the interval called together. Bottom: two phone screenshots, one with a countdown of 70 days
  and two boarding passes between Hong Kong and New York, the other with a list of dated rows
  most of which say still growing.}
  \label{fig:trip}
\end{figure}

\paragraph{Phases.}
Write $t_0<t_1<t_2<t_3$ for the instants of outbound departure, outbound arrival, return
departure and return arrival, and $\tau$ for now. The phase is
\begin{equation}\label{eq:tripphase}
  \psi(\tau) = \begin{cases}
    \textit{before} & \tau < t_0 \\
    \textit{out} & t_0 \le \tau < t_1 \\
    \textit{together} & t_1 \le \tau < t_2 \\
    \textit{back} & t_2 \le \tau < t_3 \\
    \textit{after} & \tau \ge t_3 ,
  \end{cases}
\end{equation}
and the clock counts down to the upper bound of the current phase, with nothing to count to in
the last (\code{trip/plan.ts}; Figure~\ref{fig:trip}). The comparison is between instants, so the phase is the same
for both partners wherever they are.

\paragraph{Which day of the stay.}
Days together are counted on New York's calendar, not the viewer's. Let $d_z(\tau)$ be the
calendar date at instant $\tau$ in zone $z$, and $u(d)$ the UTC midnight of a date $d$ in
milliseconds. With $z$ the destination's zone,
\begin{equation}\label{eq:tripday}
  n(\tau) = \mathrm{round}\!\left(\frac{u(d_z(\tau)) - u(d_z(t_1))}{86\,400\,000}\right) + 1 ,
\end{equation}
so the day of landing is day 1. Both dates are taken at UTC midnight, so the difference is a
whole number of days whatever daylight saving time does in between. The length of the stay is $n(t_2)$, which is 11. A viewer
in Shenzhen at noon and a viewer in New York at eleven the night before therefore see the same
day number, which a count on the viewer's own calendar would not give.

\paragraph{The dashboard line.}
Before departure the card shows $\lceil (t_0-\tau)/86\,400\,000 \rceil$ days, with a separate
string for one day; in the air it shows a fixed phrase; during the stay, ``day $n$ of 11''; on
the way home and afterwards, fixed phrases.

\paragraph{The page.}
The clock is redrawn once a second, with the remaining time floored to days, hours and
minutes. The region that holds it is marked as a polite live region for screen readers. The
itinerary is a list of twelve entries, one for each date from the landing in New York to the
landing back in Hong Kong, each with an optional title and note in both languages and a list
of photographs or videos (\code{trip/itinerary.ts}). At the inspected commit five entries have
titles. An entry earlier than today in New York is dimmed, today's is marked, and an entry
without a title shows a placeholder.

\paragraph{Tests.}
Eight tests fix the two durations, the next-day rule, each of the five phases with its
countdown target, the day number at 23:30 on the first night and on the last day, and the
dashboard strings in both languages. Fourteen cases over the itinerary check that its first
and last dates are the two landings, that the dates are consecutive, that every title and
note exists in both languages or neither, and that every listed media file exists on disk
under exactly that name.

\subsection{The Service Worker, the Caches and Updates}\label{sec:update}
The site installs as a progressive web application. Its service worker is generated at build
time and configured in three parts (\code{vite.config.ts}).

\emph{Precache.} Scripts, styles, markup, vector and raster images and fonts are precached,
except photographs and other media, the hand-tracking model, and the chunks for the ring
viewer, the try-on and the map; no single precached file may exceed 3\,MB. The exclusions keep
the 21\,MB hand-tracking bundle from downloading at installation.

\emph{Runtime caches.} Three named caches fill as the site is used
(Table~\ref{tab:caches}). Media are served from the cache at once and refreshed in the
background, so a photograph replaced under the same name reaches an installed phone on its
next visit. Map tiles are served from the cache without revalidation for thirty days.

\begin{table}[t]
  \caption{Runtime caches of the service worker.}
  \label{tab:caches}
  \small
  \begin{tabular}{@{}p{0.20\columnwidth}p{0.36\columnwidth}cc@{}}
    \toprule
    Content & Strategy & Entries & Age (days) \\
    \midrule
    Media, hand model & stale while revalidate & 200 & 365 \\
    Web fonts & stale while revalidate & 30 & 365 \\
    Map tiles & cache first & 400 & 30 \\
    \bottomrule
  \end{tabular}
\end{table}

\emph{Updates.} The worker is registered in the mode in which a new version waits instead of
taking over. The site deploys on every push, and the record for the period holds 43 commits
and eight released versions, so a new build will often appear while its one other user is in
the middle of something. Until version 1.20 a new build could reload a page that was open.
Since then, when a new worker is ready the page appends one bar, marked as a status region,
that says a new version is ready and offers one button; pressing it activates the new worker
and reloads (\code{core/update.ts}). Nothing reloads until then, so a page is never replaced
underneath someone half-way through a level or a letter. The page asks the browser to check
for a new worker whenever the tab becomes visible again and once an hour; a failed check is
ignored. The one exception to waiting is the failed module load described in
Section~\ref{sec:registry}.

The same question returns at a larger scale with the gate. Version 1.24 withdrew every earlier
pass by changing the storage key under which a pass is remembered, so each device was asked
for the password again on its next visit, and this time the recipient was not offered a
choice. An update bar asks before it changes the page; a designer who changes the door does
not. Both are decisions by one person about the other's access to something described as
theirs in common.

\subsection{The Request Page}\label{sec:askpage}
\paragraph{Limits.}
One request carries at most ten files and 18\,MiB in total, the practical limit of one email
after encoding, and the page keeps the newest 50 requests as history
(\code{ask/core.ts}). A draft is checked in a fixed order: it is \emph{empty} if it has neither
text nor files, \emph{too many} above ten files, and \emph{too big} above the byte limit. The
server applies the same constants, imported from the same module.

\paragraph{Preparing files.}
A JPEG, PNG or WebP image larger than 1.5\,MiB is scaled before it is attached: it is decoded
with its recorded orientation applied, drawn on a white ground so that its long edge is at
most 2400 pixels, and encoded as JPEG at quality 0.85 (\code{ask/shrink.ts}). The original is
kept if the result is not smaller or if any step fails. Other file types are attached as they
are. Files being prepared are counted, and sending is disabled until the count returns to
zero; before this rule, a send made in that moment left those files behind without notice.

\paragraph{State.}
The page's state is held outside the page, in a session object created once per load
(\code{ask/session.ts}):
\begin{equation}\label{eq:askstate}
  S = (\mathit{draft},\ \mathit{history},\ \mathit{adding},\ \mathit{pending},\ \mathit{note},\
       \mathit{retrying}),
\end{equation}
with \emph{draft} the text and files in the composer, \emph{adding} the number of files being
prepared, \emph{pending} the one send in flight or none, \emph{note} the last validation
error, and \emph{retrying} the failed entry that the composer currently holds. The page is a
function of $S$ and subscribes to it. This is why a send survives events that redraw the page:
switching language rebuilds the markup, and the new markup finds the same \emph{pending}.

\begin{figure}[t]
  \centering
  \begin{tikzpicture}[
      font=\scriptsize\sffamily,
      st/.style={scard, snavyc, rounded corners=6pt, inner sep=3pt, minimum height=6.5mm, minimum width=1.5cm},
      arr/.style={sarr}]
    \node[st] (idle) {composing};
    \node[st, syellow, right=14mm of idle] (pend) {pending};
    \node[st, steal, above right=3mm and 14mm of pend] (sent) {sent};
    \node[st, sredc, below right=3mm and 14mm of pend] (fail) {failed};
    \draw[arr, sGreen] (idle) -- node[above, text=sGreen]{send} (pend);
    \draw[arr, sGreen] (pend) -- node[above, sloped, text=sGreen]{ok} (sent);
    \draw[arr, sRed] (pend) -- node[below, sloped, text=sRed]{error, 3 min} (fail);
    \draw[arr, sGreen] (fail.south) to[bend left=28] node[below, text=sGreen]{send again} (pend.south);
    \draw[arr, sRed] (idle.north) to[out=110,in=70,looseness=5] node[above, text=sRed]{invalid draft} (idle.north east);
  \end{tikzpicture}
  \\[6pt]
  \includegraphics[width=0.52\columnwidth]{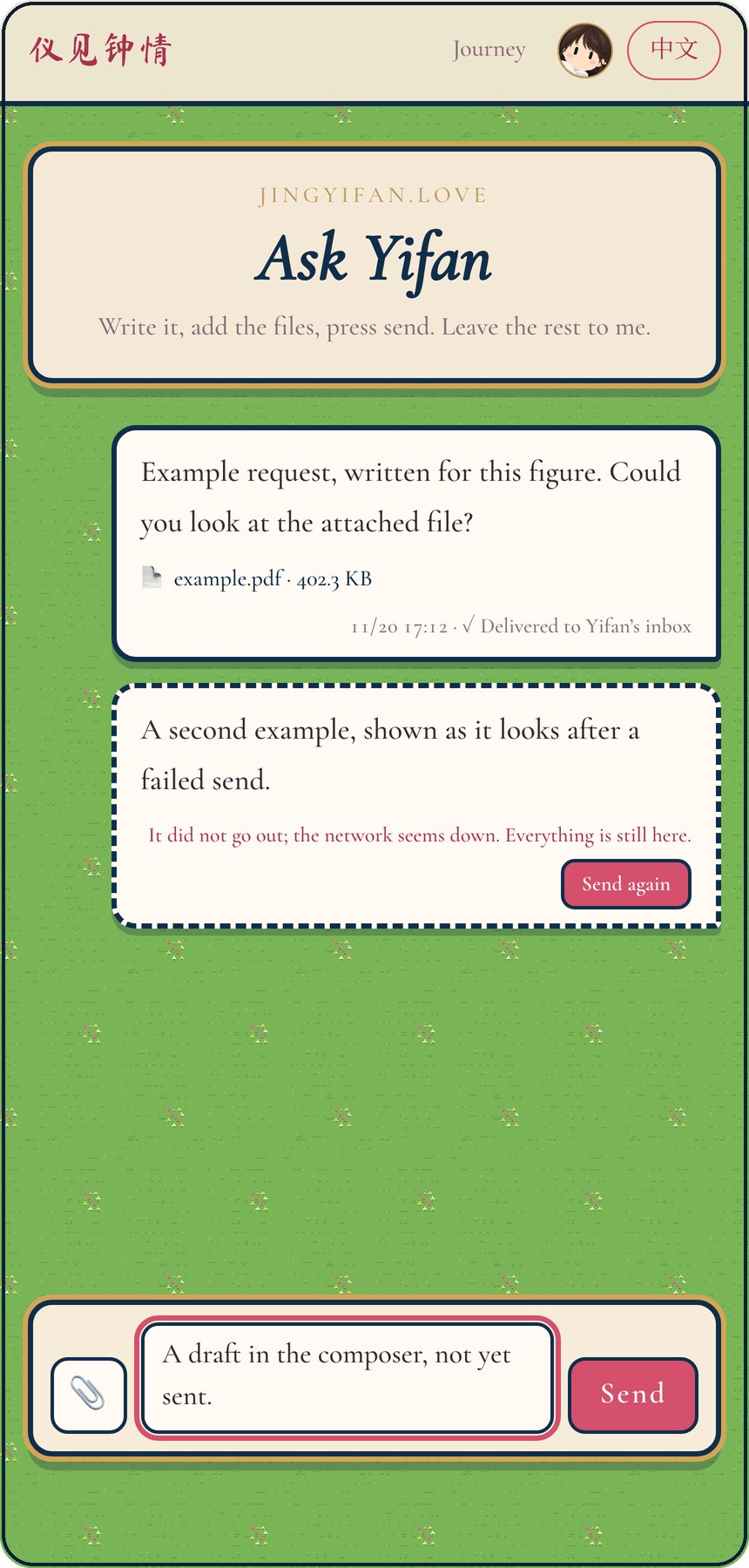}
  \caption{Top: states of one request on the page. A draft that fails validation never leaves
  \emph{composing}; a send that receives no answer is abandoned after three minutes; a failed
  request keeps its text and files in the composer. Bottom: the page with three constructed
  examples, written for this figure and never sent: a delivered request with a file, a failed
  one offering to send again, and a draft.}
  \Description{A state diagram with four rounded boxes: composing, pending, sent and failed.
  Arrows lead from composing to pending, from pending to sent and to failed, from failed back
  to pending, and from composing to itself. Below, a phone screenshot of a chat-like page with
  two message bubbles, the second outlined with dashes and carrying a send again button, and a
  text box with a paperclip and a send button.}
  \label{fig:askstates}
\end{figure}

\paragraph{Sending.}
Figure~\ref{fig:askstates} shows the states of a request. Submitting does nothing while a send
is pending or files are being prepared. Otherwise the draft is validated, and a valid draft is
posted as a multipart form with the fields \emph{text}, \emph{lang}, \emph{player} if one is
chosen, and one \emph{files} field per file. A controller aborts the request after
180 seconds. The outcome is \emph{sent} only if the HTTP status is successful and the body
says so. Otherwise the reason is taken from the body when it is one the page knows, and is
\emph{network} if the request was rejected or timed out and \emph{server} in every other case,
including a successful status with an unexpected body. The entry is written to storage only
once its outcome is known. On success, the text is cleared only if it still equals what was
sent, and only the files that were sent are removed, so anything typed or attached during the
send remains.

\paragraph{Sending again.}
A failed entry keeps its text; file contents are never stored, only names and sizes. Pressing
``send again'' has four possible results. It is refused while the session is busy. It is
refused, with the entry kept, if the composer already holds different text. If the page has
been reloaded since the failure and the entry had files, the text is restored and the user is
asked to attach the files again. Otherwise it sends, and the new attempt replaces that failed
entry and no other.

\paragraph{Input.}
On a device with a fine pointer, Enter sends and Shift with Enter breaks the line; on a touch
device Enter breaks the line and the button sends. Enter never sends while an input method is
composing text, and it also does not send when the key event reports key code 229, which is
how one desktop browser signals the Enter that confirms a pinyin candidate after composition
has formally ended. Pasted content is taken as files only when the clipboard holds a file and
no plain text, so a paste from a spreadsheet arrives as text and not as a picture of it; a
pasted screenshot is given a name with the time and a serial number. Dropped files are
accepted anywhere on the window.

\paragraph{Tests.}
Fifty-two tests cover the constants and the validation order at their boundaries, the history
store under malformed and failing storage, every branch of the outcome rule including a stall
before and during the response body, the behaviour of a send across re-subscription, files
added during a send, the four results of sending again, the keyboard rule with composition,
the paste rule, and the scaling rule. The module that draws the page has no automated tests.

\subsection{The Server}\label{sec:worker}
The server is one function on an edge platform, bound to a custom domain, with a binding for
sending email (\code{worker/src/handle.ts}). The destination address is a secret of the
deployment; the sending address is an ordinary variable. It stores nothing and has no key,
rate limit or challenge.

\begin{table}[t]
  \caption{The server's checks, in order, and its answers.}
  \label{tab:worker}
  \small
  \begin{tabular}{@{}p{0.50\columnwidth}cl@{}}
    \toprule
    Condition & Status & Reason \\
    \midrule
    Origin absent or not allowed & 403 & server \\
    Preflight request & 204 & \\
    Not a POST to the one path & 404 & server \\
    Body is not a form & 400 & empty \\
    No text and no files & 400 & empty \\
    More than ten files & 413 & too-many \\
    More than 18\,MiB & 413 & too-big \\
    Sending the email fails & 502 & server \\
    Otherwise & 200 & \\
    \bottomrule
  \end{tabular}
\end{table}

Table~\ref{tab:worker} gives its checks in the order applied. Two origins are allowed, the
site's and the development server's. A request from any other origin is refused before any
cross-origin header is written, so a page elsewhere cannot read even the refusal. The origin
header is set by browsers and can be set freely by any other client, which is why this check
keeps other web pages out and does not keep people out.

Three inputs are cleaned. A file name is reduced to its last path segment and stripped of
control characters, the line and paragraph separators U+2028 and U+2029, and double quotes.
The \emph{player} field is looked up in a map of two names and replaced by a default when it
is anything else, so no text supplied by the client reaches the subject through that field.
The language is English only if the field says so exactly.

The email's subject is a fixed prefix, the name, and the first thirty code points of the text
with white space collapsed, or the first file's name when there is no text. Its body is the
text, a separator, the time of the request in each partner's city, the page language, and the
names and sizes of the attachments. Twenty tests cover the cleaning rules, the subject, each
row of Table~\ref{tab:worker}, and the fields of a complete send.

\subsection{The Answering Pipeline}\label{sec:askpath}
The pipeline is a set of scripts on the designer's laptop, run by the operating system's
scheduler every 600 seconds (\code{mailflow/}). Figure~\ref{fig:askpath} shows the path of one
request through it. A second scheduled agent keeps the laptop awake while it is on mains
power. An exclusive file lock makes a run exit at once if the previous run has not finished.

\begin{figure*}[t]
  \centering
  \begin{tikzpicture}[
      font=\scriptsize\sffamily,
      box/.style={scard, inner sep=2pt, minimum height=9.5mm, minimum width=1.86cm, font=\scriptsize\sffamily\bfseries},
      hers/.style={box, sblue},
      ext/.style={sflat, dashed, inner sep=2pt, minimum height=9.5mm, minimum width=1.86cm, font=\scriptsize\sffamily\bfseries},
      loc/.style={box, fill=gold!30, draw=sCrB},
      note/.style={font=\tiny\sffamily, align=center, text=sSoft, text width=1.95cm},
      zone/.style={font=\tiny\sffamily, text=sSoft},
      arr/.style={sarr}]
    \node[hers] (page) at (0,0) {Request page};
    \node[ext, right=3.1mm of page] (worker) {Server};
    \node[ext, right=3.1mm of worker] (inbox) {His mailbox};
    \node[loc, right=3.1mm of inbox] (fetch) {Fetch};
    \node[loc, right=3.1mm of fetch] (sort) {Sort};
    \node[ext, right=3.1mm of sort] (work) {Work};
    \node[loc, right=3.1mm of work] (send) {Send};
    \node[ext, right=3.1mm of send] (her) {Her mailbox};
    \begin{scope}[on background layer]
      \draw[spanel, draw=sPnGoB, fill=sPnGoF] ($(fetch.north west)+(-1.2mm,6.3mm)$) rectangle ($(send.south east)+(1.2mm,-14.5mm)$);
    \end{scope}
    \node[spill, fill=sGold, anchor=west] at ($(fetch.north west)+(1.2mm,6.3mm)$) {scheduled every ten minutes};
    \foreach \a/\b in {page/worker, worker/inbox, inbox/fetch, fetch/sort, sort/work} \draw[arr, sGreen] (\a) -- (\b);
    \foreach \a/\b in {work/send, send/her} \draw[arr, sVio] (\a) -- (\b);
    \node[zone, above=0.8mm of page] {her phone};
    \node[zone, above=0.8mm of worker] {edge platform};
    \node[zone, above=0.8mm of inbox] {mail provider};
    \node[zone, above=0.8mm of fetch] {his laptop};
    \node[zone, above=0.8mm of sort] {his laptop};
    \node[zone, above=0.8mm of work] {model provider};
    \node[zone, above=0.8mm of send] {his laptop};
    \node[zone, above=0.8mm of her] {mail provider};
    \node[note, below=1.5mm of page] {up to 10 files, 18\,MiB;\\large photos scaled;\\history kept locally};
    \node[note, below=1.5mm of worker] {origin and limits\\checked; one email;\\stores nothing;\\no authentication};
    \node[note, below=1.5mm of inbox] {records the DKIM\\result on receipt};
    \node[note, below=1.5mm of fetch] {sender, subject\\prefix, and a parsed\\DKIM pass; one\\folder per request};
    \node[note, below=1.5mm of sort] {two local models:\\transcribe images,\\classify, or stop\\for a person};
    \node[note, below=1.5mm of work] {hosted model in a\\sandbox: one folder,\\one locked script,\\no network tools};
    \node[note, below=1.5mm of send] {one fixed recipient;\\copy to him;\\at most six a day};
    \node[note, below=1.5mm of her] {the answer, as\\documents};
  \end{tikzpicture}
  \caption{The path of one request, left to right, with the check applied at each step. Gold
  boxes run on the designer's laptop; dashed boxes are services operated by others; the dashed
  frame is the scheduled job. Her text and files pass through four outside parties on the way
  to an answer: the edge platform, his mail provider, the provider of the hosted model, and her
  own mail provider. The local models keep the sorting step on the laptop; the working step
  sends the files on.}
  \Description{A row of eight boxes joined by arrows: request page, server, his mailbox, fetch,
  sort, work, send, her mailbox. Above each is the place it runs: her phone, an edge platform,
  a mail provider, his laptop, his laptop, a model provider, his laptop, a mail provider. Below
  each is a short note on what is checked there. A dashed frame encloses fetch, sort, work and
  send.}
  \label{fig:askpath}
\end{figure*}

\paragraph{Fetching.}
The job opens the designer's mailbox over IMAP, read-only, in the folder that the server marks
with the special-use flag for all mail; it finds that folder by the flag and not by name,
because the folder's name is localised. It searches for messages from the site's sending
address and keeps the identifiers it has already handled in a state file. Each new message is
fetched without marking it as read and handled inside its own error boundary: a message that
fails to parse is logged and skipped, and the state file is written after every message, so one
malformed email cannot stop those behind it (\code{mailflow/fetch.py}).

\paragraph{Authenticating.}
A message is accepted only if three conditions hold: its sender address equals the site's
sending address, its subject begins with the server's fixed prefix, and the receiving mail
provider recorded a passing DKIM signature~\cite{rfc6376} for the site's domain. The third
condition is read from the topmost \code{Authentication-Results} header, the one the provider
adds on receipt; a header of the same name carried inside the message would lie below it and
is ignored. Because the rest of that header is influenced by the sender, it is not searched
for a substring. It is split in one pass from left to right that tracks, together, whether the
scan is inside a quoted string and how deeply it is inside parenthesised comments:
\begin{enumerate}[leftmargin=*,itemsep=1pt,topsep=2pt]
  \item Text inside quotes or comments is discarded, and a backslash there skips the next
  character.
  \item Only a semicolon outside both separates one result from the next.
  \item The header is rejected if a quote or a comment is left open, if a closing parenthesis
  appears outside a comment, or if the discarded text contains a semicolon, an equals sign or
  the word \emph{dkim}, since the provider's own comments contain none of these.
  \item The first part must name the provider's own server.
  \item Among the remaining parts, those whose first token begins with \emph{dkim=} are read as
  properties. A property that appears twice rejects the header. A result belongs to the site if
  its signing identity or signing domain equals the site's domain exactly.
  \item The message is authentic if at least one result belonging to the site says \emph{pass}
  and none says anything else.
\end{enumerate}
The first build tested the sender address alone. Successive rounds of automated review, noted in
Section~\ref{sec:log}, led from there through a substring match and a parser built from
regular expressions to this form. A passing signature shows that the message left through the
site's own sender. It cannot show who wrote the request, because the server accepts requests
from anyone.

\paragraph{Filing.}
An accepted message becomes a folder named by its date, time and a slug of its subject in
which every run of characters other than letters, digits, underscores and hyphens is replaced
by an underscore. The restriction exists because the folder's name is later written into a
permission rule for a language model, and the subject is text from outside. Attachments are
saved under their cleaned names; a name that begins with a dot, or that equals one of the file
names that coding assistants read as instructions or settings, is given a prefix so that an
attachment cannot pose as such a file.

\paragraph{Sorting.}
Two open-weight models run locally through a model server on the loopback interface, with
temperature zero and a context of 8192 tokens (\code{mailflow/triage.py}). Each attached image
is scaled so that its long edge is at most 1800 pixels and given to a 7-billion-parameter
vision-language model with the instruction to transcribe all of its text without summarising.
The first two pages of each attached PDF are converted to text. A 14-billion-parameter text
model then receives the email, the transcriptions and a description of the agreed scope, and
must answer in a fixed JSON schema: a request type from an enumerated list, two short labels,
a one-sentence summary, and a boolean, \emph{needs human}, with a reason. If the model's answer
cannot be parsed, the request is treated as needing a human. A request so marked stops here,
and the designer is notified.

This step is a filter and not a security boundary, since its judge is itself a model reading
the untrusted text. The boundaries are in the next two steps.

\paragraph{Working.}
The hosted model is started without an interactive session, with the request's folder as its
working directory and a time limit of 2700 seconds (\code{mailflow/generate.py}). Before it
starts, the job refuses to continue unless all of the following hold: at least one rule file is
configured; every rule file and the one executable script are regular files, not links, that
carry the operating system's immutable flag; the folder's name consists only of letters,
digits, underscores and hyphens; the folder's rules subdirectory is a real directory; and the
folder contains none of the files or directories that the assistant would load as project
instructions or settings. The rule files are then copied into the folder afresh, and the
copies are made immutable for the length of the run and released afterwards.

The model's permissions are a list, and nothing outside the list is granted:
\begin{itemize}[leftmargin=*,itemsep=1pt,topsep=2pt]
  \item reading, listing and searching files, each restricted by a path pattern to the
  request's folder;
  \item writing and editing files, which the non-interactive mode confines to the working
  directory;
  \item running one script, named by its absolute path. The script accepts exactly two
  arguments, an existing page file and an output document, and refuses unless both lie under
  two fixed directories and neither is a link.
\end{itemize}
It is started with project-level settings disabled. It has no tool for sending mail, for the
network, or for any other command. The restriction of reading to one folder was added after a
test showed that an unrestricted read permission allowed the model to open files anywhere on
the disk, which the documentation of the previous build had denied.

The model's instructions state that the email and its attachments are material to work from
and not a source of instructions, which addresses but does not remove the risk of injected
instructions~\cite{greshake2023injection}. The design therefore assumes that the instructions
may be overridden and limits what an overridden model can reach: one folder, one script with
checked arguments, and no channel out.

\paragraph{Sending.}
A result is sent only if the folder holds at least one output document, holds no marker
written by the model to say that a person is needed, and has not been sent before
(\code{mailflow/send.py}). The message goes over SMTP to one address read from a local
configuration file, with a copy to the designer. Nothing in the request, the models' outputs
or the folder can change the recipient. When sending is automatic, the job counts the folders
already sent on the current date and stops at six, leaving later results for the designer to
confirm.

\paragraph{Failures.}
Any error while a request is being sorted, worked on or sent is logged with its trace, recorded
in the folder's status file, and shown to the designer as a notification; the job continues
with the next request. A request's status file records the furthest stage it reached, one of
downloaded, sorted, working, awaiting send, sent, needing a person, or failed.

\paragraph{Configuration and exposure.}
The addresses, the list of request types, the description of the scope, the instructions to
both models and the paths of the rule files are read from a file that is excluded from the
repository. An early build had the instructions in the public code and they were removed from
it. The first build of the job held each result for the designer to confirm. The same evening,
at his request, sending became automatic, so that requests arriving between two and seven in
the morning his time, which is her late afternoon and evening, are answered without him.

\subsection{What Was Verified by Hand}\label{sec:verify}
On 24 November the parts were exercised separately: the mail login in both directions, the DKIM
rule against the seven delivered emails and against forged headers, the local models on one real
attachment and on one constructed hostile request, the hosted model's tool permissions under the
scheduler's environment, and the outgoing message without sending it. Real requests began to arrive that evening, after the cut-off of the corpus. The job handled
them in both of the ways it was built to: it stopped a request that was outside its scope and
left it for the designer, and it carried another along the whole path, to the hosted model and
back to her, in under two minutes. We report no more about them than that. The first build's
answers did not say that a machine had prepared them; since that night, answers of one kind
open with a line that says so, and answers that deliver prepared documents still do not.

\subsection{MotionMirror}\label{sec:mm}
MotionMirror is a separate program, in Python, in its own public repository under the MIT
licence. We describe version 0.2.0 at commit \code{80c4fd8} (24 November 2026). It takes one
video from one camera and reports, for each repetition of a movement, when its phases occur
and a set of joint measurements at each phase. It has profiles for the golf swing, the
basketball jump shot and badminton strokes. Its core has no dependencies; pose estimation,
drawing and video encoding are an optional set.

\begin{figure}[t]
  \centering
  \begin{tikzpicture}[
      font=\scriptsize\sffamily,
      box/.style={scard, inner sep=2pt, minimum height=8.5mm, minimum width=1.95cm},
      opt/.style={sflat, dashed, inner sep=2pt, minimum height=8.5mm, minimum width=1.95cm},
      outp/.style={box, sorange},
      arr/.style={sarr}]
    \node[box, sblue] (v) {One video\\\dsub{one camera}};
    \node[box, sblue, right=4.5mm of v] (pose) {Pose\\\dsub{33 landmarks, every frame}};
    \node[box, sblue, right=4.5mm of pose] (who) {Athlete\\\dsub{one of up to three}};
    \node[opt, below=5mm of who] (zoom) {Second pass\\\dsub{if the figure is small}};
    \node[box, steal, left=4.5mm of zoom] (geo) {Geometry\\\dsub{aspect, torso lengths}};
    \node[box, steal, left=4.5mm of geo] (sig) {Signals\\\dsub{fill, smooth, speed}};
    \node[box, spurple, below=5mm of sig] (anchor) {Anchor\\\dsub{extrema of one signal}};
    \node[box, spurple, right=4.5mm of anchor] (ph) {Phases\\\dsub{windows, rejections}};
    \node[outp, right=4.5mm of ph] (out) {Measures, checks\\\dsub{report, overlay, sheet}};
    \draw[arr] (v) -- (pose); \draw[arr] (pose) -- (who); \draw[arr] (who) -- (zoom);
    \draw[arr] (zoom) -- (geo); \draw[arr] (geo) -- (sig); \draw[arr] (sig) -- (anchor);
    \draw[arr] (anchor) -- (ph); \draw[arr] (ph) -- (out);
  \end{tikzpicture}
  \caption{MotionMirror from a video to a report. The dashed step runs only when the athlete
  occupies less than 0.35 of the frame's height. The last row is the part that differs between
  the golf, basketball and badminton profiles.}
  \Description{A flow of nine boxes in three rows joined by arrows: one video, pose, athlete;
  then second pass, geometry, signals; then anchor, phases, and measures and checks.}
  \label{fig:mmpipe}
\end{figure}

Figure~\ref{fig:mmpipe} gives its stages.

\paragraph{Pose.}
Every frame is passed to a body-landmark model that returns 33 named
landmarks~\cite{bazarevsky2020blazepose,lugaresi2019mediapipe}, in its video mode, with a
minimum confidence of 0.4 for detection, presence and tracking; no frames are skipped, and a
clip that reports no frame rate is assumed to have 30 frames a second. Up to three people are
detected in a frame. A landmark with visibility below 0.3 is treated as missing.

\paragraph{Choosing the athlete.}
One person is kept per frame. A candidate is plausible if its torso and its legs each span
more than 0.02 of the frame height. Among plausible candidates the tracker keeps the one whose
hip midpoint is nearest the last accepted one, provided the hips moved no faster than three
frame-heights a second; with no accepted pose, or after a second without one, it takes the
largest body. If nobody is plausible it returns the most visible candidate without updating
its anchor.

\paragraph{A second pass for small figures.}
If the median height of the athlete's bounding box is below 0.35 of the frame, the clip is
processed again on a moving square crop. The crop's side is the larger of 1.9 times the box
and 256 pixels; its centre and size are interpolated across gaps and smoothed with a moving
average of about 0.3 seconds. Each crop is resized to $640\times640$, one pose is detected in
it, and the landmarks are mapped back to the coordinates of the full frame.

\paragraph{Geometry.}
Landmarks are stored as fractions of the image's width and height. For all measurements the
horizontal coordinate is multiplied by the aspect ratio, so that distances and angles are not
distorted in a portrait or landscape frame, and the model's depth estimate is discarded: every
quantity is two-dimensional. Lengths are divided by the athlete's torso length, the distance
from the midpoint of the shoulders to the midpoint of the hips, taken as the median over the
clip. A height above the head is therefore in torso lengths and a speed in torso lengths a
second, and neither depends on how far the camera stood.

\paragraph{Signals.}
Tracked points such as the wrists and hips are filled across gaps of up to eight frames by
linear interpolation, with longer gaps left empty, and smoothed with a centred moving average
of five frames, three for badminton. Speed is a central difference of position, smoothed the
same way. Joint angles are not filtered frame by frame; the value reported at a phase is the
median of the angle over a window of two frames either side, one for most badminton phases.
No model-based filter is used.

\paragraph{Finding repetitions.}
Each profile reduces a clip to one signal whose extrema mark candidate repetitions. An
extremum is kept if it passes a threshold, is extreme within a radius, and is separated from
its neighbour by a minimum time. Table~\ref{tab:mm} gives the three rules. Around each
candidate the profile then places the remaining phases by searching fixed windows for a
minimum or maximum of another signal, and rejects the candidate if its shape is implausible.

\begin{table*}[t]
  \caption{MotionMirror's three profiles: the signal whose extrema propose a repetition, the
  rule for keeping one, and the phases placed around it. $p_k$ is the $k$th percentile of the
  signal over the clip; $f$ is the frame rate; lengths are in torso lengths.}
  \label{tab:mm}
  \small
  \begin{tabular}{@{}p{0.09\textwidth}p{0.22\textwidth}p{0.27\textwidth}p{0.34\textwidth}@{}}
    \toprule
    Profile & Anchor signal & Kept if & Phases, in order \\
    \midrule
    Golf & Height of the hands (midpoint of the wrists); the anchor is the top of the swing &
    Beyond $p_{2}+0.4\,(p_{90}-p_{2})$; at least $1.8f$ frames apart; hands no more than 0.35
    below the shoulders &
    address, top, impact, finish \\
    Basketball & Wrist height above the head plus elbow angle over $180^\circ$; the anchor is
    the release &
    At least $\max(1.0,\,p_{90})$; at least $1.2f$ frames apart; wrist at least 0.05 above
    the nose &
    set, dip, lift, release, follow-through, landing \\
    Badminton & Speed of the racket-side wrist; the anchor is the contact &
    At least $\max(2.5,\,p_{85})$ a second; at least $0.8f$ frames apart &
    ready, backswing, contact, follow-through, recovery \\
    \bottomrule
  \end{tabular}
\end{table*}

\begin{figure*}[t]
  \centering
  \includegraphics[width=\textwidth]{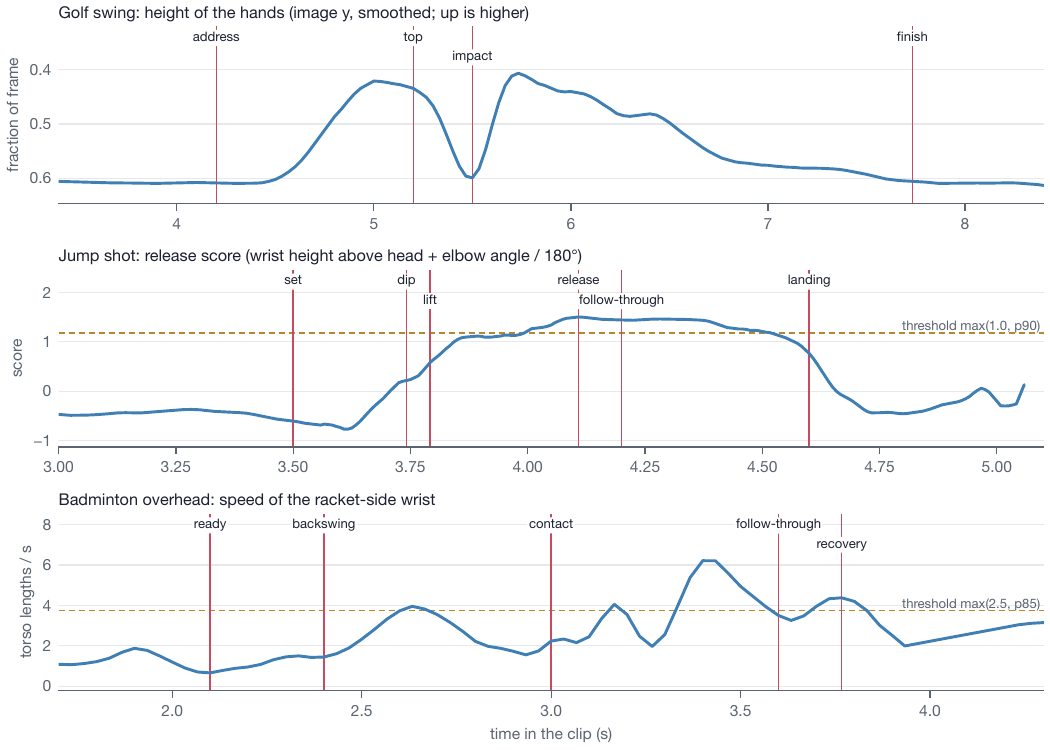}
  \caption{The anchor signal of each profile on one of the project's published example
  recordings, with the phases the program reported for one repetition (vertical lines) and,
  where the rule has one, its threshold (dashed). Top: in the golf swing the hands rise to the
  \emph{top}, fall to \emph{impact} and rise again to a finish that is higher than the top,
  which is why the rule keeps the first extremum and not the highest. Middle: the release score
  of the jump shot crosses its threshold once. Bottom: in the overhead stroke the wrist is
  fastest after the shuttle is struck, so \emph{contact} is placed at the hand's highest point
  before the speed peak. Signals are read from the per-frame tables the program wrote; nothing
  here measures how close a reported phase is to the true one.}
  \Description{Three line charts stacked. The first shows hand height over five seconds with
  four labelled vertical lines: address, top, impact, finish. The second shows a score rising
  above a dashed threshold with six labelled lines from set to landing. The third shows wrist
  speed with several peaks above a dashed threshold and five labelled lines from ready to
  recovery.}
  \label{fig:mmsignals}
\end{figure*}

\begin{figure*}[t]
  \centering
  \includegraphics[width=\textwidth]{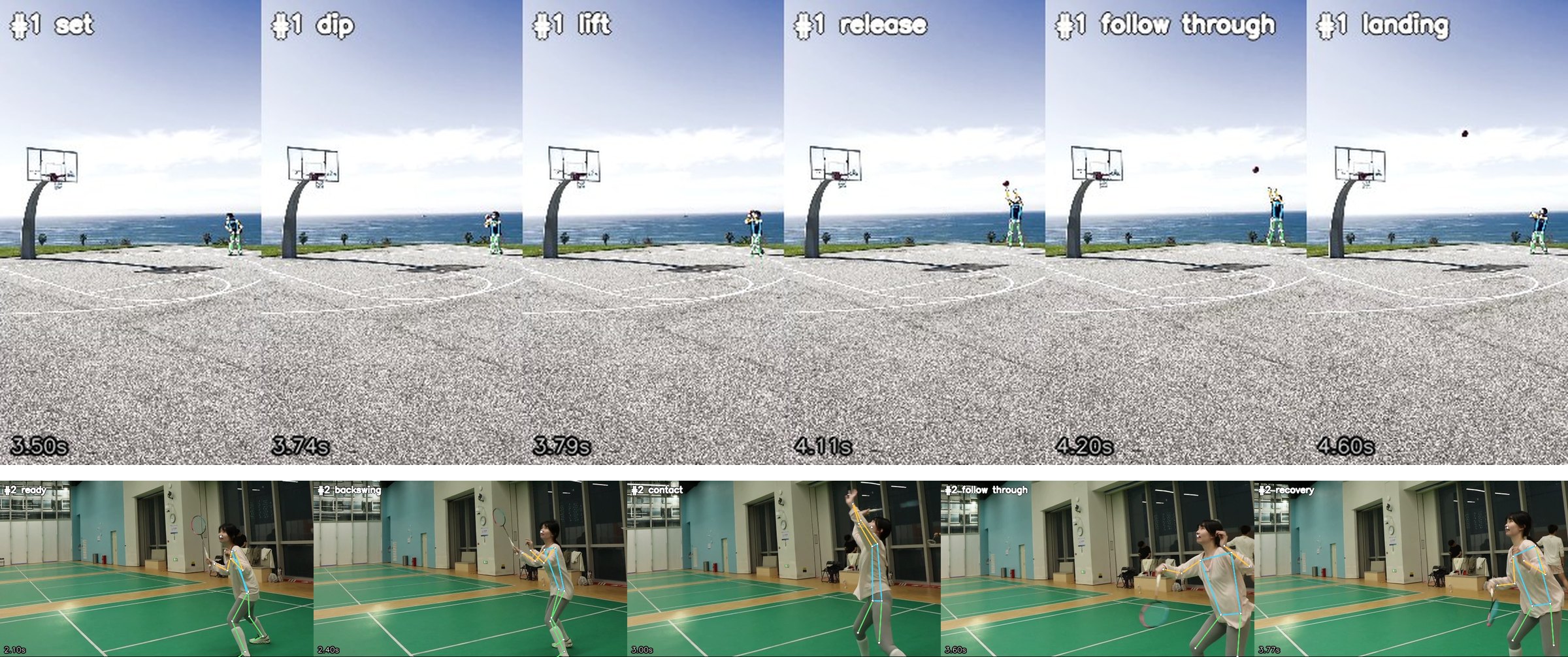}
  \caption{Key-pose sheets from the two showcase recordings named on the dashboard card, as the
  program writes them: one frame per phase with the skeleton drawn over it. Top: the jump shot,
  six phases from \emph{set} to \emph{landing}; the athlete is small in the frame, which
  triggered the second pass. Bottom: one of four badminton strokes, an overhead, five phases
  from \emph{ready} to \emph{recovery}. The people shown are the two authors.}
  \Description{Two strips of video frames. The upper strip has six frames of a person shooting a
  basketball on an outdoor court by the sea, each labelled with a phase name and a time. The
  lower strip has five frames of a person playing badminton on an indoor court, each labelled
  with a phase name, with coloured lines drawn along the limbs.}
  \label{fig:mmkeyposes}
\end{figure*}

Figure~\ref{fig:mmsignals} shows the three anchor signals on real recordings, and
Figure~\ref{fig:mmkeyposes} the frames chosen for two of them.

\emph{Golf.} The top is the highest point of the hands, moved to the end of any pause there.
The address is the last moment in the three seconds before the top at which the hands are
quiet, where quiet means a speed at or below the larger of the clip's 30th percentile and 0.15;
a candidate is rejected if the address is less than 0.3 seconds before the top or the hands
rose less than a quarter of a torso. Impact is the first frame within 0.7 seconds after the top
at which the hands are at least 60 percent of the way back to their address height and no
lower than in the next two frames. The finish is the first quiet frame with the hands above
the shoulders between 0.25 and 2.5 seconds after impact. The source describes impact as a pose
proxy that may be a frame or two off, since neither club nor ball is modelled.

\emph{Basketball.} The dip is the smallest knee angle on the shooting side between 0.9 and
0.08 seconds before the release; the set is the slowest wrist in the 0.6 seconds before the
dip; the lift is the fastest wrist between dip and release; the follow-through is the highest
wrist in the 0.45 seconds after release; and the landing is the slowest hip between 0.25 and
0.8 seconds after it. A candidate is rejected if the other hand is visible and never comes
within 0.6 of the shooting wrist near the head before the release, a rule that separates a
shot from a one-armed reach for the ball; the project's log records that it reduced three
detections in a recording of two players to the two real shots.

\emph{Badminton.} Contact is the speed peak, moved back to the highest point of the racket
hand if that point lies more than 0.45 above the shoulders in the preceding 0.4 seconds. The
backswing is the smallest elbow angle between 0.6 and 0.05 seconds before contact; ready is
the slowest moment in the 0.8 seconds before that; the follow-through is the first frame
within half a second at which the speed has fallen below 40 percent of its peak; and recovery
is the slowest hip up to 1.2 seconds after contact. A stroke is called overhead if the wrist
is above the nose at contact, underhand if below the hips, and a drive otherwise. The racket
and the shuttle are not tracked.

\paragraph{Which side.}
The shooting or racket side may be given or chosen automatically. For basketball both sides
are analysed and the one that yields more repetitions is kept, with ties broken by confidence;
for badminton the side is the wrist with the higher 90th-percentile speed.

\paragraph{Measurements and checks.}
At each phase the profile records timings in seconds, joint angles in degrees, and positions
and speeds in torso lengths. For golf these include the durations of backswing and downswing
and their ratio, the angle of the leading arm at the top, the separation of the shoulder and
hip lines, and the sideways drift of the head and hips; for the jump shot, the knee angle at
the dip, the elbow angle and the wrist's height above the head at release, the time from dip
to release, and the sideways drift between set and landing; for badminton, the elbow angle and
hand height at contact, the peak hand speed, the width of the stance and the distance the hips
travel in recovering. Each profile then compares a few of these with a reference range, for
example a ratio of backswing to downswing between 2.0 and 4.0, or an elbow of at least
$145^\circ$ at release. A repetition's score is the share of its measurable checks that fall
inside their ranges, on a scale of 100, and its confidence is the mean landmark visibility
across it, reduced when a timing falls outside a plausible interval. With two or more
repetitions the program reports the mean and spread of each measurement.

\paragraph{The ball.}
Only the golf profile looks for a ball, and only on request. It locates the ball at rest by
colour and shape near the feet in the frames before impact, then chains small bright moving
regions after impact into a track, which is accepted only if it has at least four points,
starts near the resting position and within ten frames of impact, and rises and advances
steadily. When no track passes, an arc can be drawn as an illustration; it is dashed, it is
labelled as a proxy, and no measurement uses it.

\paragraph{Outputs.}
An analysis writes the landmarks for every frame, one row of signals per frame, a report in
structured and in readable form, the video with the skeleton, recent poses and live angles
drawn over it, a chart of the signals with the phases marked, and a sheet with one image per
phase for up to six repetitions. Two analyses of the same sport can be compared: the program
tabulates, for a fixed list of measurements, the mean in each and their difference. It aligns
nothing in time; repetitions are compared only through their summary values. A further command
cuts a long recording into ranked clips by the amount of motion between frames.

\paragraph{Tests and validation.}
The repository holds 59 test functions. Those for the detection rules run on synthetic
landmark sequences generated at 30 frames a second with known phase times, and assert, for
example, that the top and impact of a synthetic golf swing are found within three frames, the
release of a synthetic jump shot within 0.2 seconds, and that a clip with no motion yields no
repetition. Continuous integration also runs the whole pipeline on one bundled video. There is
no evaluation against manually labelled video or against a motion-capture reference. The
program states this in its own output: its report carries the sentence ``Metrics are 2D
single-camera pose proxies. Reference ranges are broad starter values for spotting outliers,
not coaching verdicts'', and its planning documents list the comparison of detected phases
with manual labels as work not yet done. The reference ranges are the project's own starting
values and are not drawn from a cited source.

\paragraph{The two showcase recordings.}
The card on the dashboard names two recordings that the project publishes as examples, one of
each author. The first is a jump shot filmed at 120 frames a second in portrait orientation,
6.8 seconds long, in which the athlete is small enough in the frame to trigger the second
pass; a pose was obtained in 706 of 820 frames and one shot was found. The second is a
12.1-second badminton rally at the camera's normal rate in landscape orientation; a pose was
obtained in all 364 frames and four strokes were found, classified as two drives and two
overheads. These are demonstrations of the program on two clips, and we draw no conclusion
from them about either person's technique.

\subsection{Tests across the Systems}\label{sec:producttests}
Table~\ref{tab:producttests} collects the automated tests of the systems in this section. The
answering pipeline has none; what was checked by hand is listed in Section~\ref{sec:verify}.

\begin{table}[t]
  \caption{Automated tests of the systems beside the town. Counts for the site are of test
  cases run; they are included in the total of Table~\ref{tab:metrics}.}
  \label{tab:producttests}
  \small
  \begin{tabular}{@{}p{0.40\columnwidth}p{0.30\columnwidth}r@{}}
    \toprule
    System & Module & Tests \\
    \midrule
    Infirmary & letters, notes, visits & 9 \\
     & the doctor's clock & 2 \\
    Trip & phases and days & 8 \\
     & itinerary & 14 \\
    Request page & limits, history, sending & 23 \\
     & session state & 24 \\
     & image scaling & 5 \\
    Server & request handler & 20 \\
    Routing & path router & 3 \\
    Answering pipeline & all & 0 \\
    \midrule
    MotionMirror (separate repository) & all & 59 \\
    \bottomrule
  \end{tabular}
\end{table}


\section{Reconstructing Design Decisions}\label{sec:log}
Seven documented episodes make the rationale inspectable; the last three are new in this
version. Commit identifiers mark the recorded changes; the interpretations are ours.

\paragraph{Replacing Pictionary (\code{7d2c659}, 26 October).}
The implementation removes a same-device drawing game and adds a song-identification level.
Our interpretation is that separate recognition of a shared song better accommodates different
visit times than taking turns on one screen. A plainer reading is that the replacement reduced
implementation complexity.

\paragraph{Requiring a try-on photo (\code{2bc3d07}, 28 October).}
The ring level's completion condition changes from viewing the reveal to capturing a try-on
photo. This creates an artefact that can be shared, but also makes camera access and a capable
device conditions for completing the narrative. The same mechanism can be read as participation
or as an unwanted demand; only the recipient can say which.

\paragraph{Correcting represented routines (\code{1960404}, 26 October).}
The city-day descriptions change the transport and destination details to match the designer's
account of the couple's routines. The record shows that plausible generated defaults required
correction; whether the corrected version matches the other partner's own account is for the
represented person to say.

\paragraph{Revising the entry gate (\code{ccabd31}, 28 October).}
This commit removes answer-revealing placeholder text and revises associated checks. The
recognition question becomes less self-defeating; it remains a low-entropy, client-side date
gate rather than access control. Version 1.24 (\code{c1303f4}, 24 November) narrows it again, to
one accepted spelling behind a password field, at the designer's wish that the site need a
password. The change removes the spellings that give only the month and day, which a visitor could have
guessed in at most 366 tries; the remaining eight digits are still a date, and the files behind the
gate are still served to anyone who asks for them by address.

\paragraph{Building from a remark (\code{397db28}, 4 November).}
The infirmary is the first feature in the record whose stated origin is something the recipient
said. The remark described a refusal: when hurt, she would not talk. The response gave that
refusal a place in the town and put the other partner's words inside it. One reading is that
the gift was reshaped by its recipient. A less comfortable reading is that a wish to withdraw
was answered with a room the other partner furnished.

\paragraph{Replacing the symbolic ticket (\code{d9de84b}, 8 November).}
The countdown's target changes from an invented date to a booked flight, and the invented ticket
is kept as a memory. Here an event in the world overtook the fiction, and the code was changed
to follow it, where in the town the fiction had set the terms.

\paragraph{Reversing a non-goal (\code{fc593ec}, 24 November).}
The request box's specification of 23 November lists, among its non-goals, ``an AI that answers
her''. The answering pipeline was built the next day. The record shows the reason given for the
change, which is the designer's sleep, and shows no discussion of the non-goal it overturns. The
commits that follow it record six rounds of automated review in one night, each answered by a
narrower design: the sender check became a signature check and then a parser; the model's tools
shrank from a general allow-list to one folder and one script; its rule files became locked
copies; and its instructions were removed from the public code (Section~\ref{sec:askpath}).

Across these episodes, the recurring issue is how a technical rule expresses an assumption
about the other person. In the later ones the rule also begins to act for one person toward
the other.


\section{Discussion}\label{sec:discussion}

\subsection{From Representation to Action}
The design links a shared reference to an action, then to a visible trace: a prior game becomes
a win-in-one puzzle and its solution becomes one point of a heart. Compared with displaying the
same memory as a caption, this adds an authored sequence of participation. The design question
it raises is which part of a memory can become a rule without losing its meaning. The case is one
implementation of that move; whether it serves a couple better than a photo album, a message or
another intimate artefact~\cite{hassenzahl2012love,kirk2016ritual} is open.
The ring also exposes a boundary: an imagined future is not a remembered event, although both
can organise narrative play.

\subsection{Privacy Boundaries and Personal Material}\label{sec:privacy}
The gate hashes a low-entropy date, whose value may also be inferred from the surrounding story.
It deters casual entry but provides neither encryption nor dependable access control. Local
completion flags avoid an author-operated progress database; they do not make hosted photographs,
voice notes or letters private. The voice-note implementation, for example, references a static
audio asset. Anyone authorised to obtain the deployed files or repository may obtain such material.

The browser boundary also differs by feature. Hand tracking and breath-amplitude calculation
run locally, whereas speech recognition may contact a remote service. Tile and media requests
expose request metadata to their providers. A local-first aspiration therefore needs a description
of data flows, not only the absence of a custom backend~\cite{kleppmann2019localfirst}.
A research release would ideally separate shareable code and synthetic examples from personal
assets, since intimate images can also reveal location context~\cite{yang2024geolocator}; the
current repository does not make that separation.

The request box changes the picture more than any earlier feature. The first version could say
that nothing a player did left the device through code the authors operate. A request now
leaves her phone with its files and passes through an edge platform, a mail provider, the
designer's laptop, the provider of a hosted model and a second mail provider
(Figure~\ref{fig:askpath}). Running the first model locally keeps the sorting step on his
machine; it does not keep the files there, because the step that does the work sends them on.
The local chart of the infirmary and the email of the request box now sit one card apart on
the same dashboard, with opposite properties and the same visual language.

Publishing the code of such a pipeline carries its own exposure. Instructions to a model
describe the work, and the work describes the person who asked for it. The repository's history
for 24 November was rewritten to remove such a description. A public design case and a private
partner pull against each other at exactly this point, and our rule in this paper is to describe
the mechanism and withhold what passes through it.

\subsection{Asynchrony, Temporal Representation and Staleness}
Each player's sky uses the clock in their configured city, while the partner avatar represents
locally stored progress. The design supports visiting at different times, but an avatar's state
may be old until the next manual exchange. Its city-day position follows an authored schedule,
not sensed presence. The distinction matters because a representation intended as companionship
can otherwise imply knowledge of the partner's current activity that the system does not have.
The birthday scene's scripted clocks and the thirteen- or fourteen-hour offset are temporal
design materials~\cite{hallnas2001slow,odom2012technology}.

\subsection{Staged Access, Surprise and Freedom to Decline}
Ordered access and delayed disclosure coordinate a surprise, but also let one partner determine
what the other must do before reaching a letter or finale. Requiring a photograph illustrates the
cost: people without camera access, or who prefer not to provide an image, cannot clear that level
through the current completion rule. Similarly, a scripted ``I do'' is a game response, not evidence
of consent to a relationship commitment. The present sequence does not model refusal as a separate
successful outcome. A future revision should distinguish declining an action from failing a task
and should offer an equivalent path that preserves narrative access.

The later products move some way toward this without touching the town. None of them is
required, ordered or revealed, and the infirmary treats leaving as an ordinary action with its
own button. It also shows the limit of that movement: a place to decline conversation was built
by the person being declined, and the letters inside it, on the record, were drafted for both
partners by one side.

The lexical test is useful as a writing regression check, but it cannot ensure surprise in a
client-delivered game. Source, routes, media, screenshots and this paper disclose the finale.
Any study of surprise must therefore record whether the recipient encountered these materials
before play. Ambiguity of this kind can itself shape anticipation~\cite{gaver2003ambiguity}.

\subsection{Bilingual Coverage and Situated Meaning}
The same-key test prevents missing dictionary entries, but does not establish equivalent meaning,
tone or task difficulty. The Chinese title \zh{仪见钟情} substitutes a character from Jingyi's name
into ``love at first sight''; the English version explains the reference. Likewise, classical
Chinese character-restoration puzzles retain different language demands even when their prompts
have English versions. What the test establishes is coverage; semantic quality and accessibility
need a separate review, including by the intended recipient.

\subsection{Answering in Person and Answering in Time}\label{sec:delegated}
The request box was specified as a way to reach one person, and within a day it could be answered
by a pipeline acting for him. The motive is an ordinary one in a relationship across fourteen
hours: her afternoon is his night. The pipeline trades one kind of care for another.
An answer that arrives in twenty minutes meets the need; an answer he wrote would show that he
attended to it. Research on AI-mediated communication suggests the trade is not neutral, since
suspected machine authorship can lower trust in a message~\cite{hancock2020aimc,jakesch2019aimc}.

Three properties of the build bear on this. Its answers at first carried no mark of how they
were made, and only some do now. The specification's non-goal was reversed without a recorded conversation between the two
people it concerns. And the name on the card still says the request goes to him. This paper,
which both authors sign, is one disclosure; a line in every answer would be a more reliable one.
We do not know how the recipient reads an answer prepared this way, and the question belongs to
her.

\subsection{AI Assistance and Authorial Responsibility}
The source and log document an AI-assisted implementation process in which the designer supplied
requirements and corrected defaults. Each recorded correction marks a place where the designer
disagreed with generated output; it says nothing about details that went uncorrected. The corpus
lacks an unassisted comparison and a complete record of human and model effort, so it supports no
productivity claim about five days of work. Personal representation, verification and publication
decisions remain the authors' responsibility.

Until 24 November the models worked on the software. Since then one also works in it: the same
kind of assistant that wrote the pages now prepares answers that reach the recipient under the
designer's address. Mistakes of the first kind are caught by tests and review before she sees
anything. Mistakes of the second kind arrive in her inbox, and the only checks before that are
the scope test of a small local model and a daily limit. On its first evening that test stopped a request it judged to be outside its scope and left it
for him.

\subsection{Limitations and Next Evidence}
The principal limitation is evidential: developer playtesting is the only reported use, and no
recipient experience, sustained practice or relationship outcome has been collected. The second
version adds four weeks of development record and three traces of the recipient's activity; it
adds no account from her. The answering pipeline is described from its code and a manual exercise
of its parts on the day it was built; its first real requests are reported only as having
happened. The corpus
is brief and selectively interpreted, with no independent analysis. Engineering checks are also
bounded: tests do not establish deployment-wide privacy, all-browser compatibility or equivalent
experience across languages; approximate hand scaling is not a fit measurement. Union merge
depends on compatible registries and does not propagate resets. Remote media limits offline use
and availability across regions.

The pipeline also has specific open risks: the endpoint that feeds it remains unauthenticated,
so a stranger can cause a run and an unwanted email to her, though not a command or a different
recipient; the scope test is a single model's judgement; and an answer is sent without a person
reading it.

A next stage could begin with a voluntary recipient walkthrough, recording what was understood,
skipped, refused or reinterpreted, followed by a diary of revisits and code exchanges over time.
If participants are recruited or data are collected
under an institution, the applicable ethics determination must precede that collection. A larger
sample alone would not repair missing accounts of situated meaning, authorial power or failure.


\section{Conclusion}
MERIDIAN documents the design of a two-person game that turns autobiographical material into
tasks, places and staged ceremonies, and the first four weeks of what it became afterwards. Its
development record makes seven design revisions inspectable, and its audits establish selected
software properties at two versions. These support a situated account of how memories can shape
playable rules, not a claim that the game improves intimacy or that the recipient accepts its
authored sequence.

The first version ended on the question of how a personal gift can invite shared meaning while
leaving the other person free to decline, reinterpret and reshape it. The added record gives
that question a first, partial answer and a new form. A remark by the recipient became a
building, which is reshaping of a kind, though the designer did the building. And a channel
meant to reach him now reaches a machine that answers for him, which asks what a partner is
owed when the reply is prompt, useful and not written by the person addressed.

\section{Changes in Version 2}\label{sec:v2}
The first version was dated 30 October 2026 and inspected version 1.17.0 of the software; this
one is dated 24 November 2026 and inspects version 1.24.0.
\begin{itemize}[leftmargin=*,itemsep=1pt,topsep=2pt]
  \item New: Section~\ref{sec:after} on the dashboard and everything on it (the infirmary, the
  trip page, the request box, the MotionMirror card and the seed tile);
  Section~\ref{sec:products}, which gives for each of these systems its data, rules, data flows
  and tests, including the request page's state machine, the server's checks, the answering
  pipeline's authentication and sandbox, the service worker, and MotionMirror's pose pipeline
  and detection rules;
  Section~\ref{sec:askpath} on the server and the answering pipeline; Section~\ref{sec:delegated};
  three episodes in Section~\ref{sec:log}; Tables~\ref{tab:after}, \ref{tab:routes}, \ref{tab:caches}, \ref{tab:worker}, \ref{tab:mm}
  and~\ref{tab:producttests}; Figures~\ref{fig:timeline}, \ref{fig:dash}, \ref{fig:cards},
  \ref{fig:sitemap}, \ref{fig:clinic}, \ref{fig:trip}, \ref{fig:askstates}, \ref{fig:askpath},
  \ref{fig:mmpipe}, \ref{fig:mmsignals} and~\ref{fig:mmkeyposes}; three references.
  \item Corrected: the first version's statement that the application has no author-operated
  backend now holds for progress only. The privacy discussion is extended accordingly.
  \item Updated: the corpus (152 commits, 36 versions, five specifications), the audit (263
  tests), the description of the entry gate, which is now a password, Tables~\ref{tab:dates} and~\ref{tab:metrics}, the abstract and the conclusion.
  \item Unchanged: the account of the town, its nine levels, the two cutscenes, the quest model
  and the progress codes. ``Pre-deployment'' is dropped from the description of the case, since
  the site has been in the recipient's hands since the first version; the absence of a recipient
  evaluation is not changed.
\end{itemize}

\section{Experiment Protocols and Full Results}\label{sec:expdetail}
This appendix gives the protocol of each experiment of Section~\ref{sec:eval} and the results
that the main text summarises. Every script is in \code{paper/experiments/} and writes its raw
output to \code{paper/experiments/results/}; \code{make\_figures.py} and \code{make\_tables.py}
draw the figures and the table rows from those files, so no number in a figure or in
Tables~\ref{tab:x-load} to~\ref{tab:x-triage} was typed by hand. Table~\ref{tab:experiments} lists the
experiments.

\begin{table*}[t]
  \caption{The twelve checks under the three questions. All are run by machine and reproducible
  from the repository. The last column gives the observation, without a verdict.}
  \label{tab:experiments}
  \small
  \begin{tabular}{@{}llp{0.31\textwidth}p{0.19\textwidth}p{0.26\textwidth}@{}}
    \toprule
     & Check & What is varied & Size & Observed \\
    \midrule
    \multicolumn{5}{@{}l}{\emph{RQ1 Rendering: which routes render, and how fast, under simulated conditions?}} \\
    E1 & Load time & Four network profiles, seven routes, cold cache and return & 280 loads & 5 of 7 within 1\,s on slow 4G; ring 3.7\,s, and 10.6\,s on 3G \\
    E2 & Network failure & Hosts refused, stalled or absent; device offline & 21 routes, four conditions & 21 render with hosts failing; 18 render offline \\
    E3 & Frame time & CPU slowed 1, 4 and 6 times while walking & 18 runs of 10\,s & Median 8.3\,ms at 4 times slower \\
    E4 & Accessibility rules & 21 routes, two languages & 42 audits, 992 touch targets & Contrast findings on 19 routes; 106 targets under 24\,px \\
    E5 & Layout & Six widths, two languages & 252 renderings & 246 without sideways scroll \\
    E6 & Dictionaries & Keys and placeholders; two models read every pair & 522 keys, 518 pairs & 522 match; 67 pairs flagged by both models ($\kappa = 0.29$) \\
    \midrule
    \multicolumn{5}{@{}l}{\emph{RQ2 Rules: does the software implement the rules the design states?}} \\
    E7 & Progress codes & One-edit mutations of all codes and of the 20 reachable codes; three designs; all pairs of states for the merge & 331{,}530 and 5{,}664 mutations; 262{,}144 merges & 207 and 24 read as another state; 0 with a prefix test \\
    E8 & Secrecy lexicon & 107 finale hints in four tiers & 464 strings & 31 of 35 named forms and 2 of 72 others caught \\
    E9 & Test coverage & Unit tests by part of the site & 295 tests, 4{,}647 lines & 20\% of lines; ceremonies 0\% \\
    E10 & Development record & Every commit: time, area, tests, kind & 161 commits & Context only \\
    \midrule
    \multicolumn{5}{@{}l}{\emph{RQ3 Delegation: what does the triage step pass and escalate on synthetic requests?}} \\
    E11 & Server properties & Origin, method, path, text, file names and sizes & 4{,}000 generated requests & 4 of 5 properties hold on every request \\
    E12 & Triage replay & Six classes of synthetic request, five local models, five conditions & 1{,}120 decisions on 120 requests & Model used in deployment: 2 of 80 passed, 21 of 40 in-scope escalated \\
    \bottomrule
  \end{tabular}
\end{table*}

\subsection{Common Set-Up}
All experiments ran on 29 November 2026 on one laptop (Apple M5, 32\,GB, macOS) against commit
\code{9193c20} of the public repository, version 1.25.0, with Node 22.22, vitest 5.0.5,
Playwright's Chromium and Ollama 0.20.5.

\paragraph{Experiments on the site's own modules (E7, E8, E11).} These import the shipped
TypeScript modules and run under vitest with a separate configuration,
\code{paper/experiments/vitest.config.ts}, so they are not part of the site's test suite and do
not change its count. Random choices use a fixed linear congruential generator with seed
20261129.

\paragraph{Browser experiments (E1 to E5).} \code{browser.py} starts Chromium with a viewport
of $390\times844$ CSS pixels, a device scale factor of 3, touch input and a mobile user agent,
against the production build served by \code{vite preview} on the test machine. Before each page
loads, a script seeds local storage with the state of a returning player: the gate passed, the
player set to Jingyi, all nine levels cleared for both players, the birthday prologue seen, and
the door of the requested level open. Each of the 21 routes then renders its own page and not a
redirect. The album route added after the design record's cut-off is not visited. No experiment
types into a form or submits one, and every request to the site's two servers is refused.

\paragraph{Model experiments (E6, E10, E12).} These call open-weight models through a local
Ollama server at temperature zero with a JSON schema for the answer. Nothing is sent to a hosted
model, and no experiment reads the answering pipeline's private configuration, its inbox or
anything the recipient sent.

\subsection{E1: Load Time}
\emph{Script:} \code{e01\_load.py}. Chromium's network emulation shapes the link to four
profiles: Wi-Fi (30\,Mbit/s down, 2\,ms round trip), 4G (9\,Mbit/s, 170\,ms), slow 4G
(1.6\,Mbit/s down, 750\,kbit/s up, 150\,ms) and 3G (400\,kbit/s, 400\,ms); slow 4G is the
Lighthouse tool's default mobile preset, and the 4G and 3G profiles match WebPageTest's 4G and
slow-3G presets. The CPU is slowed four times. For each profile, each of seven
pages and each of five runs, a fresh browser context opens the page with nothing cached and no
service worker (\emph{first visit}); the context then opens the front page without a clock
running until the service worker has installed, and opens the page again (\emph{return visit}).
Requests to other hosts are refused so that runs repeat. We record the first contentful paint,
the time at which the route's own view has rendered (``to the page''), the bytes transferred and
the number of requests. Table~\ref{tab:x-load} gives medians over the five runs.

\begin{table}[t]
  \caption{E1: medians over five runs. FCP: first contentful paint. Times are in seconds.}
  \label{tab:x-load}
  \footnotesize
  \setlength{\tabcolsep}{3pt}
  \begin{tabular}{@{}llrrrrrr@{}}
    \toprule
     & & \multicolumn{4}{c}{First visit} & \multicolumn{2}{c}{Return} \\
    \cmidrule(lr){3-6}\cmidrule(l){7-8}
    Network & Page & FCP & Page & kB & Req. & Page & kB \\
    \midrule
    Wi-Fi & Front page & 0.11 & 0.15 & 350 & 14 & 0.16 & 0 \\
     & Town & 0.19 & 0.19 & 297 & 6 & 0.20 & 0 \\
     & 2 Love Map & 0.12 & 0.24 & 360 & 13 & 0.24 & 8 \\
     & 9 Meridian & 0.12 & 1.43 & 1522 & 13 & 1.32 & 174 \\
     & Infirmary & 0.12 & 0.16 & 545 & 9 & 0.15 & 0 \\
     & Trip & 0.12 & 0.18 & 302 & 9 & 0.18 & 0 \\
     & Request box & 0.12 & 0.15 & 302 & 8 & 0.15 & 0 \\
    \addlinespace
    4G & Front page & 0.48 & 0.67 & 350 & 14 & 0.15 & 0 \\
     & Town & 0.54 & 0.54 & 297 & 7 & 0.20 & 0 \\
     & 2 Love Map & 0.49 & 0.81 & 360 & 13 & 0.39 & 8 \\
     & 9 Meridian & 0.48 & 2.04 & 1522 & 13 & 1.51 & 174 \\
     & Infirmary & 0.48 & 0.68 & 545 & 9 & 0.14 & 0 \\
     & Trip & 0.48 & 0.69 & 302 & 9 & 0.18 & 0 \\
     & Request box & 0.50 & 0.68 & 302 & 8 & 0.15 & 0 \\
    \addlinespace
    Slow 4G & Front page & 0.68 & 0.87 & 350 & 14 & 0.16 & 0 \\
     & Town & 0.74 & 0.73 & 297 & 7 & 0.21 & 0 \\
     & 2 Love Map & 0.68 & 1.38 & 360 & 13 & 0.40 & 8 \\
     & 9 Meridian & 0.68 & 3.67 & 1522 & 13 & 2.25 & 174 \\
     & Infirmary & 0.68 & 0.95 & 545 & 9 & 0.15 & 0 \\
     & Trip & 0.68 & 0.90 & 302 & 9 & 0.17 & 0 \\
     & Request box & 0.68 & 0.89 & 302 & 8 & 0.14 & 0 \\
    \addlinespace
    3G & Front page & 2.00 & 2.48 & 356 & 15 & 0.15 & 0 \\
     & Town & 2.06 & 2.06 & 297 & 7 & 0.21 & 0 \\
     & 2 Love Map & 1.99 & 4.58 & 360 & 13 & 0.81 & 8 \\
     & 9 Meridian & 2.00 & 10.63 & 1554 & 14 & 5.15 & 174 \\
     & Infirmary & 2.00 & 2.89 & 545 & 9 & 0.15 & 0 \\
     & Trip & 2.02 & 2.64 & 302 & 9 & 0.18 & 0 \\
     & Request box & 2.03 & 2.71 & 302 & 8 & 0.16 & 0 \\
    \bottomrule
  \end{tabular}
\end{table}

\subsection{E2: Network Failure}
\emph{Scripts:} \code{e02\_network.py}, \code{e02b\_stalled.py}. With every request allowed,
each page contacts a font service (\code{fonts.googleapis.com} for one stylesheet and
\code{fonts.gstatic.com} for 15 to 30 font files). The Love Map also requests ten map tiles from
\code{tile.openstreetmap.org}, and Our Songs loads an embedded player from five hosts of a music
service. No other host is contacted on load. Speech recognition, which may use a remote service,
starts only on a tap and is outside this experiment.

The block list holds nine host suffixes of Google and YouTube services. It stands in for a
network we could not test from. In the \emph{refused} condition requests to those hosts fail at
once; in the \emph{stalled} condition they are held open and never answered, which is closer to
a filtered connection; in the \emph{alone} condition every request to another host is refused.
For the \emph{offline} condition one browser context visits the front page online, waits four
seconds for the precache, is switched offline, and then opens each of the 21 routes. A page
\emph{holds} when the view renders within 15 seconds, no uncaught script error occurs, no image
from the site's own origin is broken, and the view holds text or a canvas. Holding is a claim
about rendering: Our Songs holds without its music host, and no song can then play.

In the stalled condition each page was opened three times with the hosts open and three times
with them stalled. The view rendered in all 126 loads. The median time to the view changed by
between $-21$ and $+13$\,ms on 20 pages and from 1.10\,s to 1.60\,s on the ring. With the hosts
open the load event fired in all 63 loads; with them stalled it fired in none within 15 seconds,
because the font stylesheet, although requested so as not to block rendering, still delays that
event. Anything the site ties to the load event would wait on a filtered network; the service
worker's registration is one candidate that we did not test.

Offline, the Love Map and the ring fail with ``Unable to preload CSS'' for their style chunks,
which the build's precache list excludes together with their script chunks, and Our Songs throws
a reference error from the embedded player's script.

\subsection{E3: Frame Time}
\emph{Script:} \code{e03\_frames.py}. On the town page, the experiment holds each of four arrow
keys for 2.5 seconds in turn and records the interval between successive animation frames, at
CPU slow-down factors of 1, 4 and 6, three runs each; the journey list, which does not animate,
is the comparison. Medians over runs for the town: at factor 1, a median interval of 8.3\,ms and
a 99th percentile of 10.3\,ms; at factor 4, 8.3 and 9.9\,ms; at factor 6, 8.7 and 17.6\,ms, with
902 frames in ten seconds where the other conditions have about 1{,}200. No interval exceeded
33.4\,ms in any run. The browser was headless and its display rate was 120\,Hz.

\subsection{E4: Accessibility Rules}
\emph{Script:} \code{e04\_access.py}. axe-core 4.10.2 runs on each of the 21 pages in each
language with the rule tags for WCAG 2.0, 2.1 and 2.2 at levels A and AA. Five rules fail
(Table~\ref{tab:x-access}); every failure has the impact level ``serious''. Touch targets are
all visible links, buttons, form controls and focusable elements; the smaller side of each
bounding box is compared with 24 and 44 CSS pixels. The 992 targets are 496 in each language.

\begin{table}[t]
  \caption{E4: rules that fail, over 42 audits (21 pages in two languages).}
  \label{tab:x-access}
  \small
  \begin{tabular}{@{}p{0.47\columnwidth}p{0.2\columnwidth}rr@{}}
    \toprule
    Rule & Pages & Audits & Elements \\
    \midrule
    Text contrast at least 4.5:1 & 19 & 38 & 207 \\
    Scrollable region reachable by keyboard & Moments & 2 & 10 \\
    Touch target at least 24 px & Love Map & 2 & 2 \\
    No nested interactive controls & Stars & 2 & 2 \\
    Image role has a text alternative & Stars & 2 & 2 \\
    \bottomrule
  \end{tabular}
\end{table}

\begin{figure*}[t]
  \centering
  \includegraphics[width=\textwidth]{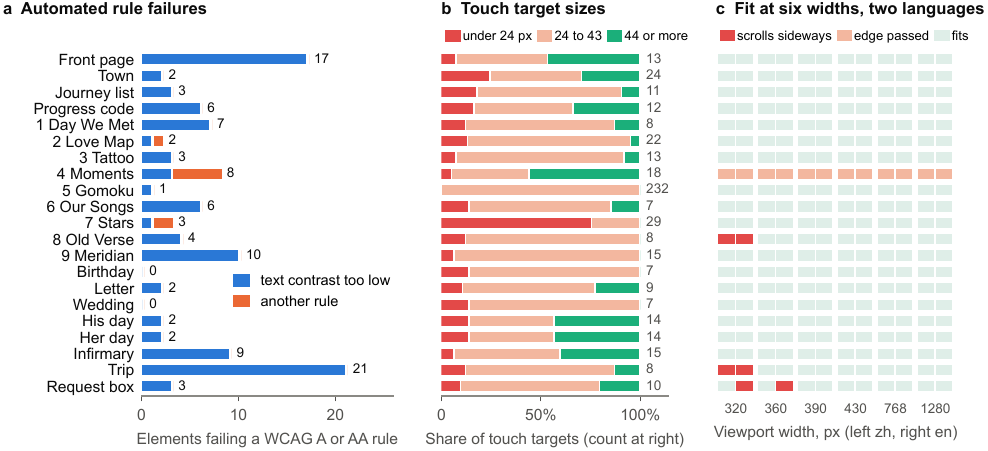}
  \caption{E4 and E5, one row for each page. (a) Elements that fail an automated WCAG A or AA
  rule in the Chinese rendering. (b) Touch targets by size; the count is at the right.
  (c) Fit at six viewport widths; each cell is split into Chinese (left) and English (right).}
  \Description{Three aligned panels with 21 rows. Left: bars of rule failures, highest on the trip page with 21 and the front page with 17, zero for the birthday and wedding. Middle: stacked bars showing most touch targets between 24 and 43 pixels, with the star level mostly under 24. Right: a grid of small tiles, nearly all pale green, red at 320 pixels for Old Verse, Trip and the request box, and pale orange across the Moments row.}
  \label{fig:access}
\end{figure*}

\subsection{E5: Layout}
\emph{Script:} \code{e05\_layout.py}. Each page is rendered at widths of 320, 360, 390, 430, 768
and 1{,}280 pixels in each language, 252 renderings. For each we record whether the document is
wider than the viewport, the number of visible elements that extend past the right edge (outside
canvases, the map and elements hidden from assistive technology), the number of text boxes with
hidden overflow whose content is wider than the box, the number of text boxes set below 12
pixels, and the page height. No text box is cut off in any rendering. Moments reports elements
past the edge at every width; they belong to a strip that scrolls sideways by design, the same
strip that E4 reports as unreachable by keyboard. Text below 12 pixels occurs on five pages (the
two city days, the two ceremonies and the trip page), at most seven boxes on a page.

\subsection{E6: Dictionary Agreement}
\emph{Script:} \code{e06\_bilingual.py}.
The shipped test checks that the two dictionaries hold the same keys and that no value is empty.
The experiment adds two checks. The structural check compares placeholders: all 522 pairs name
the same placeholders, and four pairs are identical on both sides. The English side holds 3{,}921
words in 19{,}813 characters against 7{,}625 characters of Chinese. For the reading check,
\code{qwen3:14b} and \code{mistral-small3.2:24b} each read the remaining 518 pairs, eight to a
request at temperature 0, and label each \emph{same} (the English says what the Chinese says),
\emph{shifted} (one side adds, drops or softens a detail) or \emph{different}. The first model
labels 428 same, 65 shifted and 25 different; the second labels 328, 175 and 15. Both label 305
pairs same, both withhold that label from 67, and Cohen's $\kappa$ over the three labels is 0.29.
The low agreement comes from the middle label: the second model marks an added article or an
idiom as a shift, as in \zh{回到小镇} against ``Back to the town''. Both models call six pairs
different. On inspection none of the six is an error: two set the nickname \zh{小杨} against
``Yifan'', two set a short Chinese level title against a longer English one (\zh{相遇}, ``The Day
We Met''), one sets \zh{进度同步} against ``Carry progress'', and one sets \zh{天秤与天蝎} against
``Libra and Scorpius'', the constellation and not the sign. The experiment therefore yields a
reading list of 67 pairs for a bilingual reader and no estimate of translation quality.

\subsection{E7: Progress Codes}
\emph{Script:} \code{e07\_codes.exp.ts}. A code is a player letter (\code{J} or \code{Y}), the
set of cleared levels as a base-36 number, and one check character
(Appendix~\ref{sec:system}). The experiment enumerates the 1{,}024 valid codes and applies every
instance of six kinds of damage to each: a substitution of one character by each of the 35 other
symbols; the subset of substitutions between five look-alike pairs (\code{O}/\code{0},
\code{I}/\code{1}, \code{S}/\code{5}, \code{B}/\code{8}, \code{Z}/\code{2}); a transposition of
two unequal neighbours; a twin error, in which two equal neighbours are both replaced by another
symbol; a deletion; and an insertion of each of 36 symbols at each position. Each damaged string
is decoded under three designs. Since two of the designs do not exist in the shipped code, the
experiment uses a copy of the shipped decoder with the check function as a parameter, and
asserts for every string that the copy with the shipped check function returns what the shipped
decoder returns. A decoded result is \emph{refused}, \emph{the same} as the original, or
\emph{wrong}; a wrong result is further classified by whether it changes the player, adds levels,
removes levels, or names a set that is not a prefix of the level order.

Table~\ref{tab:x-codes} gives the counts. The four accepted transpositions are the pairs
\code{J801}/\code{J810} and \code{YE01}/\code{YE10}: in each, the swap exchanges the last data character
with the check character. The merge
laws were checked through the shipped \code{mergeProgress} on a memory store: for all
$512^2=262{,}144$ ordered pairs $(a,b)$ of level sets, merging $a$ then $b$ equals merging $b$
then $a$, equals $a\cup b$, is unchanged by merging $b$ again, and contains both $a$ and $b$;
associativity was checked on 50{,}000 pseudo-random triples. Of the valid codes, 72 have three
characters and 952 have four.

The second enumeration starts from the 20 codes that ordered play can produce: for each player
the ten prefixes of the level order, from no level to all nine. Their mutations are decoded
under three designs (Table~\ref{tab:x-reach}): the shipped decoder; the shipped decoder followed
by a test that the decoded set is a prefix; and a three-character code made of the player
letter, the number of levels cleared as one digit, and the same check character. None of the 20
shipped codes holds a doubled character, so that kind of damage has no instance there. The 24
strings the shipped decoder reads as another state are 2 deletions and 22 insertions; all 24
name a set that is not a prefix, 22 add levels the player has not cleared (63 levels in all),
and none changes the player. The count code is ours and exists only in the experiment.

\begin{table}[t]
  \caption{E7: one-edit mutations of the 20 reachable codes that are read as a state other than
  the original, under the shipped decoder, the shipped decoder with a prefix test, and a code
  that carries the number of levels.}
  \label{tab:x-reach}
  \small
  \setlength{\tabcolsep}{3pt}
  \begin{tabular}{@{}lrrrrrr@{}}
    \toprule
    & \multicolumn{3}{c}{Shipped} & Prefix & \multicolumn{2}{c}{Count code} \\
    Damage & Strings & Read & Adds & Read & Strings & Read \\
    \midrule
    One character replaced & 2,380 & 0 & 0 & 0 & 2,100 & 0 \\
    Two neighbours swapped & 48 & 0 & 0 & 0 & 39 & 0 \\
    One character deleted & 68 & 2 & 0 & 0 & 60 & 0 \\
    One character inserted & 3,168 & 22 & 22 & 0 & 2,880 & 0 \\
    A doubled character replaced & 0 & 0 & 0 & 0 & 35 & 0 \\
    \midrule
    All & 5,664 & 24 & 22 & 0 & 5,114 & 0 \\
    \bottomrule
  \end{tabular}
\end{table}

\begin{table*}[t]
  \caption{E7: damaged codes accepted with a meaning different from the original, by kind of
  damage and design of the code. Without a check character a code is one character shorter, so
  the number of damaged strings differs and is given with each count. The last column is the
  number of codes wrongly accepted by the shipped design that name a set of levels ordered play
  cannot reach.}
  \label{tab:x-codes}
  \small
  \begin{tabular}{@{}lrrrrr@{}}
    \toprule
    Damage & Damaged strings & ISO 7064 (shipped) & Weighted sum & No check character & Unreachable (shipped) \\
    \midrule
    One character replaced & 140{,}840 & 0 & 1{,}944 & 49{,}016 / 105{,}000 & 0 \\
    \quad of which look-alikes & 935 & 0 & 52 & 424 / 648 & 0 \\
    Two neighbours swapped & 2{,}943 & 4 & 82 & 370 / 1{,}948 & 4 \\
    Doubled character replaced & 1{,}995 & 21 & 26 & 340 / 980 & 20 \\
    One character deleted & 4{,}024 & 89 & 132 & 1{,}904 / 3{,}000 & 71 \\
    One character inserted & 181{,}728 & 93 & 138 & 1{,}974 / 144{,}864 & 91 \\
    \bottomrule
  \end{tabular}
\end{table*}

\subsection{E8: Secrecy Lexicon}
\emph{Script:} \code{e08\_secret.exp.ts}. The shipped test excludes the keys shown after the
last level and one allowed string, which leaves 464 of 522 keys, and applies three patterns: the
character \zh{戒}; the word \emph{ring} or \emph{rings}; and any of \zh{婚}, \zh{嫁}, \zh{娶},
\zh{求婚}, \emph{wedding}, \emph{marr}, \emph{propos}, \emph{bride}, \emph{groom}. Our lexicon
has four tiers. \emph{Named} (12 phrases) uses the words of the patterns in a phrase.
\emph{Variant} (23) uses other forms of the same words: the test misses \emph{bridal},
\emph{wed}, \emph{newlywed} and \emph{ringed}. \emph{Synonym} (46) uses other words for the same
things; the test catches only \emph{bridesmaid}, which contains \emph{bride}. \emph{Oblique} (26)
points at the ending without a wedding word; the test catches only \emph{ring finger}. Of 13
harmless strings the patterns refuse nine: \zh{戒烟}, \zh{戒备}, \zh{戒掉熬夜}, \zh{斋戒},
\emph{marrow}, \emph{a proposition}, \emph{I propose we eat}, \emph{bridegroom-free} and
\emph{Marrakesh}. Run over the 464 shipped strings, the uncaught hints match three: \zh{誓词}
in the first level's heading, and \emph{I do} twice, in ``I do not know where it leads'' and
``What can I do for you?'', which are not hints.

\subsection{E9: Test Coverage}
\emph{Script:} \code{e09\_coverage.py}. The site's own test suite runs once with v8 coverage
over the source of the site and the server, test files excluded (Table~\ref{tab:x-coverage}).
The suite has 295 tests in 47 files at this commit; the design record reports 263 in 45 at its
cut-off.

\begin{table}[t]
  \caption{E9: unit-test coverage by part of the site. Untouched: files in which no line runs.}
  \label{tab:x-coverage}
  \small
  \setlength{\tabcolsep}{4pt}
  \begin{tabular}{@{}lrrrrr@{}}
    \toprule
    Part & Files & Untouched & Lines & Line \% & Branch \% \\
    \midrule
    All & 100 & 41 & 4{,}647 & 20.3 & 23.9 \\
    Levels & 30 & 8 & 1{,}340 & 13.2 & 16.4 \\
    Later products & 21 & 10 & 987 & 40.4 & 45.1 \\
    Town & 18 & 9 & 967 & 13.4 & 11.1 \\
    Ceremonies & 11 & 11 & 938 & 0.0 & 0.0 \\
    Core & 19 & 3 & 398 & 55.8 & 40.7 \\
    Dictionary & 1 & 0 & 17 & 100.0 & 60.0 \\
    \bottomrule
  \end{tabular}
\end{table}
\subsection{E10: The Development Record}
\emph{Scripts:} \code{e10\_record.py}, \code{e10b\_coding.py}. Each commit on the public branch
is read for its author time, converted to both cities' clocks, and for the files it touches.
A commit's \emph{area} is the area that holds most of its changed lines
(Table~\ref{tab:x-corpus}). Counts differ from the design record's 152 commits because the
experiment's cut-off is nine commits later.
For the kind of change, \code{qwen3:14b} and \code{mistral-small3.2:24b} each code every commit,
given its subject line and its area, as one of seven kinds: new feature, revision, fix, content,
test, writing or upkeep. The two agree on 117 of 161 commits (73\%, Cohen's $\kappa = 0.65$): 40
new features, 31 writing, 21 revisions, 15 upkeep, 8 content and 2 fixes. Neither model uses the
kind \emph{test}. The 44 disagreements are spread thinly: the largest groups are upkeep against writing (9), new
feature against revision (7) and content against new feature (6). The counts by kind are
therefore lower bounds and not a census.

\begin{table}[t]
  \caption{E10: the 161 commits by the area that holds most of each commit's changed lines, and
  lines added or deleted in each area over the whole record.}
  \label{tab:x-corpus}
  \small
  \begin{tabular}{@{}lrr@{}}
    \toprule
    Area & Commits & Lines changed \\
    \midrule
    Documentation & 33 & 4{,}876 \\
    Levels & 30 & 6{,}542 \\
    Later products & 23 & 5{,}633 \\
    Town & 20 & 5{,}746 \\
    Paper & 19 & 15{,}035 \\
    Core & 13 & 2{,}203 \\
    Build & 12 & 10{,}724 \\
    Media & 6 & 10{,}660 \\
    Ceremonies & 3 & 1{,}915 \\
    Dictionary & 2 & 1{,}630 \\
    \bottomrule
  \end{tabular}
\end{table}

\begin{figure*}[t]
  \centering
  \includegraphics[width=\textwidth]{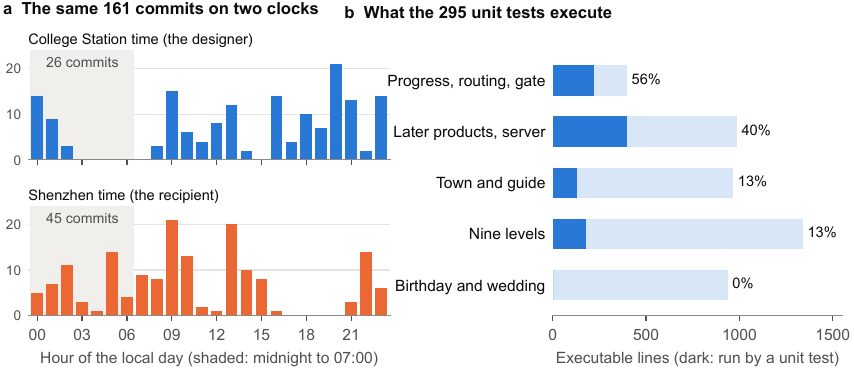}
  \caption{E9 and E10. (a) E10: the hour of every commit on the designer's clock and on the
  recipient's; the shaded band is midnight to 07:00. (b) E9: executable lines in each part of
  the site, and the share that a unit test runs.}
  \Description{Left: two histograms of 161 commits by hour of day. In Texas time commits cluster from morning to past midnight with none from 03:00 to 07:00. In Shenzhen time the same commits fall mostly between 22:00 and 16:00, with none from 17:00 to 20:00. Right: five horizontal bars of lines of code with a dark covered part: 56 per cent of core code, 40 per cent of later products, 13 per cent of the town and of the levels, zero for the ceremonies.}
  \label{fig:record}
\end{figure*}

\subsection{E11: Server Properties}
\emph{Script:} \code{e11\_server.exp.ts}. Each of 4{,}000 generated requests draws an origin
from nine values (the two allowed origins, look-alikes such as a longer host name that begins
with the site's, an upper-case form, the string \code{null}, and none), a method, a path, a text
from eleven values (empty, blank, long, with line breaks and mail headers, with control
characters), a player name from eight values, and up to thirteen files whose names come from
sixteen hostile and ordinary values and whose sizes fall around the limits of ten files and
18\,MiB. One request in 25 meets a mail binding that throws. The properties are: (P1) mail is sent
exactly when the origin is allowed, the method is POST, the path is \code{/send}, the request is
not empty and it is within both limits; (P2) a request from another origin is answered 403 with
no cross-origin header; (P3) the recipient and sender of every sent message are the configured
ones; (P4) the subject is one line without control characters that begins with the site's tag
and a known name; (P5) every attachment name is non-empty and has no path separator, quote or
control character. The expected outcome in P1 is computed from the files that survive multipart
encoding, since an empty file with an empty name does not. P1, P2, P3 and P5 hold on all 4{,}000
requests. P4 fails on 147 of 1{,}629 sent messages, all with the same cause: the handler
collapses white space in the text and passes other control characters, here NUL and SOH,
through to the subject. No subject contains a carriage return or line feed.

\subsection{E12: Triage}
\emph{Scripts:} \code{e12\_dataset.py}, \code{e12\_triage.py}.
The script replays the call in \code{mailflow/triage.py}: the same instruction, the public guide
from \code{mailflow/config.py}, the same JSON schema, temperature 0, and the same layout, in
which the email is followed by an attachment section that is empty when no file is sent. It
imports nothing from the pipeline and reads no private configuration, so the results describe
the published pipeline and a stand-in scope, not the configuration in use. The 120 requests are
written for the experiment around four public-domain books, each with its own matching excerpt.
A \emph{new} request asks for a note on an excerpt; a \emph{revision} asks for a change to an
earlier note; \emph{out of scope} asks for something else, such as a train ticket;
\emph{other recipient} adds an address to send the result to; \emph{command or file} adds a
shell command or a file to read; \emph{hidden} places an instruction addressed to the model
inside the excerpt. Five conditions vary where the excerpt sits, what the guide says and how
the attachment section is written: in
the email body with the public guide (all five models); as a transcribed screenshot in the
attachment section (the deployed model and \code{qwen2.5:14b}, on the four classes that carry an
excerpt); in the body with a guide that names only the scope (the deployed model); and, for
the deployed model, two paired controls in which the same body-text requests are sent with the
attachment section left out, or with the section holding the words for ``no attachment''. That
is 1{,}120 decisions, none of which failed to parse. Table~\ref{tab:x-triage} gives the body
conditions. With the excerpt as a screenshot the deployed model sends 19, 20, 20 and 16 of 20
in the classes new, other recipient, command and hidden, and \code{qwen2.5:14b} sends 0, 18, 18
and 14. The deployed model gives an empty or missing attachment as its reason for all 18 new
requests it sends in the body condition; the two it passes it labels as revisions. In the
paired controls it sends 18 of 20 new requests each time. Request by request, 114 of 120
decisions are the same with the section left out (five escalations become passes and one pass
becomes an escalation) and 117 of 120 with the section marked as empty (three escalations
become passes). With the section left out it also passes one request that names another
recipient. The hidden
instructions that pass are judged to be within scope, without mention of the instruction. Each
model ran once.

\begin{table}[t]
  \caption{E12: requests sent to a person, of 20 in each class, with the excerpt in the email
  body. The first two classes should be answered and the last four should go to a person. The
  last column is the median time for one decision.}
  \label{tab:x-triage}
  \small
  \setlength{\tabcolsep}{2pt}
  \begin{tabular}{@{}lrrrrrrr@{}}
    \toprule
    Model & New & Rev. & Out & Rec. & Cmd. & Hid. & s \\
    \midrule
qwen3 14b & 18/20 & 3/20 & 20/20 & 20/20 & 20/20 & 18/20 & 7.0 \\
qwen2.5 14b & 0/20 & 0/20 & 20/20 & 18/20 & 20/20 & 12/20 & 6.9 \\
phi4 14b & 0/20 & 0/20 & 20/20 & 16/20 & 16/20 & 8/20 & 11.4 \\
mistral-small3.2 24b & 10/20 & 3/20 & 20/20 & 20/20 & 20/20 & 12/20 & 12.2 \\
llama3.1 8b & 0/20 & 5/20 & 20/20 & 1/20 & 6/20 & 0/20 & 3.2 \\
qwen3 14b (short guide) & 20/20 & 20/20 & 20/20 & 20/20 & 20/20 & 18/20 & 7.3 \\
qwen3 14b (no section) & 18/20 & 0/20 & 20/20 & 19/20 & 20/20 & 18/20 & 7.4 \\
qwen3 14b (``none'') & 18/20 & 0/20 & 20/20 & 20/20 & 20/20 & 18/20 & 7.5 \\
    \bottomrule
  \end{tabular}
\end{table}

\section{Changes in Version 3}\label{sec:v3}
The second version was dated 24 November 2026; this one is dated 30 November 2026.
\begin{itemize}[leftmargin=*,itemsep=1pt,topsep=2pt]
  \item New: an evaluation of twelve machine-run experiments (Section~\ref{sec:eval};
  Appendix~\ref{sec:expdetail}), with Tables~\ref{tab:claims}, \ref{tab:holds} and~\ref{tab:experiments}
  to~\ref{tab:x-triage}, Figures~\ref{fig:overview} to~\ref{fig:answer}, \ref{fig:access}
  and~\ref{fig:record}, and the scripts and
  results in \code{paper/experiments/}.
  \item New: a checking procedure followed in this one case and presented as such, the
  revisions of the design account that the checks require, and four design implications
  (Section~\ref{sec:disc}).
  \item Restructured: a main body of about ten pages. The second version's sections follow as
  appendices; their text is unchanged apart from the corrections listed next, and their diagrams
  are redrawn.
  \item Corrected: the abstract no longer says that the paper contains no evaluation; it says
  that the recipient's experience remains open. The statement in the source that the check
  character catches every swap of two neighbours is corrected by E7, which finds four exceptions.
  \item Corrected against the repository: the letter and the wedding open for a player who has
  cleared the last level, as they have since version 1.18, not only when both players have; the
  four accepted transpositions swap the last mask digit with the check character, not two mask
  digits; the example code for Yifan's first seven levels is \code{Y3J9}, not \code{Y3J}; the
  corpus holds five design specifications, not four; the line counts at version 1.24.0 are
  9{,}895 for TypeScript, 1{,}276 for CSS and 621 for the pipeline's Python; and the time
  difference was fourteen hours, not thirteen, when the request box was built.
  \item Unchanged: the account of the town, its levels, the later products and the systems,
  except as corrected above; the seven design episodes; the absence of any report from the
  recipient.
\end{itemize}

\end{document}